\documentclass[useAMS,usenatbib]{mnras}
\usepackage{graphicx}
\usepackage{amsmath, amssymb}
\usepackage[T1]{fontenc}
\usepackage[dvipsnames]{xcolor}
\usepackage{bm}
\usepackage[capitalize]{cleveref}
\usepackage{acro}
\usepackage{caption}
\usepackage{subcaption}
\usepackage{booktabs}
\usepackage{colortbl}
\usepackage{arydshln}
\usepackage{makecell}
\usepackage{multirow}
\usepackage{hyperref}
\usepackage{orcidlink}
\usepackage{siunitx}

\hypersetup{breaklinks=true}

\newcommand{\kmsecMpc}{\ensuremath{\mathrm{km}\,\mathrm{s}^{-1}\,\mathrm{Mpc}^{-1}}}
\newcommand{\kpc}{\mathrm{kpc}}
\newcommand{\mas}{\mathrm{mas}}
\newcommand{\uas}{\ensuremath{\mu\mathrm{as}}}
\newcommand{\magn}{\mathrm{mag}}
\newcommand{\dex}{\mathrm{dex}}

\newcommand{\mWH}{m^W_H}
\newcommand{\MWH}{M^W_{H,1}}
\newcommand{\AH}{A_H}
\newcommand{\bW}{b_W}
\newcommand{\ZW}{Z_W}
\newcommand{\pihat}{\hat{\varpi}}
\newcommand{\piobs}{\varpi_{\rm obs}}
\newcommand{\deltapi}{\delta_\varpi}
\newcommand{\dmin}{d_{\min}}
\newcommand{\dmax}{d_{\max}}

\makeatletter
\newcommand*\rel@kern[1]{\kern#1\dimexpr\macc@kerna}
\newcommand*\widebar[1]{%
  \begingroup
  \def\mathaccent##1##2{%
    \rel@kern{0.8}%
    \overline{\rel@kern{-0.8}\macc@nucleus\rel@kern{0.2}}%
    \rel@kern{-0.2}%
  }%
  \macc@depth\@ne
  \let\math@bgroup\@empty \let\math@egroup\macc@set@skewchar
  \mathsurround\z@ \frozen@everymath{\mathgroup\macc@group\relax}%
  \macc@set@skewchar\relax
  \let\mathaccentV\macc@nested@a
  \macc@nested@a\relax111{#1}%
  \endgroup
}
\makeatother

\DeclareAcronym{CMB}{short = CMB, long  = cosmic microwave background}
\DeclareAcronym{LOS}{short = LOS, long  = line-of-sight}
\DeclareAcronym{HST}{short = {\textit{HST}}, long = Hubble Space Telescope}
\DeclareAcronym{MW}{short = MW, long = Milky Way}
\DeclareAcronym{C22}{short = C22, long = Cycle~22}
\DeclareAcronym{C27}{short = C27, long = Cycle~27}
\DeclareAcronym{LMC}{short = LMC, long = Large Magellanic Cloud}
\DeclareAcronym{CDF}{short = CDF, long = cumulative distribution function}
\DeclareAcronym{BBC}{short = BBC, long = BEAMS with Bias Corrections}
\DeclareAcronym{PPC}{short = PPC, long = posterior predictive check}
\DeclareAcronym{SBI}{short = SBI, long = simulation-based inference}

\title[Forward-modelling Milky Way Cepheids]{Forward-modelling Milky Way Cepheids: selection effects and physical priors in the \textit{Gaia}--\textit{HST} calibration}
\author[R. Stiskalek et al.]{Richard Stiskalek$^{1}$\orcidlink{0000-0002-0986-314X}\thanks{\href{mailto:richard.stiskalek@physics.ox.ac.uk}{richard.stiskalek@physics.ox.ac.uk}},
Adam G. Riess$^{2,3}$\orcidlink{0000-0002-6124-1196},
Harry Desmond$^{4}$\orcidlink{0000-0003-0685-9791},
Guilhem Lavaux$^{5}$\orcidlink{0000-0003-0143-8891}\newauthor
and Dan Scolnic$^{6}$\orcidlink{0000-0002-4934-5849}
\\
$^{1}$Astrophysics, University of Oxford, Denys Wilkinson Building, Keble Road, Oxford, OX1 3RH, UK\\
$^{2}$Space Telescope Science Institute, 3700 San Martin Drive, Baltimore, MD 21218, USA\\
$^{3}$Department of Physics and Astronomy, Johns Hopkins University, Baltimore, MD 21218, USA\\
$^{4}$Institute of Cosmology \& Gravitation, University of Portsmouth, Dennis Sciama Building, Portsmouth, PO1 3FX, UK\\
$^{5}$CNRS \& Sorbonne Universit\'{e}, UMR 7095, Institut d'Astrophysique de Paris, 98 bis boulevard Arago, F-75014 Paris, France\\
$^{6}$Department of Physics, Duke University, Durham, NC 27708, USA\\
}

\date{Accepted XXX. Received YYY; in original form ZZZ}
\pubyear{2025}

\defcitealias{Riess_2021}{R21}
\defcitealias{Riess_2022}{R22}
\defcitealias{Hogas_2026}{HM26}

\begin{document}\label{firstpage}
\pagerange{\pageref{firstpage}--\pageref{lastpage}}
\maketitle

\begin{abstract}
The advent of high-precision \textit{Gaia} parallaxes for Milky Way Cepheids enables per cent-level calibration of the local distance ladder and the Hubble constant $H_0$.
We revisit the Milky Way Cepheid calibration from \textit{Gaia} EDR3 parallaxes using a fully forward-modelled Bayesian framework that simultaneously infers the period--luminosity relation, the \textit{Gaia} parallax zero-point offset, and individual stellar distances while explicitly incorporating the disc geometry of the Galaxy
through the distance prior and the selection functions specified in two \textit{HST} SH0ES campaigns.
We derive an analytic treatment of the detection probability that accounts for magnitude, parallax, period, and extinction cuts and reduces it to a tractable integral over distance and sky position.
Posterior predictive checks show that this generative model matches the observed distributions of parallaxes, magnitudes, and periods.
Modelling Galactic structure and survey truncation self-consistently in a Bayesian framework yields period--luminosity parameters that agree with the SH0ES maximum-likelihood values at the ${<}0.5\,\sigma$ level, a consequence of the small intrinsic scatter of the Cepheid period--luminosity relation.
Adopting the uniform-in-volume prior recently advocated by~\citet{Hogas_2026}, without simultaneously accounting for selection, leads to a ${\sim}\,0.05~\magn$ bias in the period--luminosity zero-point and posterior predictive distributions incompatible with the observed data; this shift is mostly driven by the omission of the selection model, and produces an apparent and unjustified shift in $H_0$ that reflects this mismodelling.
A consistent Bayesian treatment of Galactic structure and selection effects reinforces the local distance-ladder determination of $H_0$, and hence the Hubble tension with early-Universe inferences.
\end{abstract}



\section{Introduction}\label{sec:intro}

The ${\sim}\,5\sigma$ discrepancy between local distance-ladder and \ac{CMB}-calibrated inferences of the Hubble constant $H_0$, known as the Hubble tension, is one of the most pressing problems in cosmology~\citep[e.g.][]{Riess_2022, Planck_2020, Louis_2025, Camphuis_2025, Freedman_2025, DiValentino_2025}.
While there are many routes to local $H_0$, which contribute to the overall tension (see~\citealt{H0DN_2025} for their covariance-weighted combination), the sharpest local constraint comes from the specific combination of \textit{Gaia} EDR3 parallaxes, Cepheids, and Type~Ia supernovae.
This motivates a careful examination of the statistical assumptions entering the \ac{MW} Cepheid calibration.

Here we examine the \ac{MW} Cepheid calibration at the level of individual stars.
We construct the first Bayesian forward model of the \ac{MW} Cepheid population that explicitly accounts for the sample selection function, incorporates the thin-disc geometry of the Galaxy as a distance prior, and infers the intrinsic population properties jointly with the period--luminosity relation and the \textit{Gaia} parallax zero-point offset.
In particular, we show that the reduced Hubble tension reported by~\citet[hereafter \citetalias{Hogas_2026}]{Hogas_2026}, who adopted a uniform-in-volume distance prior without selection modelling, is an artefact of that omission.
We show that when Galactic disc geometry and survey truncation are modelled consistently, the inferred period--luminosity parameters agree well with the maximum-likelihood values obtained by SH0ES from their parallax-space regression --- a consequence, we show, of the small intrinsic scatter of the Cepheid period--luminosity relation.
Adopting a uniform-in-volume prior without selection modelling induces a bias.
The prior describes the characteristics of the underlying population while the selection function determines which of those objects enter the sample; ignoring the latter while specifying the former generates mock populations inconsistent with the observed sample.

The Cepheid distance scale has been extensively tested against potential astrophysical systematics.
JWST observations have ruled out unrecognised photometric crowding in extragalactic hosts as a significant contributor to the tension, with analyses of more than \num{1000} Cepheids in N4258 and up to 19 SN~Ia hosts finding no evidence for bias in \ac{HST} photometry~\citep{Riess_2024, Riess_2025}.
A second key uncertainty for the parallax-based Cepheid calibration is the \textit{Gaia} parallax zero-point offset, $\deltapi$.
While the EDR3 calibration models $\deltapi$ as a function of magnitude, colour, and ecliptic latitude~\citep{Lindegren_2021}, it remains weakly constrained for bright stars like \ac{MW} Cepheids ($G < 9~\magn$), a brightness range where EDR3 lacks calibration sources.
The SH0ES analysis~\citep[hereafter \citetalias{Riess_2021}]{Riess_2021} combined high- and low-parallax samples of Cepheids with \ac{HST} photometry to jointly measure the offset (which is additive) and the Cepheid luminosity (which is multiplicative), yielding $\deltapi = -14 \pm 6~\uas$.
Independent validations using asteroseismology, photometric parallaxes, red clump stars, eclipsing binaries, and orbital parallaxes generally favour residual offsets between $0$ and $-20~\uas$, suggesting that, for bright stars, the EDR3 parallax-bias recipe of~\citet{Lindegren_2021} (the $Z_5/Z_6$ correction evaluated from source magnitude, colour or effective wavenumber, ecliptic latitude, and astrometric-solution type) may overcorrect the zero-point by ${\sim}\,15~\uas$~\citep{Zinn_2021, Groenewegen_2021, Huang_2021, Fabricius_2021, Bhardwaj_2021, Stassun_2021, Groenewegen_2023, Ding_2025}.
An alternative route to Cepheid calibration that largely avoids the uncertainty of the parallax offset calibrates Cepheids in open clusters where the \textit{Gaia} DR3 parallaxes are measured from fainter cluster stars ($12 < G < 18~\magn$), well within the range of the~\citet{Lindegren_2021} calibration, and achieves a 0.9 per cent luminosity-scale precision consistent with that of the field Cepheids~\citep{Riess_2022a,CruzReyes_2023}.
The metallicity dependence of the period--luminosity relation has also been explored extensively~\citep{Breuval_2022, Molinaro_2023}, although its magnitude remains debated, with recent claims of a negligible effect~\citep{Madore_2025, Madore_2026} challenged by evidence favouring the standard slope of ${\sim}-0.2 \pm 0.1~\magn\,\dex^{-1}$~\citep{Breuval_2025}.
However, this term has little leverage on $H_0$, with a change of ${\sim}\,0.2~\kmsecMpc$ for a $0.1~\magn\,\dex^{-1}$ change in slope due to the consistency of Cepheid metallicities along the distance ladder.

While many potential astrophysical systematics have been investigated, increasing attention has also focused on the statistical framework used to infer the Cepheid calibration and the treatment of selection effects.
Several studies have explored the sensitivity of the distance ladder to modelling assumptions.
For example,~\citet{Efstathiou_2021} examined alternative implementations of the SH0ES analysis and discussed the impact of specific modelling choices;~\citet{Kushnir_2025} removed the requirement of a Cepheid period--luminosity relation;~\citet{Bidenko_2023} refit using a Gaussian process;~\citet{Mortsell_2022} explored colour selection for Cepheids; see~\citet{Verde_2024} for a review.
Subsequent analyses by the SH0ES team~\citep[hereafter \citetalias{Riess_2022}]{Riess_2022} have addressed these issues in updated treatments.
Collectively, this body of work underscores that per cent-level inference in the distance ladder requires careful and transparent treatment of statistical assumptions.
On the Bayesian side,~\citet{Cardona_2017} marginalised over hyperparameters that down-weight outliers,~\citet{Delgado_2019} constructed a hierarchical model for period--luminosity relations from \textit{Gaia} parallaxes, and~\citet{Feeney_2018} built a hierarchical model of the full distance ladder, inferring $H_0$ end-to-end but adopting a uniform prior on distance modulus (thus failing to capture the intrinsically uniform-in-volume distribution of galaxies) and neglecting the selection function.
None of these Bayesian approaches, however, incorporate a principled treatment of sample selection.
\citet{Stiskalek_2026} applied rigorous selection modelling to the geometric-anchor and Cepheid-host rungs of the distance ladder, but compressed the \ac{MW} calibration into a single Gaussian constraint on the period--luminosity zero-point.
\citet{Desmond_2025} showed, in a toy model and directly on the CosmicFlows-4 dataset~\citep{Tully_2023}, that neglecting selection within a Bayesian framework can significantly bias the inference of $H_0$, and that the distance prior must be correspondingly restricted even for nominally volume-limited samples---a toy case in which selection is a sharp cutoff in true distance, possible only in simulations.
For the Cepheid samples below, selection is instead imposed on observed quantities such as magnitude or period, so the analogous correction is the selection normalisation that marginalises over both those observables and the latent Cepheid properties.

We describe the \ac{MW}, \ac{LMC}, and N4258 Cepheid data in~\cref{sec:data}, and develop the Bayesian forward model in~\cref{sec:methods}.
\Cref{sec:results} presents the baseline posteriors and model validation, while \cref{sec:discussion} discusses the implications for $H_0$. \Cref{sec:conclusion} concludes.
Throughout, $\mathcal{N}(x \mid \mu,\, \sigma^2)$ denotes the univariate normal density with mean $\mu$ and variance $\sigma^2$, generalised to $\mathcal{N}(\bm{x} \mid \bm{\mu},\, \bm{\Sigma})$ for the multivariate case with covariance matrix $\bm{\Sigma}$; $x \hookleftarrow \mathcal{N}(\mu,\, \sigma^2)$ indicates that $x$ is drawn from the corresponding distribution; $\Phi$ is the standard normal \ac{CDF}; and all logarithms are base-$10$.


\section{Geometric and Cepheid data}\label{sec:data}

We use the \ac{MW} Cepheids compiled by~\citetalias{Riess_2021}, with \ac{HST} photometry in the F555W, F814W, and F160W bands and \textit{Gaia} EDR3 parallaxes~\citep{GaiaEDR3_2021}.
Of the $75$ Cepheids in the~\citetalias{Riess_2021} catalogue, seven lack \textit{Gaia} EDR3 parallaxes that pass the RUWE or GOF quality cuts (CY~Aur, DL~Cas, RW~Cam, SV~Per, SY~Nor, RX~Cam, and U~Aql) and are excluded.
We further exclude S~Vul and SV~Vul, which were flagged as possible outliers in~\citetalias{Riess_2021}. Both exhibit significant secular period evolution, and as the longest-period, most massive Cepheids in the sample, their large angular diameters may also cause a shifting \textit{Gaia} photocentre.
This leaves $66$ stars in the \ac{MW} sample, the same sample analysed by~\citetalias{Riess_2021} and~\citetalias{Hogas_2026}.
Magnitudes are standardised via the reddening-free Wesenheit magnitude $\mWH$~\citep{Madore_1982}.

\begin{figure*}
    \centering
    \includegraphics[width=\textwidth]{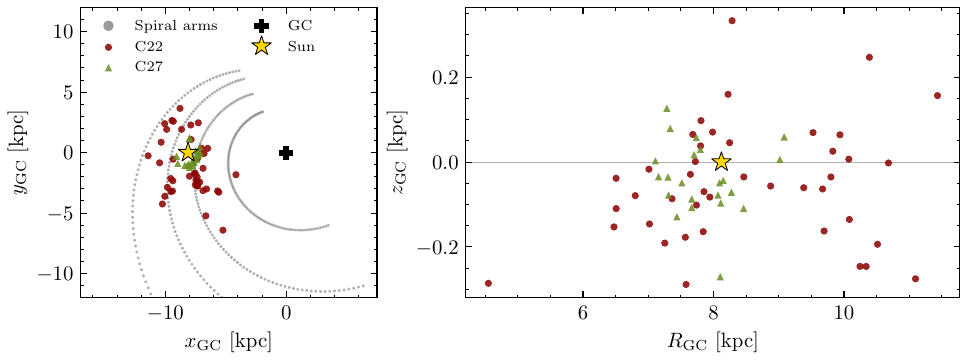}
    \caption{Galactocentric distribution of the \ac{MW} Cepheids (\ac{C22}: dark red circles; \ac{C27}: olive triangles), with distances from parallax inversion assuming $\deltapi = -0.01~\mas$.
    \textit{Left}: face-on projection with~\citet{Drimmel_2025} spiral arm traces (grey); the Sun and Galactic Centre (GC) positions are marked. The sample spans $R_{\rm GC} \approx 7$--$11~\kpc$, concentrated in the solar neighbourhood.
    \textit{Right}: edge-on view, confirming that the Cepheids trace the Galactic thin disc with height above the midplane $|z_{\rm GC}| \lesssim 0.2~\kpc$.}
    \label{fig:cepheid_spatial}
\end{figure*}

The Cepheids were observed across two \ac{HST} campaigns with distinct selection criteria given in~\citetalias{Riess_2021}: \ac{C22} targeted long-period Cepheids ($P > 8~{\rm days}$) within a distance range of ${\sim}\,6.6~\kpc$, while \ac{C27} targeted nearby Cepheids at distances ${<}\,1.25~\kpc$, with $V > 6~\magn$ to avoid \textit{Gaia} saturation.
\ac{C27} was undertaken after the discovery of the \textit{Gaia} parallax offset anomaly and the subsequent need for a broader range of Cepheid parallaxes to break degeneracies.
\Cref{fig:cepheid_spatial} shows the Galactocentric distribution of the combined sample, confirming that it traces the thin disc.
Later in the analysis, we include additional Cepheid samples with geometric constraints: the \ac{LMC} and N4258, taken from the SH0ES compilation of~\citetalias{Riess_2022}, as well as their host galaxy geometric distance measurements~\citep{Pietrzynski_2019, Reid_2019}.
\textit{Gaia} parallaxes are subject to a global zero-point offset~\citep{Lindegren_2021}, which is a function of source magnitude, colour, and angular coordinate. $\deltapi$ denotes the \textit{residual} parallax offset, i.e.\ that which may not be fully corrected by~\citet{Lindegren_2021}, who refer to an uncertainty in their correction of several \uas. For the regime of interest for Cepheids, the bright end of the~\citet{Lindegren_2021} calibrated range, we estimate the uncertainty in the correction to be $\sigma \sim 10~\uas$. The reported parallax is $\piobs = 1/d - \deltapi$, where $d$ is the distance; we treat $\deltapi$ as a single residual offset shared by all \ac{MW} Cepheids and infer it as a free parameter throughout.

For each \ac{MW} Cepheid, the~\citetalias{Riess_2021} catalogue provides the Wesenheit apparent magnitude $\mWH$ with measurement uncertainty $\sigma_m$, the pulsation period $P$, a direct spectroscopic metallicity $[{\rm Fe/H}]$, and the \textit{Gaia} EDR3 parallax $\piobs$ with uncertainty $\sigma_\varpi$.
We adopt the same $[{\rm Fe/H}]$ values as~\citetalias{Riess_2021} and~\citetalias{Hogas_2026} for consistency, but note the modest revisions reported by~\citet{Bhardwaj_2023}. In~\cref{tab:observables} we summarise the observables that enter our model.
The Wesenheit magnitude is constructed from the F555W, F814W, and F160W \ac{HST} bands to cancel reddening to first order under the assumed extinction law~\citep{Madore_1982}, so the per-star apparent-magnitude model carries no distance-dependent extinction term $\AH$ that would otherwise need to be inferred or read from a dust map.
For the \ac{C22}\,+\,\ac{C27} sample used here, the per-star Wesenheit colour term is $F160W - \mWH = R\,(F555W - F814W) = 0.50 \pm 0.13~\magn$, combining intrinsic Cepheid colour and reddening, with the intrinsic colour expected to dominate.
The Wesenheit construction suppresses this scatter.
Were it not applied, a parallax-based recalibration of the period--luminosity zero-point would absorb only the mean reddening, leaving the $0.13~\magn$ as a per-star residual in distance modulus.
Because $\mWH$ is computed from the per-star photometry, the correction does not couple to the adopted distance prior; the only such coupling enters via the optional \ac{C22} extinction selection cut $\AH < 0.4~\magn$, which is evaluated from the three-dimensional dust map at the latent distance and propagated jointly with the disc distance prior in the selection-normalisation integral~(Eq.~\ref{eq:prob_detection};~\cref{sec:discussion_limitations}).
Periods are determined from decades of high-cadence monitoring and are effectively noise-free, especially in the \ac{MW} and \ac{LMC}, with uncertainties at the 1--2 per cent level in $\log P$ for more distant hosts.
Metallicity uncertainties are dominated by systematic methodological differences rather than statistical errors, with a typical scatter of ${\sim}\,0.06~\dex$.

The \ac{C22} parent population comprises known \ac{MW} Cepheids drawn from historical variability catalogues such as the General Catalogue of Variable Stars~\citep{Samus_2017}, OGLE~\citep{Udalski_1992}, and~\citet{Tammann_2003}, which impose an implicit completeness limit of $V \lesssim 15~\magn$.
From this parent sample, three explicit cuts were applied for \ac{HST} observation by SH0ES: a period cut $P > 8~{\rm days}$, an extinction cut $\AH < 0.4~\magn$, and a saturation cut $V > 6~\magn$ to avoid \textit{Gaia} detector saturation.
\Cref{tab:data_summary} summarises the four datasets.
The period cut is distance-independent, whereas the extinction and brightness cuts are distance-dependent.
Of the ${\sim}\,70$ Cepheids in~\citet{Tammann_2003} satisfying these criteria, \ac{HST} snapshot scheduling randomly selected $50$ for observation.
We assume this scheduling to be equivalent to random subsampling, which therefore requires no additional modelling.
The $V > 6~\magn$ cut excludes only the nearest Cepheids; only a single object (T~Mon, $V \approx 6.07~\magn$) lies near this boundary.
The extinction cut $\AH < 0.4~\magn$ preferentially removes distant Cepheids along dusty sightlines.
To evaluate this extinction cut at arbitrary positions in the \ac{MW}, we use a combination of the \texttt{Bayestar19} three-dimensional dust map~\citep{Green_2019}, which covers declinations $\delta > -30^\circ$, and the~\citet{Marshall_2006} map for the inner \ac{MW} where \texttt{Bayestar19} lacks coverage.\footnote{Dust maps are queried using the \texttt{dustmaps} Python package~\citep{Green_2018_dustmaps}.}
\texttt{Bayestar19} reddening is converted to $\AH$ using $R_H = 0.469$~\citep{Green_2019}, while~\citet{Marshall_2006} $A_{K_s}$ values are converted via $\AH = 1.55\, A_{K_s}$.
The formal implementation of this cut within the selection function is described in~\cref{sec:selection_model}.

\ac{C27} targets nearby, high-parallax Cepheids to complement the more distant \ac{C22} sample, selecting those with photometric parallax $\varpi_{\rm phot} > 0.8~\mas$ and $\AH < 0.6~\magn$.
The photometric parallax used for this pre-\ac{HST} target selection was computed from pre-existing photometry, periods, and metallicities,
\begin{equation}\label{eq:pi_pred}
\begin{split}
    \log \frac{\varpi_{\rm phot}}{1~\mas} = -0.2\bigg[&\mWH - \tilde{M}^W_{H,1} - \tilde{b}_W \left(\log \frac{P}{1~{\rm day}} - 1\right) \\
    &- \tilde{Z}_W\,[{\rm Fe/H}] - 10\bigg],
\end{split}
\end{equation}
where $\mWH$ here denotes the pre-\ac{HST} estimate from existing photometry and $\tilde{M}^W_{H,1}$, $\tilde{b}_W$, and $\tilde{Z}_W$ are the period--luminosity zero-point, slope, and metallicity coefficient from~\citet{Riess_2016}.
This cut was applied to the predicted photometric parallax rather than the \textit{Gaia} trigonometric parallax because the \ac{C27} proposal was written before \textit{Gaia} DR3; the DR2 parallax uncertainties were roughly twice as large, making the photometric estimate more precise.
However, we shall later treat this selection as a smooth lower limit on $\piobs$.

\begin{figure*}
    \centering
    \includegraphics[width=\textwidth]{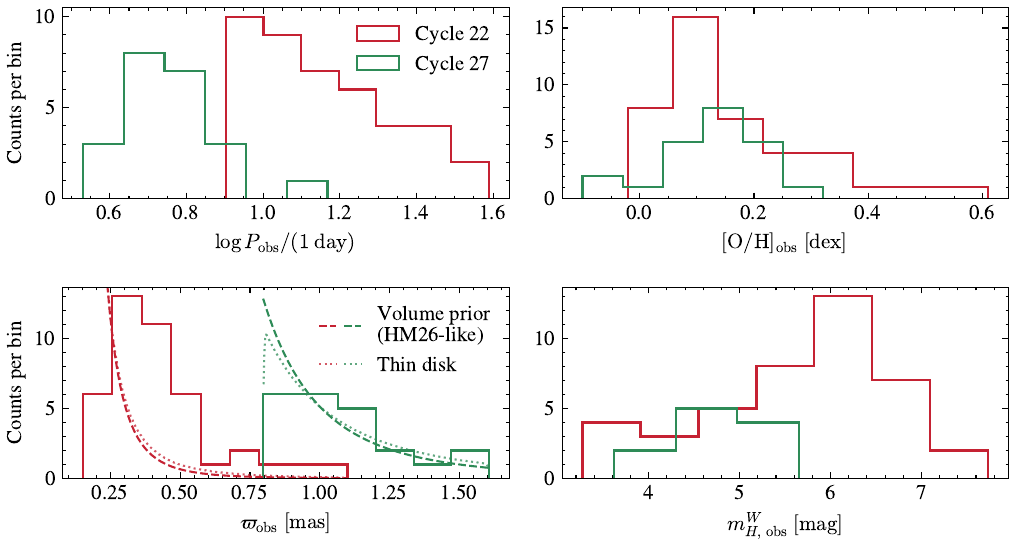}
    \caption{Distributions of observed period ($\log P_{\rm obs}$), metallicity ($[{\rm O/H}]_{\rm obs}$), \textit{Gaia} EDR3 parallax ($\piobs$), and Wesenheit apparent magnitude ($m^W_{H,{\rm obs}}$) for the \ac{C22} (red) and \ac{C27} (green) Cepheid samples.
    Dashed lines in the parallax panel show the expected counts per bin from the uniform-in-volume prior adopted by~\protect\citetalias{Hogas_2026}, $p(\piobs) \propto (\piobs+\deltapi)^{-4}$, assuming $\deltapi = -0.01~\mas$; each curve is normalised over the observed parallax range of that campaign and then multiplied by the sample size and parallax bin width.
    Dotted lines show the corresponding unselected prediction from the thin-disc prior of~\cref{eq:disc_prior}, normalised in the same way.
    \protect\citetalias{Hogas_2026} assumed the uniform-in-volume prior without selection modelling, yielding a misspecified generative model for the observed Cepheid samples.}
    \label{fig:data_distributions}
\end{figure*}

As an additional constraint on the \textit{Gaia} parallax zero-point and the period--luminosity relation, we include two galaxies with independent geometric distance measurements. \citet{Pietrzynski_2019} measured a distance modulus to the \ac{LMC} of $18.477 \pm 0.026~\magn$ from detached eclipsing binaries, while~\citet{Reid_2019} measured a distance modulus to N4258 of $29.398 \pm 0.032~\magn$ from water megamaser kinematics.
Both are adopted as geometric anchors for the Cepheid calibration, following the SH0ES programme~\citepalias{Riess_2022}.
We use $69$ \ac{LMC} and $443$ N4258 Cepheids with \ac{HST} photometry, observed using the same three-filter system as the \ac{MW} sample, enabling a consistent calibration of the period--luminosity relation across these distance rungs.
The~\citetalias{Riess_2022} catalogue provides \ac{LMC} and N4258 Cepheid Wesenheit apparent magnitudes together with per-galaxy covariance matrices $\bm{\Sigma}_j$, where $j \in \{\mathrm{LMC},\, {\rm N4258}\}$ labels the anchor galaxy throughout, that account for crowding-induced photometric biases from sky background estimation and systematic uncertainties in the metallicity scale~(section~2.1 of~\citetalias{Riess_2022}).
For \ac{MW} Cepheid metallicities we convert $[{\rm Fe/H}]$ to $[{\rm O/H}] = [{\rm Fe/H}] + 0.06$, following the~\citetalias{Riess_2022} prescription based on~\citet{Romaniello_2022}; the \ac{LMC} and N4258 metallicities are already reported as $[{\rm O/H}]$.
The \ac{LMC} sample uses a single metallicity value in accord with the finding by~\citet{Romaniello_2022} that all of the SH0ES Cepheids were consistent with the mean spectroscopic metallicity of the \ac{LMC} field.

\Cref{fig:data_distributions} compares the distributions of $\log P_{\rm obs}$, $[{\rm O/H}]_{\rm obs}$, $\piobs$, and $m^W_{H,{\rm obs}}$ for the \ac{C22} and \ac{C27} samples.
The \ac{C27} selection on predicted parallax yields a sample of nearby Cepheids with systematically shorter periods and larger parallaxes, whereas the \ac{C22} sample extends to longer periods and smaller parallaxes.
Both samples span a similar metallicity range, $[{\rm O/H}] \approx 0.0$ to $0.4~\dex$ with a median of ${\sim}\,0.1~\dex$.
Each sample is a subset of known classical Cepheids from the GCVS catalogue.
Both are confined to the \ac{MW} disc and, even within the \ac{MW}, cover only a limited range of distances due to selection: the parallax distributions in~\cref{fig:data_distributions} lack Cepheids at small $\piobs$ relative to the expectation from a volume-complete survey.
The dashed lines in~\cref{fig:data_distributions} show the misspecified \citetalias{Hogas_2026}-like uniform-in-volume reference model, while the dotted lines show the thin-disc prediction without any selection applied.
The two no-selection distributions are similar, so the observed parallax histograms disagree mainly because the target selection removes low-parallax objects from the parent population.
The discrepancy is particularly pronounced for \ac{C22}.
\ac{C27}, by contrast, more closely resembles a volume-limited sample, albeit local, because of its parallax cut, which imposes an effective upper distance limit.
We shall compare to their work in~\cref{sec:discussion_hogas}.
This incompleteness---driven by the selection criteria described above---must be accounted for in the inference, and motivates the forward-modelling approach of~\cref{sec:methods}.

\begin{table}
    \centering
    \begin{tabular}{lp{4.2cm}l}
    \toprule
    Symbol & Description & Populations \\
    \midrule
    $\mWH$, $\sigma_m$ & Wesenheit Cepheid magnitude and uncertainty & All \\
    $\log P_{\rm obs}$ & Cepheid pulsation period & All \\
    $[{\rm O/H}]_{\rm obs}$ & Cepheid (spectroscopic) metallicity & All \\
    $\piobs$, $\sigma_\varpi$ & \textit{Gaia} EDR3 parallax and its uncertainty & MW \\
    $\ell$, $b$ & Cepheid Galactic longitude and latitude & MW \\
    $\tilde{\mu}_j$, $\sigma_{\mu,j}$ & Host galaxy geometric distance modulus and its uncertainty & LMC, N4258 \\
    $\bm{\Sigma}_j$ & SH0ES Cepheid magnitude covariance matrix & LMC, N4258 \\
    \bottomrule
    \end{tabular}
    \caption{Summary of observables. ``All'' populations refer to all four datasets: \ac{C22}, \ac{C27}, \ac{LMC}, and N4258.}
    \label{tab:observables}
\end{table}

\begin{table}
    \centering
    \setlength{\tabcolsep}{6pt}
    \begin{tabular}{llcl}
    \toprule
    Sample & Source & $N_{\rm Ceph}$ & Selection / anchor \\
    \midrule
    \ac{C22} & \makecell[l]{\ac{HST}\,+\,\textit{Gaia} \\ EDR3} & 44 & \makecell[l]{$P > 8~{\rm days}$, $V > 6~\magn$, \\ $\AH < 0.4~\magn$, $D \leq 6~\kpc$} \\
    \ac{C27} & \makecell[l]{\ac{HST}\,+\,\textit{Gaia} \\ EDR3} & 22 & \makecell[l]{$\varpi_{\rm phot} > 0.8~\mas$, \\ $\AH < 0.6~\magn$, $V > 6~\magn$} \\
    \midrule
    \ac{LMC} & \ac{HST} (SH0ES) & 69 & \makecell[l]{DEB: $\mu = 18.477$ \\ $\pm\, 0.026~\magn$} \\
    N4258 & \ac{HST} (SH0ES) & 443 & \makecell[l]{Megamaser: $\mu = 29.398$ \\ $\pm\, 0.032~\magn$} \\
    \bottomrule
    \end{tabular}
    \caption{Summary of datasets. $N_{\rm Ceph}$ is the number of Cepheids in each sample. \ac{C22} and \ac{C27} are from~\protect\citetalias{Riess_2021}; \ac{LMC} and N4258 are from~\protect\citetalias{Riess_2022}. Counts are after excluding 9 stars (7 lacking \textit{Gaia} EDR3 parallaxes passing GOF/RUWE quality cuts, 2 outliers) from the original 75. For the \ac{LMC} and N4258, only \ac{HST} photometry is used. Geometric anchors: \ac{LMC} detached eclipsing binary distance from~\protect\citet{Pietrzynski_2019}; N4258 megamaser distance from~\protect\citet{Reid_2019}.}
    \label{tab:data_summary}
\end{table}


\section{Bayesian forward model}\label{sec:methods}

\begin{figure*}
    \centering
    \includegraphics[width=0.98\textwidth]{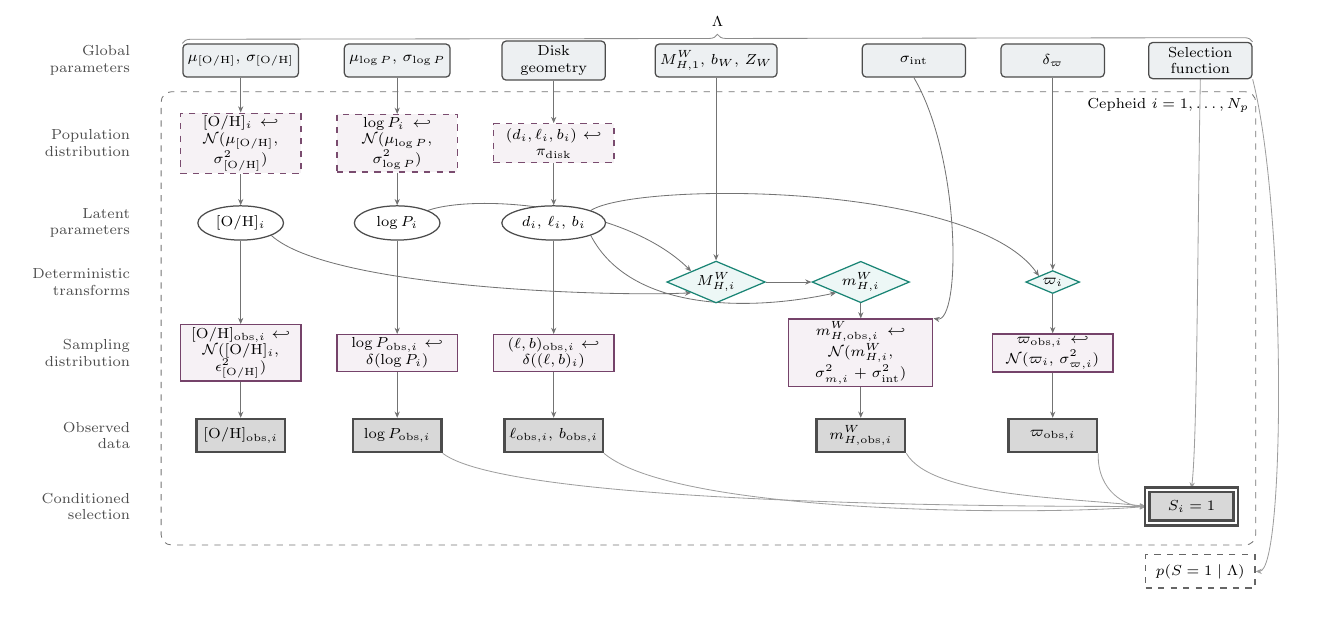}
    \caption{Directed acyclic graph of the forward model for a single \ac{MW} campaign.
    Rows separate global parameters, population distributions, per-Cepheid latent variables, deterministic transformations, sampling distributions, observed data, and conditioned selection.
    The Cepheid plate denotes conditional independence across stars given the global and population-level variables.
    The selection plate denotes the detection fraction entering the campaign normalisation described in~\cref{sec:selection_model}.
    The disc prior generates $(d_i,\, \ell_i,\, b_i)$, with observed angles drawn through delta-function sampling distributions.
    The period and metallicity hyperparameters define population distributions for $\log P_i$ and $[{\rm O/H}]_i$.
    The latent period maps directly to $\log P_{{\rm obs},i}$, while $[{\rm O/H}]_i$ maps to $[{\rm O/H}]_{{\rm obs},i}$ and enters the magnitude path through $M^W_{H,i}$ and $m^W_{H,i}$.
    When multiple campaigns are modelled jointly, the period--luminosity parameters $(\MWH,\, \bW,\, \ZW)$ and parallax offset $\deltapi$ are shared across all populations, while the period and metallicity hyperparameters and intrinsic scatter are independent per campaign.
    The \ac{LMC} and N4258 anchor-galaxy distance nodes and geometric-distance sampling distributions are not shown because this diagram represents a single \ac{MW} campaign.}
    \label{fig:model_DAG}
\end{figure*}

We construct a Bayesian forward model to jointly infer the period--luminosity relation parameters, the parallax zero-point offset $\deltapi$, and per-star distances, broadly following the forward-modelling framework of~\citet{Stiskalek_2026}.
Our approach differs from that of~\citetalias{Riess_2021}, who optimise in parallax space using model (i.e., photometric) parallaxes derived from observed magnitudes and periods, in that we forward-model the Cepheid observables directly by sampling per-star distances as latent variables.
The data consist of Wesenheit apparent magnitudes $\mWH$ from \ac{HST}, pulsation periods $P$, spectroscopic metallicities $[{\rm O/H}]$, and \textit{Gaia} EDR3 parallaxes $\varpi$ for the $66$ \ac{MW} Cepheids described in~\cref{sec:data}, together with the host geometric distance and Cepheid photometry in the \ac{LMC} and N4258~(\cref{tab:observables}).
The model comprises four Cepheid populations, all consistently measured with \ac{HST}: the two \ac{MW} campaigns \ac{C22} and \ac{C27}, and the \ac{LMC} and N4258.
All four share the period--luminosity relation parameters $(\MWH,\, \bW,\, \ZW)$, while the two \ac{MW} campaigns share the parallax zero-point offset $\deltapi$. Each population $p$ has independent period distribution hyperparameters, and all populations except the \ac{LMC} additionally have independent metallicity distribution hyperparameters~(Eq.~\ref{eq:logP_OH_prior}).
For the \ac{LMC}, the catalogued metallicities are used as fixed covariates~(\cref{sec:likelihoods}).
The two \ac{MW} campaigns are additionally modelled with independent intrinsic scatters $\sigma_{\rm int}^{\rm C22}$ and $\sigma_{\rm int}^{\rm C27}$; for the \ac{LMC} and N4258, an intrinsic scatter of $0.06~\magn$ is already included in the SH0ES-reported covariance matrices $\bm{\Sigma}_j$, and we do not recalibrate it. As we find, this $0.06~\magn$ scatter closely matches our inferred values for the \ac{MW} campaigns.
This population-level independence is a flexible modelling choice rather than an assertion that the intrinsic Cepheid populations must differ; for simplicity, we treat the population hyperparameters independently.
The period population prior has little direct leverage because period measurement uncertainties are negligible, so the likelihood effectively fixes each Cepheid's period to its observed value, and the period prior enters only through the population normalisation and selection terms.
Similarly, the metallicity population prior has limited leverage because the metallicity measurement uncertainty is comparable to the width of the observed metallicity distribution, and the period--luminosity relation is only weakly sensitive to metallicity.
For the intrinsic scatter, we explicitly test shared-scatter variants in~\cref{sec:results} and find that the period--luminosity parameters are robust.
We denote the full set of global parameters $\bm{\Lambda}$; the directed acyclic graph of the model is shown in~\cref{fig:model_DAG}.

\subsection{Priors}\label{sec:priors}

\Cref{tab:parameters} summarises all inferred and fixed parameters.
We adopt broad normal priors on the period--luminosity parameters $\MWH$, $\bW$, and $\ZW$, centred approximately on the~\citetalias{Riess_2021} baseline values.
Since both the priors and posteriors are near-Gaussian, the fractional information contributed by the prior is $\sigma_{\rm post}^2 / \sigma_{\rm prior}^2$, where $\sigma_{\rm post}$ is the posterior width reported in~\cref{sec:results}.
This evaluates to at most 1 per cent for $\MWH$, $\bW$, and $\ZW$ in the baseline model, so any double-counting with~\citetalias{Riess_2021} is negligible.
Following~\citetalias{Riess_2021}, we sample the zero-point offset $\deltapi$ from a normal prior.
Similarly, the intrinsic scatters $\sigma_{\rm int}^{\rm C22}$ and $\sigma_{\rm int}^{\rm C27}$ are each sampled independently from a truncated normal prior.

For the $i$\textsuperscript{th} \ac{MW} Cepheid, the distance $d_i$ is drawn from a Galactic disc prior that models Cepheids as tracing a thin disc. The joint density over position $(d,\, \ell,\, b)$ is
\begin{equation}\label{eq:disc_prior}
    \pi_{\rm disc}(d,\, \ell,\, b) = \frac{1}{\mathcal{Z}_V}\, d^2 \exp\!\left(-\frac{R_{\rm GC}(d,\, \ell,\, b)}{R_d}\right) \exp\!\left(-\frac{|z(d,\, b)|}{z_d}\right),
\end{equation}
where $R_{\rm GC}$ is the Galactocentric radius, $z$ the height above the midplane, $R_d$ the disc scale length, $z_d$ the scale height, and $\mathcal{Z}_V = \int {\rm d}d\, {\rm d}\ell\, {\rm d}b\, \cos b\; d^2 \exp(-R_{\rm GC}/R_d)\exp(-|z|/z_d)$ is the volume prior normalisation.
The factor $d^2$ arises from the volume element in spherical coordinates.
Galactocentric coordinates follow from heliocentric distance $d$ and Galactic longitude and latitude $(\ell, b)$ as
\begin{align}
    R_{\rm GC} &= \sqrt{(R_\odot - d \cos b \cos \ell)^2 + (d \cos b \sin \ell)^2}, \label{eq:galactic_coords_R} \\
    z &= d \sin b, \label{eq:galactic_coords_z}
\end{align}
where $R_\odot = 8.122~\kpc$ is the solar Galactocentric distance~\citep{Gravity_2019}.

The true pulsation period and metallicity of each Cepheid are treated as latent variables drawn from population-specific Gaussian distributions,
\begin{align}\label{eq:logP_OH_prior}
    \log P &\hookleftarrow \mathcal{N}(\mu_{\log P}^p,\, (\sigma_{\log P}^p)^2), \nonumber \\
    [{\rm O/H}] &\hookleftarrow \mathcal{N}(\mu_{[{\rm O/H}]}^p,\, (\sigma_{[{\rm O/H}]}^p)^2),
\end{align}
where $p$ denotes the population to which the Cepheid belongs.
Each population has an independent set of period hyperparameters $(\mu_{\log P}^p,\, \sigma_{\log P}^p)$; the metallicity hyperparameters $(\mu_{[{\rm O/H}]}^p,\, \sigma_{[{\rm O/H}]}^p)$ are sampled independently for all populations except the \ac{LMC}, for which the catalogued metallicities are treated as fixed inputs since the SH0ES catalogue reports only a single metallicity value for \ac{LMC} Cepheids~(\cref{sec:likelihoods}).
The means are sampled with uniform priors and the standard deviations with half-normal priors.
The period and metallicity distributions are assumed independent.
Such Gaussian hyperpriors have been found to yield unbiased regression even in cases where the population distribution is non-Gaussian~\citep{MNR}.
In the remainder of this section, we suppress the population superscript $p$ for brevity; all population-dependent quantities are understood to carry it.

We adopt $R_d = 2.5~\kpc$ and $z_d = 0.1~\kpc$, values inspired by $R_d = 2.30 \pm 0.09~\kpc$ measured by~\citet{Bobylev_2021} for young ($< 120~{\rm Myr}$) Cepheids and by~\citet{Nunnari_2025}; fixing instead to the central~\citet{Bobylev_2021} values, including $z_d = 75 \pm 5~{\rm pc}$, has no appreciable effect on the inferred parameters.
The prior is truncated to $[\dmin,\, \dmax]$ and normalised numerically per sightline via Simpson's rule, where $\dmin = 0.1~\kpc$ and $\dmax = 8.5~\kpc$ for \ac{C22} and $\dmax = 2.0~\kpc$ for \ac{C27}.
The same spatial prior enters both the per-star likelihood and the selection-normalisation integral, so the likelihood and selection terms describe the same underlying parent population.
These upper limits are never reached by the posterior and have no impact on the inference, provided they are sufficiently large.
For the \ac{LMC} and N4258, we sample the distance $d_j$ to each host from a uniform-in-volume prior, $\pi(d_j) \propto d_j^2$, and apply the geometric distance modulus measurement as a likelihood constraint in~\cref{eq:anchor_geo}.
Thus, for each anchor galaxy, the distance posterior contains the factor $\pi(d_j)\,\mathcal{N}(\tilde{\mu}_j \mid \mu(d_j),\, \sigma_{\mu,j}^2)$, where $\tilde{\mu}_j \pm \sigma_{\mu,j}$ is the geometric distance modulus measurement.
There is no double counting of volume factors: for example, the N4258 megamaser analysis of~\citet{Reid_2019} used a prior uniform in distance rather than uniform in volume.

The angular positions of observed Cepheids are known, but computing the detection probability of~\cref{sec:selection_model} requires marginalising over the angular position $\bm{\Omega} = (\ell,\, b)$ on the sky, since the normalisation integral extends over all positions where a Cepheid \emph{could} have been detected.
The conditional prior on distance given sky position is $\pi_{\rm disc}(d \mid \bm{\Omega}) = \pi_{\rm disc}(d,\, \bm{\Omega}) / \pi_{\rm disc}(\bm{\Omega})$, where the population angular distribution follows from marginalising~\cref{eq:disc_prior} over distance,
\begin{equation}\label{eq:sightline_prior}
    \pi_{\rm disc}(\bm{\Omega}) = \frac{\mathcal{Z}(\bm{\Omega})}{\mathcal{Z}_V},
\end{equation}
where
\begin{equation}\label{eq:Z_Omega}
    \mathcal{Z}(\bm{\Omega}) \equiv \int {\rm d}d\; d^2 \exp\!\left(-\frac{R_{\rm GC}}{R_d}\right) \exp\!\left(-\frac{|z|}{z_d}\right),
\end{equation}
which depends on the disc model parameters $(R_d,\, z_d)$ and is largest in directions where the disc column density is highest.

\subsection{Deterministic transformations}\label{sec:transformations}

Given the sampled global parameters, per-star distances, and latent true periods and metallicities, we compute the true observables deterministically.
The period--luminosity relation gives the absolute Wesenheit magnitude of the $i$\textsuperscript{th} Cepheid as
\begin{equation}\label{eq:PL_relation}
    M^W_{H,i} = \MWH + \bW \left(\log \frac{P_i}{1~{\rm day}} - 1\right) + \ZW\,[{\rm O/H}]_i,
\end{equation}
where $P_i$ is the latent true period and $[{\rm O/H}]_i$ the latent true metallicity; the wavelength dependence of $\ZW$ is discussed in~\citet{Breuval_2022}.
The true apparent magnitude then follows from the distance modulus,
\begin{align}
    m^W_{H,i}(d_i,\,\bm{\Lambda}) &= M^W_{H,i} + \mu(d_i), \label{eq:m_pred} \\
    \mu(d) &= 5\log_{10}(d / \kpc) + 10,
\end{align}
and the true parallax is
\begin{equation}\label{eq:pi_pred_model}
    \varpi_i = \frac{1~\mas}{d_i / \kpc} - \deltapi,
\end{equation}
where $\deltapi$ is the \textit{Gaia} parallax zero-point offset.
For the \ac{LMC} and N4258, the true apparent magnitudes follow from the same period--luminosity relation, with all Cepheids in galaxy $j$ placed at distance $d_j$.

\subsection{Likelihoods}\label{sec:likelihoods}

For each \ac{MW} Cepheid, the observed Wesenheit magnitude and \textit{Gaia} EDR3 parallax are compared to their true values.
The magnitude likelihood for the $i$\textsuperscript{th} Cepheid, conditioned on the true period $\log P_i$ and metallicity $[{\rm O/H}]_i$, is
\begin{equation}\label{eq:mW_likelihood}
\begin{split}
    &\mathcal{L}(m^W_{H,{\rm obs},i} \mid d_i,\, \log P_i,\, [{\rm O/H}]_i,\, \bm{\Lambda}) \\
    &\qquad = \mathcal{N}\!\left(m^W_{H,{\rm obs},i} \mid m^W_{H,i},\, \sigma_{m,i}^2 + (\sigma_{\rm int}^{c_i})^2\right),
\end{split}
\end{equation}
where $\sigma_{m,i}$ is the photometric uncertainty and $\sigma_{\rm int}^{c_i}$ the intrinsic scatter for campaign $c_i \in \{{\rm C22},\, {\rm C27}\}$.
The parallax likelihood is
\begin{equation}\label{eq:pi_likelihood}
    \mathcal{L}(\varpi_{{\rm obs},i} \mid d_i,\, \bm{\Lambda}) = \mathcal{N}\!\left(\varpi_{{\rm obs},i} \mid \varpi_i,\, \sigma_{\varpi,i}^2\right),
\end{equation}
where $\sigma_{\varpi,i}$ is the reported \textit{Gaia} parallax uncertainty.

The period measurement uncertainties are negligibly small, so the period likelihood is effectively a delta function,
\begin{equation}\label{eq:logP_likelihood}
    \mathcal{L}(\log P_{{\rm obs},i} \mid \log P_i) \approx \delta(\log P_{{\rm obs},i} - \log P_i).
\end{equation}
For the metallicity, individual measurement uncertainties are rarely reported; the dominant errors are methodological and approximately constant across the sample.
We therefore adopt a Gaussian likelihood,
\begin{equation}\label{eq:OH_likelihood}
    \mathcal{L}([{\rm O/H}]_{{\rm obs},i} \mid [{\rm O/H}]_i) = \mathcal{N}([{\rm O/H}]_{{\rm obs},i} \mid [{\rm O/H}]_i,\, \epsilon_{[{\rm O/H}]}^2),
\end{equation}
with a fixed measurement uncertainty $\epsilon_{[{\rm O/H}]} = 0.06~\dex$, distinct from the population scatter $\sigma_{[{\rm O/H}]}$ of~\cref{eq:logP_OH_prior}.
This value is consistent with the ${\sim}\,0.05~\dex$ systematic scale uncertainty adopted by~\citet{Gieren_2018}, the ${\sim}\,0.07~\dex$ propagated measurement error reported by~\citet{Romaniello_2022}, and the ${\lesssim}\,0.05~\dex$ precision achieved in the homogeneous reanalysis of~\citet{daSilva_2022}.
The exact value has negligible impact on the inference, as it enters only through the precision-weighted combination with $\sigma_{[{\rm O/H}]}$.

Similarly, the sky position of each Cepheid is measured effectively without error, so the likelihood of the observed angular position is a delta function, $\mathcal{L}(\bm{\Omega}_{\rm obs} \mid \bm{\Omega}) = \delta(\bm{\Omega} - \bm{\Omega}_{\rm obs})$.
The full prior on the star's three-dimensional position is $\pi_{\rm disc}(d,\, \bm{\Omega}) = \pi_{\rm disc}(d \mid \bm{\Omega})\, \pi_{\rm disc}(\bm{\Omega})$---\cref{eq:disc_prior,eq:sightline_prior}; integrating over the true sky position against the delta function evaluates $\pi_{\rm disc}(\bm{\Omega})$ at $\bm{\Omega}_{\rm obs}$.
Since $\pi_{\rm disc}(d \mid \bm{\Omega}) = \pi_{\rm disc}(d,\, \bm{\Omega}) / \pi_{\rm disc}(\bm{\Omega})$, the factor $\pi_{\rm disc}(\bm{\Omega}_{\rm obs})$ from the prior cancels with the denominator of the conditional, leaving simply $\pi_{\rm disc}(d,\, \bm{\Omega}_{\rm obs}) = \varrho(d,\, \bm{\Omega}_{\rm obs}) / \mathcal{Z}_V$, where
\begin{equation}\label{eq:varrho}
    \varrho(d,\, \bm{\Omega}) \equiv d^2 \exp\!\left(-\frac{R_{\rm GC}}{R_d}\right) \exp\!\left(-\frac{|z|}{z_d}\right)
\end{equation}
is the unnormalised disc density. The remaining global normalisation $\mathcal{Z}_V$ cancels against the selection term, which we will introduce in~\cref{sec:selection_model}.

The true period and metallicity are nuisance latent parameters that can be marginalised analytically.
Combining the measurement likelihoods with the Gaussian population prior of~\cref{eq:logP_OH_prior}, the joint marginalised likelihood is
\begin{multline}\label{eq:marg_likelihood_full}
    \mathcal{L}^{\rm marg}(m^W_{H,{\rm obs},i},\, \log P_{{\rm obs},i},\, [{\rm O/H}]_{{\rm obs},i} \mid d_i,\, \bm{\Lambda}) \\
    = \int {\rm d}\log P_i\, {\rm d}[{\rm O/H}]_i\; \mathcal{L}(m^W_{H,{\rm obs},i} \mid d_i,\, \log P_i,\, [{\rm O/H}]_i,\, \bm{\Lambda}) \\
    \times \mathcal{L}(\log P_{{\rm obs},i} \mid \log P_i)\, \mathcal{L}([{\rm O/H}]_{{\rm obs},i} \mid [{\rm O/H}]_i) \\
    \times \pi(\log P_i \mid \bm{\Lambda})\, \pi([{\rm O/H}]_i \mid \bm{\Lambda}).
\end{multline}
The delta-function period likelihood of~\cref{eq:logP_likelihood} collapses the integral over $\log P_i$, setting $\log P_i = \log P_{{\rm obs},i}$ and leaving behind the period prior $\pi(\log P_{{\rm obs},i} \mid \bm{\Lambda})$.
The true metallicity $[{\rm O/H}]_i$ enters the magnitude likelihood linearly through the $\ZW\,[{\rm O/H}]_i$ term of the period--luminosity relation, while both the metallicity likelihood and population prior are Gaussian; the three factors are jointly Gaussian in $[{\rm O/H}]_i$ and integrate analytically to give
\begin{multline}\label{eq:marg_likelihood}
    \mathcal{L}^{\rm marg}(m^W_{H,{\rm obs},i},\, \log P_{{\rm obs},i},\, [{\rm O/H}]_{{\rm obs},i} \mid d_i,\, \bm{\Lambda}) \\
    = \mathcal{N}\!\left(m^W_{H,{\rm obs},i} \mid m^W_{H,i}(d_i,\, \log P_{{\rm obs},i},\, [{\rm O/H}]_{\star,i}),\, \sigma_{1,i}^2\right) \\
    \times \mathcal{N}\!\left(\log P_{{\rm obs},i} \mid \mu_{\log P},\, \sigma_{\log P}^2\right) \\
    \times \mathcal{N}\!\left([{\rm O/H}]_{{\rm obs},i} \mid \mu_{[{\rm O/H}]},\, \epsilon_{[{\rm O/H}]}^2 + \sigma_{[{\rm O/H}]}^2\right),
\end{multline}
where $[{\rm O/H}]_{\star,i}$ is the effective metallicity, obtained as the precision-weighted mean of the observed value and the population mean,
\begin{equation}\label{eq:OH_posterior}
    [{\rm O/H}]_{\star,i} = \tilde{\sigma}_{[{\rm O/H}]}^2 \left(\frac{[{\rm O/H}]_{{\rm obs},i}}{\epsilon_{[{\rm O/H}]}^2} + \frac{\mu_{[{\rm O/H}]}}{\sigma_{[{\rm O/H}]}^2}\right),
\end{equation}
with posterior precision $\tilde{\sigma}_{[{\rm O/H}]}^{-2} = \epsilon_{[{\rm O/H}]}^{-2} + \sigma_{[{\rm O/H}]}^{-2}$.
The total variance is $\sigma_{1,i}^2 = \sigma_{m,i}^2 + (\sigma_{\rm int}^{c_i})^2 + \ZW^2\, \tilde{\sigma}_{[{\rm O/H}]}^2$.
The second factor in~\cref{eq:marg_likelihood} is the period prior evaluated at the observed period, which constrains $\mu_{\log P}$ and $\sigma_{\log P}$; the third is the marginalised metallicity likelihood, which constrains $\mu_{[{\rm O/H}]}$ and $\sigma_{[{\rm O/H}]}$.
The marginalised per-star likelihood is therefore
\begin{equation}\label{eq:perstar_likelihood}
\begin{split}
    \mathcal{L}_i = &\;\mathcal{L}^{\rm marg}(m^W_{H,{\rm obs},i},\, \log P_{{\rm obs},i},\, [{\rm O/H}]_{{\rm obs},i} \mid d_i,\, \bm{\Lambda}) \\
    &\times \mathcal{N}\!\left(\varpi_{{\rm obs},i} \mid \varpi_i,\, \sigma_{\varpi,i}^2\right).
\end{split}
\end{equation}

For $j \in \{{\rm LMC},\, {\rm N4258}\}$, the geometric distance modulus constrains the galaxy distance via
\begin{equation}\label{eq:anchor_geo}
    \mathcal{L}(\tilde{\mu}_j \mid \mu_j) = \mathcal{N}(\tilde{\mu}_j \mid \mu_j,\, \sigma_{\mu,j}^2),
\end{equation}
where $\tilde{\mu}_j$ is the measured geometric distance modulus, $\sigma_{\mu,j}$ its uncertainty, and $\mu_j = \mu(d_j)$ is the distance modulus corresponding to the sampled galaxy distance $d_j$.
The anchor Cepheid photometry is modelled as a multivariate normal:
\begin{equation}\label{eq:anchor_cepheid}
    \mathcal{L}(\bm{m}^W_{H,j} \mid \mu_j,\, \bm{\Lambda}) = \mathcal{N}(\bm{m}^W_{H,j} \mid \bm{M}^W_{H,j} + \mu_j,\, \bm{\Sigma}_j),
\end{equation}
where $\bm{M}^W_{H,j}$ is the vector of true absolute magnitudes from the period--luminosity relation and $\bm{\Sigma}_j$ is the SH0ES-reported covariance matrix.
As for the \ac{MW} Cepheids, the delta-function period likelihood sets $\log P_k = \log P_{{\rm obs},k}$ in each true absolute magnitude, leaving behind the period prior factors $\pi(\log P_{{\rm obs},k} \mid \bm{\Lambda})$.
For N4258, which has spectroscopic $[{\rm O/H}]$ measurements, the true metallicities are marginalised as in~\cref{eq:marg_likelihood}, adopting $\epsilon_{[{\rm O/H}]} = 0.06~\dex$.
Each Cepheid's true absolute magnitude is evaluated at the effective metallicity $[{\rm O/H}]_{\star,k}$ of~\cref{eq:OH_posterior}, and the marginalisation adds $\ZW^2 \tilde{\sigma}_{[{\rm O/H}]}^2$ to the diagonal of the covariance.
Because each Cepheid's magnitude depends linearly on its own metallicity, and the per-star metallicity posteriors are conditionally independent given the population hyperparameters, the off-diagonal structure of $\bm{\Sigma}_j$ is unchanged.
The marginalised anchor Cepheid likelihood is therefore
\begin{multline}\label{eq:anchor_cepheid_marg}
    \mathcal{L}^{\rm marg}_j = \mathcal{N}\!\Big(\bm{m}^W_{H,j} \mid \bm{M}^W_{H,j}(\log \bm{P}_{{\rm obs}},\, [{\rm O/H}]_{\star,j}) + \mu_j, \\
    \bm{\Sigma}_j + \ZW^2 \tilde{\sigma}_{[{\rm O/H}]}^2 \bm{I}\Big) \\
    \times \prod_{k=1}^{N_j} \mathcal{N}\!\left(\log P_{{\rm obs},k} \mid \mu_{\log P},\, \sigma_{\log P}^2\right) \\
    \times \mathcal{N}\!\left([{\rm O/H}]_{{\rm obs},k} \mid \mu_{[{\rm O/H}]},\, \epsilon_{[{\rm O/H}]}^2 + \sigma_{[{\rm O/H}]}^2\right),
\end{multline}
where $N_j$ is the number of Cepheids in galaxy $j$, and $\log \bm{P}_{{\rm obs}}$ and $[{\rm O/H}]_{\star,j}$ denote the vectors of observed periods and effective metallicities for those Cepheids.
For the \ac{LMC}, the catalogue reports a single spectroscopic metallicity value~(\cref{sec:data}); we therefore treat the central value as a fixed covariate in the true absolute magnitudes and do not infer the population hyperparameters $\mu_{[{\rm O/H}]}$ and $\sigma_{[{\rm O/H}]}$ for the \ac{LMC}.
An uncertainty on this shared value would induce a fully correlated metallicity covariance $\epsilon_{\rm LMC}^2\,\bm{1}\bm{1}^{\rm T}$ between the \ac{LMC} Cepheids, where $\bm{1}$ is a vector of ones with one entry for each \ac{LMC} Cepheid.
Since metallicity enters the period--luminosity relation linearly, this would propagate into the magnitude likelihood as $\ZW^2\epsilon_{\rm LMC}^2\,\bm{1}\bm{1}^{\rm T}$, rather than as the diagonal term $\ZW^2 \tilde{\sigma}_{[{\rm O/H}]}^2 \bm{I}$ in~\cref{eq:anchor_cepheid_marg}.
For $\epsilon_{\rm LMC}=0.06~\dex$ and $|\ZW|\simeq0.22~\magn\,\dex^{-1}$, the induced common-mode magnitude uncertainty is $0.013~\magn$, smaller than the $0.026~\magn$ geometric distance uncertainty of the \ac{LMC}; we therefore assume its effect on the inferred period--luminosity parameters is negligible.

\begin{table*}
    \centering
    \begin{tabular}{llll}
    \toprule
    Symbol & Description & Populations & Prior / Value \\
    \midrule
    \multicolumn{4}{c}{\textbf{Shared inferred parameters}} \\
    \midrule
    $\MWH$ & Period--luminosity relation zero-point at $\log P = 1$ & All & $\mathcal{N}(-5.9,\, 0.5^2)$ \\
    $\bW$ & Period--luminosity relation slope & All & $\mathcal{N}(-3.3,\, 0.5^2)$ \\
    $\ZW$ & Metallicity coefficient & All & $\mathcal{N}(-0.2,\, 0.5^2)$ \\
    $\deltapi$ & \textit{Gaia} parallax zero-point offset & C22, C27 & $\mathcal{N}(0,\, 10^2),\; \deltapi \in [-100,\, 100]~\uas$ \\
    \midrule
    \multicolumn{4}{c}{\textbf{Per-population inferred parameters}} \\
    \midrule
    $\sigma_{\rm int}^p$ & Intrinsic scatter & C22, C27 & $\mathcal{N}(0.06,\, 0.03^2),\; \sigma_{\rm int}^p > 0.01~\magn$ \\
    $\mu_{\log P}^p$ & Period distribution mean & All & $\mathcal{U}(-2.5,\, 2.5)$ \\
    $\sigma_{\log P}^p$ & Period distribution std.\ dev. & All & $\mathcal{N}(0,\, 1),\; \sigma_{\log P}^p > 0$ \\
    $\mu_{[{\rm O/H}]}^p$ & Metallicity distribution mean & C22, C27, N4258 & $\mathcal{U}(-0.5,\, 0.5)$ \\
    $\sigma_{[{\rm O/H}]}^p$ & Metallicity distribution std.\ dev. & C22, C27, N4258 & $\mathcal{N}(0,\, 1),\; \sigma_{[{\rm O/H}]}^p > 0$ \\
    $d_j$ & Host galaxy distance & LMC, N4258 & $\propto d_j^2$ \\
    $\varpi_{\min}$ & Effective parallax lower limit & C22 & $\mathcal{U}(0.1,\, 0.6)~\mas$ \\
    $m^W_{H,\max}$ & Effective magnitude upper limit & C22 & $\mathcal{U}(6,\, 10)~\magn$ \\
    $m^W_{H,\min}$ & Effective magnitude lower limit & C27 & $\mathcal{U}(2,\, 4)~\magn$ \\
    \midrule
    \multicolumn{4}{c}{\textbf{Per-star inferred parameters}} \\
    \midrule
    $d_i$ & Distance to each Cepheid & C22, C27 & Disc prior~(Eq.~\ref{eq:disc_prior}) \\
    \midrule
    \multicolumn{4}{c}{\textbf{Fixed parameters}} \\
    \midrule
    $\varpi_{\min}$ & Parallax lower limit & C27 & $0.8~\mas$ \\
    $w_\varpi$ & Transition width & C22, C27 & C22: $0.1~\mas$; C27: $0.05~\mas$ \\
    $w_m$ & Transition width & C22, C27 & C22: $0.5~\magn$; C27: $0.1~\magn$ \\
    $\log P_{\min}$ & Period lower limit & C22 & $0.90$\\
    $w_{\log P}$ & Transition width & C22 & $0.01$ \\
    ${\AH}_{,\max}$ & Extinction upper limit (optional) & C22 & $0.4~\magn$ \\
    $w_{\AH}$ & Transition width (optional) & C22 & $0.1~\magn$ \\
    $\log P_{\rm cut}$ & Period truncation & LMC, N4258 & $-0.3$ \\
    $\epsilon_{[{\rm O/H}]}$ & Metallicity measurement uncertainty & All & $0.06~\dex$ \\
    $R_d$, $z_d$ & Disc scale length and height~(Eq.~\ref{eq:disc_prior}) & C22, C27 & $2.5$, $0.1~\kpc$ \\
    \bottomrule
    \end{tabular}
    \caption{Model parameters. ``All'' refers to \ac{C22}, \ac{C27}, \ac{LMC}, and N4258. Per-population quantities (superscript $p$) are sampled independently for each population. In total, the model has $91$ inferred parameters: $4$ shared, $21$ per-population (the \ac{LMC} contributes only period hyperparameters; the effective selection thresholds contribute $3$), and $66$ latent per-star distances. Normal priors are written $\mathcal{N}(\mu,\, \sigma^2)$; bounds denote truncation.}
    \label{tab:parameters}
\end{table*}

\subsection{Selection modelling}\label{sec:selection_model}

Both the \ac{C22} and \ac{C27} samples are subject to selection criteria.
We account for these following the framework of~\citet{Kelly_2008} (recently applied and extended by~\citealt{Stiskalek_2026}), which introduces two terms: a per-star selection weight and a normalisation factor.
We model each selection criterion as a smooth cut on some observable $x$, with transition width $w_x$ (the sharp period cut of~\cref{eq:anchor_period_sel} is recovered in the $w_x \to 0$ limit).
The selection function for a lower or upper threshold is, respectively,
\begin{align}\label{eq:sel_generic}
    \mathcal{S}_{\rm low}(x) &= \Phi\!\left(\frac{x - x_{\min}}{w_x}\right), \nonumber \\
    \mathcal{S}_{\rm high}(x) &= \Phi\!\left(\frac{x_{\max} - x}{w_x}\right).
\end{align}
The selection function enters the posterior of~\cref{eq:full_posterior} through two terms.
The first is the per-star selection weight, given by~\cref{eq:sel_generic} evaluated at the observed catalogue values.
For example, for a star with observed magnitude $m$ and an upper limit at $m_{\max}$, the per-star selection weight is
\begin{equation}\label{eq:sel_perstar}
    \mathcal{S}(m) = \Phi\!\left(\frac{m_{\max} - m}{w_m}\right).
\end{equation}
The second term is a normalisation factor $[p(S = 1 \mid \bm{\Lambda})]^{-n_p}$, where $n_p$ is the number of observed stars and $p(S = 1 \mid \bm{\Lambda})$ is the marginalised detection probability.
This is obtained by integrating the selection function weighted by the likelihood and prior over all observed and latent variables:
\begin{equation}\label{eq:prob_detection_generic}
    p(S = 1 \mid \bm{\Lambda}) = \int {\rm d}\bm{x}\, {\rm d}\bm{\theta}\; \mathcal{S}(\bm{x})\, \mathcal{L}(\bm{x} \mid \bm{\theta},\, \bm{\Lambda})\, \pi(\bm{\theta} \mid \bm{\Lambda}),
\end{equation}
where $\bm{x}$ denotes the observed quantities, $\bm{\theta}$ the per-source latent parameters (such as distance), $\mathcal{L}(\bm{x} \mid \bm{\theta},\, \bm{\Lambda})$ the likelihood of the observables, and $\pi(\bm{\theta} \mid \bm{\Lambda})$ the prior, which may depend on some of the global parameters.

For the Cepheid problem, we consider the per-star selection weight of~\cref{eq:sel_perstar} to comprise cuts on up to four observables: the Wesenheit magnitude $m^W_{H,{\rm obs}}$, the parallax $\piobs$, the pulsation period $\log P_{\rm obs}$, and the $H$-band extinction $\AH$.
We assume no selection on metallicity, though one could be straightforwardly included.
We model the magnitude selection as an effective cut in the Wesenheit band; the underlying photometric criteria are not strictly in Wesenheit, but the Wesenheit magnitude is directly predictable from the period--luminosity relation, making this considerably easier in practice (see~\cref{sec:discussion_limitations} for further discussion).
The corresponding marginalised detection probability of~\cref{eq:prob_detection_generic} is
\begin{multline}\label{eq:prob_detection}
    p(S = 1 \mid \bm{\Lambda}) = \int
    {\rm d}\bm{\Omega}\, {\rm d}d\, {\rm d}m\, {\rm d}\varpi \\
    \times {\rm d}\log P\, {\rm d}\log P_{\rm obs}\, {\rm d}[{\rm O/H}]\, {\rm d}[{\rm O/H}]_{\rm obs} \\
    \times \mathcal{S}(m,\, \varpi,\, \log P_{\rm obs},\, \AH(d,\, \bm{\Omega})) \\
    \times \mathcal{L}(m \mid d,\, \log P,\, [{\rm O/H}],\, \bm{\Lambda})\; \mathcal{L}(\varpi \mid d,\, \bm{\Lambda}) \\
    \times \mathcal{L}(\log P_{\rm obs} \mid \log P)\; \mathcal{L}([{\rm O/H}]_{\rm obs} \mid [{\rm O/H}]) \\
    \times \pi_{\rm disc}(d,\, \bm{\Omega})\, \pi(\log P,\, [{\rm O/H}] \mid \bm{\Lambda}),
\end{multline}
where $\bm{\Omega}$ denotes the angular position on the sky, $m \equiv \mWH$ the Wesenheit magnitude, $\varpi$ the parallax, and $\AH(d,\, \bm{\Omega})$ the \ac{LOS} $H$-band extinction model described in~\cref{sec:data}.
The magnitude and parallax likelihoods are Gaussian, centred on the values predicted by the period--luminosity relation and $1/d - \deltapi$, respectively; the period and metallicity likelihoods $\mathcal{L}(\log P_{\rm obs} \mid \log P)$ and $\mathcal{L}([{\rm O/H}]_{\rm obs} \mid [{\rm O/H}])$ relate the observed values to the true underlying quantities.
The remaining terms are the joint position prior $\pi_{\rm disc}(d,\, \bm{\Omega})$ of~\cref{eq:disc_prior} and the prior $\pi(\log P,\, [{\rm O/H}] \mid \bm{\Lambda})$ on Cepheid properties.
In principle, the integral should include both the true sky position $\bm{\Omega}$ and the observed position $\bm{\Omega}_{\rm obs}$ with the delta-function likelihood of~\cref{sec:likelihoods}; since this collapses the integral over $\bm{\Omega}_{\rm obs}$, we write a single $\bm{\Omega}$ throughout.
Assuming that the selection cuts are independent, the joint selection function factorises as
\begin{multline}\label{eq:sel_joint}
    \mathcal{S}(m,\, \varpi,\, \log P_{\rm obs},\, \AH) \\
    = \mathcal{S}(m)\; \mathcal{S}(\varpi)\; \mathcal{S}(\log P_{\rm obs})\; \mathcal{S}(\AH),
\end{multline}
where each factor takes the $\mathcal{S}_{\rm low}$ or $\mathcal{S}_{\rm high}$ form of~\cref{eq:sel_generic} depending on whether the cut imposes a lower or upper threshold.
This nine-dimensional integral must be evaluated for each observational campaign, but the choice of Gaussian population priors and likelihoods, combined with modelling the selection cuts as normal \acp{CDF}, permits a sequence of analytic marginalisations that reduce it to a tractable form.
With the following steps, we reduce the selection normalisation to a three-dimensional numerical integral over $(d,\, \ell,\, b)$, with a closed-form integrand built from Gaussian \acp{CDF}.
The analytic steps below are the standard Gaussian-linear marginalisations used in hierarchical standard-candle models, including those used in SN~Ia cosmology~\citep[e.g.][]{March_2011,Rubin_2015,Shariff_2016}.
We first marginalise the Gaussian population variables.
First, because there is no selection on metallicity and the likelihood is normalised, the integral over $[{\rm O/H}]_{\rm obs}$ evaluates to unity and $[{\rm O/H}]_{\rm obs}$ drops out.
The true metallicity $[{\rm O/H}]$, however, remains through the magnitude likelihood.
Since selection does not depend on the observed metallicity, integrating over $[{\rm O/H}]_{\rm obs}$ removes the metallicity measurement likelihood.
Only the true metallicity remains, with its parent-population prior, and integrating it out of the product of this prior and the magnitude likelihood gives
\begin{multline}\label{eq:OH_marginalisation}
    \int {\rm d}[{\rm O/H}]\; \mathcal{L}(m \mid d,\, \log P,\, [{\rm O/H}],\, \bm{\Lambda})\, \pi([{\rm O/H}] \mid \bm{\Lambda}) \\
    = \mathcal{N}\!\left(m \mid \widebar{m}(d,\, \log P),\, \sigma_1^2\right),
\end{multline}
where $\sigma_1^2 = \sigma_m^2 + \sigma_{\rm int}^2 + \ZW^2 \sigma_{[{\rm O/H}]}^2$ (analogous to $\sigma_{1,i}$ of~\cref{eq:marg_likelihood} but with $\sigma_{[{\rm O/H}]}$ replacing $\tilde{\sigma}_{[{\rm O/H}]}$, since no per-star measurement enters); here $\sigma_m$ and $\sigma_{\rm int}$ are set to the median values across the campaign rather than per-star quantities---a simplification, since in principle the per-star uncertainties should be sampled within the normalisation integral. The model mean magnitude $\widebar{m}(d,\, \log P) = \MWH + \bW(\log P - 1) + \ZW\, \mu_{[{\rm O/H}]} + \mu(d)$ is evaluated at the prior mean metallicity $\mu_{[{\rm O/H}]}$.

Second, as in the per-star likelihood, the delta-function period likelihood of~\cref{eq:logP_likelihood} collapses the integral over $\log P_{\rm obs}$, setting $\log P_{\rm obs} = \log P$ everywhere.
By the same argument, $\widebar{m}$ is linear in $\log P$ through the slope $\bW$, so the product of~\cref{eq:OH_marginalisation} with the Gaussian period prior is again Gaussian in $\log P$.
The integral over $\log P$ with the period selection then yields
\begin{multline}\label{eq:logP_marginalisation}
    \int {\rm d}\log P\; \mathcal{S}(\log P)\, \mathcal{N}(m \mid \widebar{m},\, \sigma_1^2)\, \pi(\log P \mid \bm{\Lambda}) \\
    = \mathcal{N}(m \mid \hat{m}(d),\, \sigma_2^2) \\
    \times \Phi\!\left(\frac{\tilde{\mu}_{\log P}(m,\, d) - \log P_{\min}}{\sqrt{w_{\log P}^2 + \tilde{\sigma}_{\log P}^2}}\right),
\end{multline}
where $\hat{m}(d) = \MWH + \bW(\mu_{\log P} - 1) + \ZW\, \mu_{[{\rm O/H}]} + \mu(d)$ is the model mean magnitude at the prior means and $\sigma_2^2 = \sigma_1^2 + \bW^2 \sigma_{\log P}^2$ the total variance.
The quantities $\tilde{\mu}_{\log P}(m,\, d)$ and $\tilde{\sigma}_{\log P}^2$ are the mean and variance of the posterior Gaussian in $\log P$ obtained from multiplying the magnitude likelihood (which constrains $\log P$ through the slope $\bW$) with the period prior.
Explicitly,
\begin{align}
    \tilde{\sigma}_{\log P}^{-2} &= \bW^2 \sigma_1^{-2} + \sigma_{\log P}^{-2}, \label{eq:logP_posterior_var} \\
    \tilde{\mu}_{\log P}(m,\, d) &= \mu_{\log P} + \frac{\bW\, \tilde{\sigma}_{\log P}^2}{\sigma_1^2}\left(m - \hat{m}(d)\right), \label{eq:logP_posterior}
\end{align}
analogous to the metallicity expressions of~\cref{eq:OH_posterior}.

After both marginalisations, the detection probability reduces to an integral over $(\bm{\Omega},\, d)$ and the Gaussian observables $(m,\, \varpi)$.
Selection factors that depend deterministically on $(d,\, \bm{\Omega})$---such as the extinction cut, evaluated from the three-dimensional dust map---separate from those involving these Gaussian variables.
Each remaining selection factor is a smooth threshold on a linear function of the observables: the magnitude and parallax selections act directly on $m$ and $\varpi$, while the period selection inherited from~\cref{eq:logP_marginalisation} depends on $m$ through $\tilde{\mu}_{\log P}(m,\, d)$.

The remaining integral over Gaussian observables can then be written as a multivariate-normal \ac{CDF}.
Let $n$ denote the number of observables with Gaussian likelihoods (among $m$ and $\varpi$) that enter at least one selection cut.
Standardising these into a vector $\bm{z} \in \mathbb{R}^n$, where each component $z_i = (x_i - \hat{x}_i) / \sigma_i$ is the standardised residual of the $i$\textsuperscript{th} observable, the product of Gaussian likelihoods becomes $\phi_n(\bm{z})$, with $\phi_n$ the $n$-variate standard normal density.
For example, when both magnitude and parallax selections are applied ($n = 2$),
\begin{equation}\label{eq:z_vector}
    \bm{z} = \left(\frac{m - \hat{m}}{\sigma_2},\, \frac{\varpi - \pihat}{\sigma_\varpi}\right)^T,
\end{equation}
where $\sigma_\varpi$ is set to the median parallax uncertainty of the campaign.
Under this substitution, each selection factor takes the form $\Phi(\alpha_k + \bm{\gamma}_k^T \bm{z})$, where $\alpha_k$ is a scalar absorbing the selection threshold and the predicted observable, and $\bm{\gamma}_k \in \mathbb{R}^n$ captures the ratio of the likelihood width to the selection width along each observable.
Continuing the $n = 2$ case, an upper magnitude cut and a lower parallax cut yield
\begin{align}\label{eq:sel_example}
    \text{magnitude:} &\quad \alpha_1 = \frac{m_{\max} - \hat{m}}{w_m}, &\quad \bm{\gamma}_1 = \left(-\frac{\sigma_2}{w_m},\, 0\right)^T, \nonumber \\
    \text{parallax:} &\quad \alpha_2 = \frac{\pihat - \varpi_{\min}}{w_\varpi}, &\quad \bm{\gamma}_2 = \left(0,\, \frac{\sigma_\varpi}{w_\varpi}\right)^T,
\end{align}
where each $\bm{\gamma}_k$ has a zero entry for the observable that does not enter that cut.
The integral over $\bm{z}$ evaluates analytically:
\begin{equation}\label{eq:mvn_cdf_identity}
    \int {\rm d}\bm{z}\; \prod_{k=1}^{K} \Phi(\alpha_k + \bm{\gamma}_k^T \bm{z})\, \phi_n(\bm{z}) = \Phi_K(\bm{h};\, \bm{R}),
\end{equation}
where $\Phi_K$ is the $K$-dimensional multivariate normal \ac{CDF} with zero mean and correlation matrix $\bm{R}$, and
\begin{align}\label{eq:hk_Rjk}
    h_k &= \frac{\alpha_k}{\sqrt{1 + \lVert\bm{\gamma}_k\rVert^2}}, \nonumber \\
    R_{jk} &= \frac{\bm{\gamma}_j \cdot \bm{\gamma}_k}{\sqrt{(1 + \lVert\bm{\gamma}_j\rVert^2)(1 + \lVert\bm{\gamma}_k\rVert^2)}},
\end{align}
and $R_{kk} = 1$.
The arguments $h_k$ absorb the distance-dependent predicted observables and selection thresholds, while $R_{jk}$ captures correlations between selection criteria that share dependence on the same underlying observables.

Applying this identity gives the aforementioned three-dimensional selection normalisation,
\begin{equation}\label{eq:prob_detection_final}
\begin{split}
    p(S = 1 \mid \bm{\Lambda}) = \int {\rm d}\bm{\Omega}\, {\rm d}d &\; \mathcal{S}_{\rm ext}(d,\, \bm{\Omega})\, \Phi_K\!\left(\bm{h}(d);\, \bm{R}(d)\right) \\
    &\times \pi_{\rm disc}(d,\, \bm{\Omega}),
\end{split}
\end{equation}
where $\mathcal{S}_{\rm ext}(d,\, \bm{\Omega}) = \Phi(({\AH}_{,\max} - \AH(d,\, \bm{\Omega})) / w_{\AH})$ is the extinction selection factor evaluated from the dust map (set to unity when no extinction cut is applied), and $\Phi_K(\bm{h}(d);\, \bm{R}(d))$ encapsulates all $K$ selection cuts on the Gaussian observables.
The dimension $K$ and the explicit forms of $\bm{h}(d)$ and $\bm{R}(d)$ for each campaign are derived in~\cref{sec:selection_C27,sec:selection_C22}.

Thus, the original nine-dimensional integral of~\cref{eq:prob_detection} reduces to a numerical integral over $(d,\, \ell,\, b)$.
The distance integral is evaluated via Simpson's rule on a fine grid for each sightline,\footnote{Fixed Gaussian quadrature is another possible choice, since it avoids specifying a regular distance grid while remaining straightforward to vectorise.} while the integral over $\bm{\Omega} = (\ell,\, b)$ is approximated by Monte Carlo sampling.
For a uniform-in-volume prior with no extinction cut, the integrand would be independent of $\bm{\Omega}$ and the detection probability would reduce to a single one-dimensional integral over $d$; the disc geometry of~\cref{eq:disc_prior} and the sightline-dependent extinction, however, require the full angular integration.
Expanding $\pi_{\rm disc}(d,\, \bm{\Omega}) = \varrho(d,\, \bm{\Omega}) / \mathcal{Z}_V$ from~\cref{eq:disc_prior}, where $\varrho(d,\, \bm{\Omega})$ is the unnormalised disc density of~\cref{eq:varrho}, the factor $\mathcal{Z}_V^{-1}$ appears identically in the per-star prior and in the detection probability and cancels in the posterior~(\cref{sec:likelihoods}).
Only the unnormalised integral over $\varrho$ is therefore needed:
\begin{equation}\label{eq:prob_detection_mc}
\begin{split}
    &p(S = 1 \mid \bm{\Lambda}) \\
    &\approx \frac{4\pi}{\mathcal{Z}_V\, N_{\rm los}} \sum_{j=1}^{N_{\rm los}} \int_{\dmin}^{\dmax} {\rm d}d \; \mathcal{S}_{\rm ext}(d,\, \bm{\Omega}_j)\, \Phi_K\!\left(\bm{h}(d);\, \bm{R}(d)\right) \\
    &\hphantom{\approx \frac{4\pi}{\mathcal{Z}_V\, N_{\rm los}} \sum_{j=1}^{N_{\rm los}} \int_{\dmin}^{\dmax} {\rm d}d \;} \times\, \varrho(d,\, \bm{\Omega}_j),
\end{split}
\end{equation}
where the $\bm{\Omega}_j$ are sampled uniformly on the sky and the distance integral along each sightline is evaluated via Simpson's rule on a fine grid.
In practice, we importance-sample the $\bm{\Omega}_j$ from $\pi_{\rm disc}(\bm{\Omega})$~(Eq.~\ref{eq:sightline_prior}) to concentrate draws near the Galactic plane, where $\varrho$ is largest.

\subsubsection{C27 selection}\label{sec:selection_C27}

The \ac{C27} sample targets nearby Cepheids, selected primarily on the photometry-predicted parallax $\varpi_{\rm phot} > 0.8~\mas$ computed following~\cref{eq:pi_pred}.
Because this selection was applied using a fixed set of baseline period--luminosity parameters, modelling the criterion exactly would couple the selection function to those assumed values.
We instead approximate this as a smooth lower cut on the observed parallax at $\varpi_{\min}$ with a transition width $w_\varpi = 0.05~\mas$, which is independent of $\bm{\Lambda}$.
We similarly assume a lower cut on the Wesenheit magnitude at $m^W_{H,\min}$ to exclude the nearest, brightest Cepheids. The effective threshold $m^W_{H,\min}$ is inferred with a uniform prior.
Although \ac{C27} also applied an extinction cut ($\AH < 0.6~\magn$), we treat this as negligible for the nearby \ac{C27} sample.
The per-star selection weight is therefore
\begin{equation}\label{eq:C27_perstar}
    \mathcal{S}(\bm{x}_{{\rm obs},i}) = \Phi\!\left(\frac{\varpi_{{\rm obs},i} - \varpi_{\min}}{w_\varpi}\right) \Phi\!\left(\frac{m^W_{H,{\rm obs},i} - m^W_{H,\min}}{w_m}\right),
\end{equation}
depending only on the observed parallax and magnitude.

For the marginalised detection probability, the $[{\rm O/H}]$ and $\log P$ integrals proceed as in~\cref{eq:OH_marginalisation,eq:logP_marginalisation}, but without a period selection cut; the $\log P$ marginalisation therefore yields a Gaussian in $m$ with variance $\sigma_2^2$ and mean $\hat{m}(d)$, without the $\Phi$ factor of~\cref{eq:logP_marginalisation}.
The two remaining selection factors act on separate Gaussian observables ($K = 2$, $n = 2$): the magnitude cut on $m$ and the parallax cut on $\varpi$.
Since these are independent given the distance, the correlation matrix $\bm{R}$ is diagonal and $\Phi_2(\bm{h};\, \bm{R})$ factorises as $\Phi(h_1)\,\Phi(h_2)$.
Setting $\mathcal{S}_{\rm ext} = 1$ in~\cref{eq:prob_detection_final}, the \ac{C27} detection probability is
\begin{equation}\label{eq:C27_prob_detection}
\begin{split}
    p(S = 1 \mid \bm{\Lambda}) = \int {\rm d}\bm{\Omega}\, {\rm d}d &\; \Phi(h_1(d))\, \Phi(h_2(d)) \\
    &\times \pi_{\rm disc}(d,\, \bm{\Omega}),
\end{split}
\end{equation}
where
\begin{align}\label{eq:C27_hk}
    h_1(d) &= \frac{\hat{m}(d) - m^W_{H,\min}}{\sqrt{w_m^2 + \sigma_2^2}}, \nonumber \\
    h_2(d) &= \frac{\pihat(d) - \varpi_{\min}}{\sqrt{w_\varpi^2 + \sigma_\varpi^2}}.
\end{align}
Following~\cref{eq:prob_detection_mc}, we evaluate~\cref{eq:C27_prob_detection} with $\mathcal{S}_{\rm ext} = 1$, $N_{\rm los} = 5000$ sightlines, $\dmin = 0.1~\kpc$, and $\dmax = 2.0~\kpc$.
Because the analytic marginalisation reduces the problem to a set of one-dimensional integrals over distance, the detection probability is computationally tractable even on-the-fly during sampling.

\subsubsection{C22 selection}\label{sec:selection_C22}

The \ac{C22} sample is subject to four selection criteria.
The original photometric selection was not performed in the Wesenheit band; however, we approximate it as an effective upper cut on the Wesenheit magnitude, $\mWH < m^W_{H,\max}$, noting that the magnitude threshold is sufficiently broad that the sensitivity to the choice of band is minimised.
We similarly model an effective lower cut on the observed parallax, $\piobs > \varpi_{\min}$, and a lower cut on pulsation period, $\log P > \log P_{\min}$; an extinction cut, $\AH < {\AH}_{,\max}$, is optionally included but disabled in the baseline model~(\cref{sec:discussion_limitations}). The effective thresholds $\varpi_{\min}$ and $m^W_{H,\max}$ are jointly inferred with uniform priors.
The per-star selection weight is
\begin{multline}\label{eq:C22_perstar}
    \mathcal{S}(\bm{x}_{{\rm obs},i},\, d_i) = \Phi\!\left(\frac{m^W_{H,\max} - m^W_{H,{\rm obs},i}}{w_m}\right) \Phi\!\left(\frac{\varpi_{{\rm obs},i} - \varpi_{\min}}{w_\varpi}\right) \\
    \times \Phi\!\left(\frac{\log P_{{\rm obs},i} - \log P_{\min}}{w_{\log P}}\right) \\
    \times \Phi\!\left(\frac{{\AH}_{,\max} - \AH(d_i,\, \ell_i,\, b_i)}{w_{\AH}}\right),
\end{multline}
where the extinction factor depends on the latent distance $d_i$ through the three-dimensional dust map.

The marginalised detection probability follows~\cref{eq:prob_detection_mc}.
The period selection enters through the $\Phi$ factor of~\cref{eq:logP_marginalisation}, which depends on $m$ via $\tilde{\mu}_{\log P}(m,\, d)$.
Combined with the magnitude cut, this gives $K = 3$ selection factors acting on $n = 2$ Gaussian observables ($m$ and $\varpi$):
\begin{align}\label{eq:C22_hk}
    h_1(d) &= \frac{m^W_{H,\max} - \hat{m}(d)}{\sqrt{w_m^2 + \sigma_2^2}}, \nonumber \\
    h_2(d) &= \frac{\pihat(d) - \varpi_{\min}}{\sqrt{w_\varpi^2 + \sigma_\varpi^2}}, \nonumber \\
    h_3 &= \frac{\mu_{\log P} - \log P_{\min}}{\sqrt{w_{\log P}^2 + \sigma_{\log P}^2}}.
\end{align}
Because the parallax cut depends only on $\varpi$ while the magnitude and period cuts depend only on $m$, the correlation matrix has $R_{12} = R_{23} = 0$ and a single non-zero off-diagonal entry
\begin{equation}\label{eq:C22_R13}
    R_{13} = \frac{-\bW\, \sigma_{\log P}^2}{\sqrt{(w_m^2 + \sigma_2^2)(w_{\log P}^2 + \sigma_{\log P}^2)}},
\end{equation}
which is positive (since $\bW < 0$) and captures the correlation between the magnitude and period cuts through their shared dependence on the latent period.
The trivariate \ac{CDF} therefore factorises as $\Phi_3(\bm{h};\, \bm{R}) = \Phi(h_2)\, \Phi_2(h_1,\, h_3;\, R_{13})$.
The bivariate \ac{CDF} is evaluated via the integral representation
\begin{equation}\label{eq:phi2_integral}
    \Phi_2(x_1,\, x_2;\, \rho) = \Phi(x_1)\,\Phi(x_2) + \int_0^\rho \phi_2(x_1,\, x_2;\, t)\, {\rm d}t,
\end{equation}
where $\phi_2(x_1,\, x_2;\, t)$ is the standard bivariate normal density with correlation $t$.
The integrand is smooth and the integral is computed to high accuracy with a low-order Gauss--Legendre quadrature rule, making the evaluation of $\Phi_2$ essentially as fast as that of the univariate $\Phi$.
When the extinction cut is enabled, the extinction selection $\mathcal{S}_{\rm ext}(d,\, \bm{\Omega}) = \Phi(({\AH}_{,\max} - \AH(d,\, \bm{\Omega})) / w_{\AH})$ is included in the integrand; in the baseline model $\mathcal{S}_{\rm ext} = 1$.
We use $N_{\rm los} = 5000$ sightlines, $\dmin = 0.1~\kpc$, and $\dmax = 8.5~\kpc$.

\subsubsection{LMC and N4258 selection}\label{sec:selection_anchors}

For the anchor samples, the only selection modelled here is the period truncation.
Both anchor catalogues are truncated at $\log P_{\rm cut} = -0.3$ (i.e.\ $P_{\rm cut} \approx 0.5~{\rm day}$), excluding short-period Cepheids.
Following the same selection framework as for the \ac{MW} campaigns~(\cref{sec:selection_model}), the selection-adjusted anchor likelihood is~\cref{eq:anchor_cepheid_marg} divided by the detection probability---the probability that a Cepheid drawn from the population prior satisfies the period cut.
The per-star period prior in~\cref{eq:anchor_cepheid_marg} remains the full Gaussian $\mathcal{N}(\log P_{{\rm obs},k} \mid \mu_{\log P},\, \sigma_{\log P}^2)$; the per-star selection weight is unity for all observed Cepheids (since they all satisfy $\log P_{{\rm obs},k} > \log P_{\rm cut}$), and only the normalisation factor contributes, equivalent to the normalisation of the period prior truncated at $\log P_{\rm cut}$.
Because there is no assumed magnitude selection for the \ac{LMC} and N4258, the integral of the multivariate magnitude likelihood over $\bm{m}$ evaluates to unity irrespective of the covariance $\bm{\Sigma}_j$, and the detection probability reduces to
\begin{align}\label{eq:anchor_period_sel}
    p(S = 1 \mid \bm{\Lambda})
    &= \int_{\log P_{\rm cut}}^{\infty} \mathcal{N}(\log P \mid \mu_{\log P},\, \sigma_{\log P}^2)\, {\rm d}\log P \nonumber \\
    &= \Phi\!\left(\frac{\mu_{\log P} - \log P_{\rm cut}}{\sigma_{\log P}}\right).
\end{align}
The selection-adjusted anchor likelihood is therefore~\cref{eq:anchor_cepheid_marg} divided by $p(S = 1 \mid \bm{\Lambda})^{N_j}$.
Beyond this, we do not model further selection effects for the \ac{LMC} and N4258 Cepheids.
A complete treatment would additionally address magnitude-dependent selection within each galaxy.
For the \ac{LMC} ($\mu \approx 18.5$), Cepheids at the period cut have $\mWH \approx 17~\magn$, far brighter than any \ac{HST} detection limit; magnitude selection is therefore irrelevant.
For N4258 ($\mu \approx 29.4$), the faintest Cepheids reach $\mWH \approx 28~\magn$, still brighter than but approaching the \ac{HST} WFC3/IR detection limit in crowded fields.
Modelling this would require integrating the multivariate likelihood of~\cref{eq:anchor_cepheid_marg} over the selection region and marginalising over the magnitudes of undetected Cepheids, whose covariance with the observed sample is not available.

\subsection{Full posterior and inference}\label{sec:inference}

Combining the priors, likelihoods, and selection modelling, the full posterior is
\begin{multline}\label{eq:full_posterior}
    \mathcal{P}(\bm{\Lambda},\, \{d_i\},\, \{d_j\} \mid \bm{D}) \propto \pi(\bm{\Lambda}) \\
    \times \prod_{p \in \{{\rm C22,\, C27}\}} \Bigg[ \left[p_p(S = 1 \mid \bm{\Lambda})\right]^{-n_p} \\
    \times \prod_{i \in p} \mathcal{L}_i\, \mathcal{S}(\bm{x}_{{\rm obs},i},\, d_i)\, \pi_{\rm disc}(d_i,\, \ell_i,\, b_i)\Bigg] \\
    \times \prod_{j \in \{{\rm LMC,\, N4258}\}} \left[p_j(S = 1 \mid \bm{\Lambda})\right]^{-N_j} \mathcal{L}^{\rm marg}_j\, \mathcal{L}(\tilde{\mu}_j \mid \mu_j)\, \pi(d_j).
\end{multline}
The first product runs over the \ac{MW} campaigns, where $n_p$ is the number of Cepheids in campaign $p$, $\mathcal{L}_i$ the marginalised per-star likelihood of~\cref{eq:perstar_likelihood}, $\mathcal{S}(\cdot)$ the per-star selection weight of~\cref{eq:C27_perstar,eq:C22_perstar}, $p_p(S = 1 \mid \bm{\Lambda})$ the campaign-level detection probability of~\cref{eq:prob_detection}, and $\pi_{\rm disc}(d_i,\, \ell_i,\, b_i)$ the joint position prior of~\cref{eq:disc_prior}.
The second product runs over the \ac{LMC} and N4258, where $p_j(S = 1 \mid \bm{\Lambda})$ is the per-star period detection probability of~\cref{eq:anchor_period_sel}, $\mathcal{L}^{\rm marg}_j$ the marginalised host Cepheid likelihood of~\cref{eq:anchor_cepheid_marg}, $\mathcal{L}(\tilde{\mu}_j \mid \mu_j)$ the geometric distance constraint of~\cref{eq:anchor_geo}, and $\pi(d_j)$ a uniform-in-volume distance prior.
All Cepheids within a given host galaxy are placed at a common distance $d_j$, which is a free parameter of the model.

The sampled parameters are therefore the global parameters $\bm{\Lambda}$---comprising the shared period--luminosity relation parameters, the parallax offset, the per-campaign intrinsic scatters, and the population hyperparameters---together with the distance $d_i$ to each \ac{MW} Cepheid and the distance $d_j$ to each host galaxy.
\Cref{tab:observables} lists the observables entering the model, and~\cref{tab:parameters} collects all inferred and fixed parameters alongside the populations to which they apply; in total the model has $91$ sampled parameters.
We implement the model in \texttt{JAX} and sample with the No-U-Turn Sampler~\citep[NUTS;][]{Hoffman_2011} via \texttt{numpyro}\footnote{\href{https://num.pyro.ai/}{\texttt{num.pyro.ai}}}~\citep{Phan_2019}.
We run four independent chains, each with \num{1000} warm-up and \num{5000} sampling steps, and verify convergence by requiring $\hat{R} < 1.01$ for all parameters; we collect at least \num{5000} effective samples per parameter.

For comparison, we also implement the~\citetalias{Riess_2021} parallax-space model, which avoids sampling per-star distances.
For each Cepheid, we compute a photometric parallax from the observed Wesenheit magnitude and the period--luminosity relation,
\begin{equation}\label{eq:pi_phot}
    \varpi_{\rm phot} = 10^{-0.2\,(\mWH - M^W_H - 10)},
\end{equation}
and the $\chi^2$ per star is
\begin{equation}\label{eq:R21_chi2}
    \chi^2_i = \frac{(\piobs - \varpi_{\rm phot} + \deltapi)^2}{\tilde{\sigma}_{\varpi,i}^2},
\end{equation}
where
\begin{equation}\label{eq:R21_sigma}
    \tilde{\sigma}_{\varpi,i}^2 = (0.2\ln 10\, \varpi_{\rm phot}\, \sigma_{m,{\rm tot}})^2 + (\alpha\, \sigma_{\varpi,{\rm EDR3}})^2
\end{equation}
combining the propagated magnitude uncertainty with the \textit{Gaia} parallax error inflated by a factor $\alpha = 1.1$. This model applies no additional selection modelling and fixes the intrinsic scatter to $0.06~\magn$\footnote{We note that~\citetalias{Riess_2021} unintentionally fixed the parameters in the $\varpi_{\rm phot}$ expression in~\cref{eq:pi_phot} when computing the error terms in~\cref{eq:R21_sigma}, but we allow them to vary here, which may contribute to the ${\sim}\,0.01~\magn$ difference between~\citetalias{Riess_2021} as reported and our emulation as shown in~\cref{tab:results}.}.


\section{Results}\label{sec:results}

The baseline inference combines the \ac{C22} and \ac{C27} samples with the disc distance prior~(Eq.~\ref{eq:disc_prior}) and selection modelling applied to both \ac{MW} campaigns; the \ac{LMC} and N4258 are optionally included as additional geometric calibrators.
The term ``\ac{MW} Cepheids'' denotes the joint \ac{C22}\,+\,\ac{C27} sample; individual campaign results appear in~\cref{app:individual_results}.
Subsequent variants omit the selection modelling or replace the disc prior with a uniform-in-volume prior.

The~\citetalias{Riess_2021} $\chi^2$ method applied to the \ac{MW} Cepheids yields results closely matching our forward model. In~\cref{app:similarity} we show that this agreement is driven by the small intrinsic scatter of the period--luminosity relation, and in~\cref{app:bias_tests} we validate the inference methods on mock catalogues mimicking the \ac{C22} and \ac{C27} campaigns.

\subsection{Baseline posteriors}\label{sec:baseline}

\begin{figure*}
    \centering
    \includegraphics[width=\textwidth]{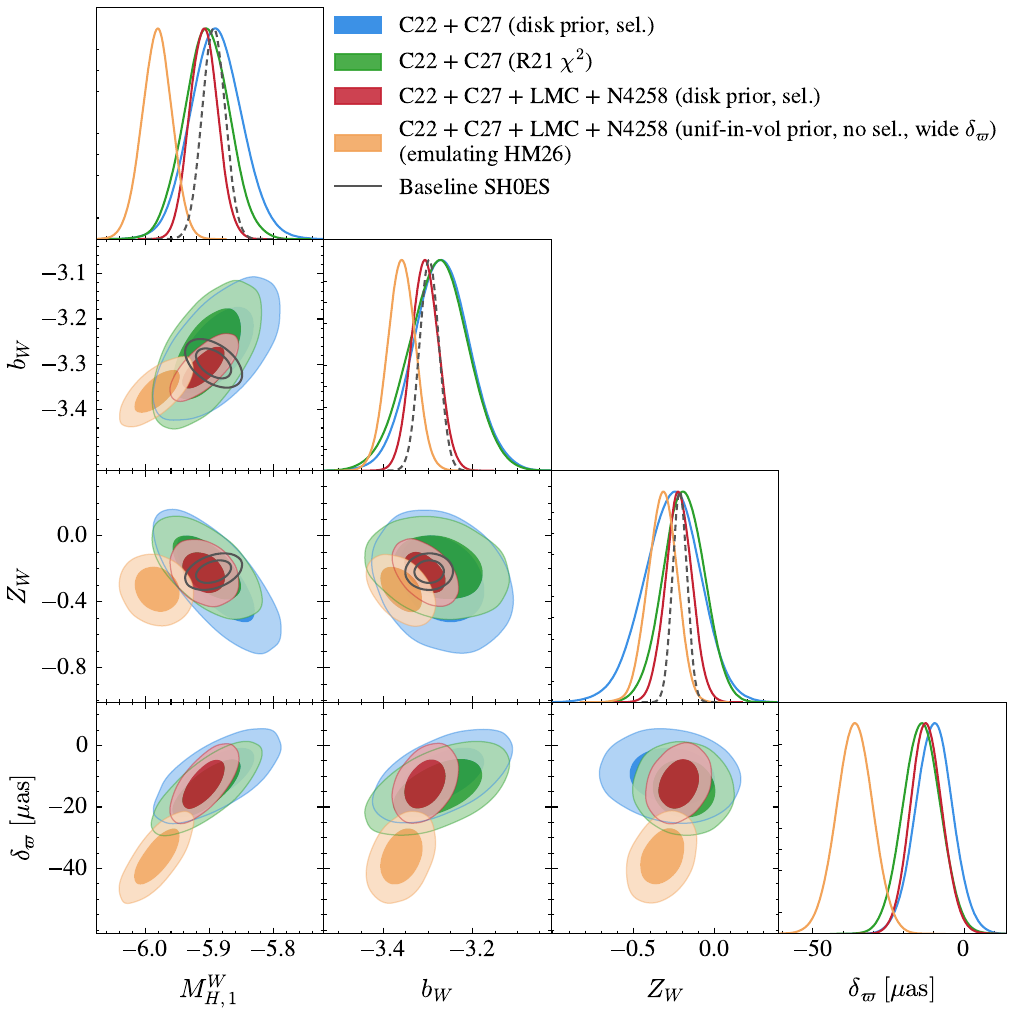}
    \caption{Marginalised posterior of the period--luminosity parameters $(\MWH,\, \bW,\, \ZW)$ and the parallax offset $\deltapi$ for three model configurations: \ac{C22} and \ac{C27} (\ac{MW} only) with a \ac{MW} disc distance prior and selection (blue, filled), the baseline model including the \ac{LMC} and N4258 with the same \ac{MW} disc prior and selection (red, filled), and all four populations with a uniform-in-volume \ac{MW} distance prior, no selection, and a wide $\deltapi$ prior, emulating~\protect\citetalias{Hogas_2026} (orange).
    The $0.7\sigma$ agreement in $\MWH$ with the baseline SH0ES value translates to a shift in $H_0$ of less than $0.2~\kmsecMpc$~(\cref{sec:discussion_H0}).
    The baseline SH0ES contours from~\protect\citetalias{Riess_2022} are shown in black, and the \protect\citetalias{Riess_2021} $\chi^2$ contours for \ac{C22}\,+\,\ac{C27} are shown in green.
    The baseline SH0ES (\protect\citetalias{Riess_2022}) constraints derive from the full three-rung fit (geometric anchors, Cepheids, and Type~Ia supernovae), while the \protect\citetalias{Riess_2021} $\chi^2$ uses only the \ac{MW} Cepheid parallaxes. For visual clarity, the SH0ES constraints are dashed in the panels along the diagonal. All contours show the $1\sigma$ and $2\sigma$ credible regions.}
    \label{fig:corner_baseline}
\end{figure*}

\Cref{fig:corner_baseline} shows the marginalised posterior of the period--luminosity parameters ($\MWH$, $\bW$, $\ZW$) and the parallax offset $\deltapi$ for three model configurations: (i) \ac{C22} and \ac{C27} with the disc prior and selection modelling but without the \ac{LMC} and N4258, (ii) the same with the \ac{LMC} and N4258 Cepheid populations and their geometric distance calibrations, and (iii) a model including all four populations that neglects selection modelling, adopts a uniform-in-volume \ac{MW} distance prior, and uses a wide $\deltapi$ prior, designed to emulate the analysis of~\citetalias{Hogas_2026}.

The \ac{MW}-only posterior is substantially wider, by factors of $1.8$, $2.1$, and $2.1$ in $\MWH$, $\bW$, and $\ZW$, respectively, but only by a factor of $1.2$ in $\deltapi$.
Both $\bW$ and $\ZW$ are strongly correlated with $\MWH$ in the \ac{MW}-only inference; including the \ac{LMC} and N4258 Cepheids partially breaks these degeneracies and tightens the constraints.
Since $\deltapi$ enters only through the \ac{MW} parallaxes, the \ac{LMC} and N4258 do not substantially tighten the constraint; the baseline posterior gives $\deltapi = -12.4 \pm 5.3~\uas$, compared with $-9.9 \pm 6.1~\uas$ from the \ac{MW}-only inference.
For comparison,~\citet{CruzReyes_2023} report $\deltapi = -13 \pm 5~\uas$ from their near-infrared \ac{HST} Wesenheit calibration using $15$ open-cluster Cepheids and the $67$-star \ac{HST} Cepheid sample, agreeing with our posterior.
The addition of the \ac{LMC} and N4258 introduces no tension with the \ac{MW}-only result. We also compare these posteriors to the baseline SH0ES contours from the joint three-rung fit reported in table~5 of~\citetalias{Riess_2022}, combining geometric calibrators, \ac{MW} and extragalactic Cepheids, and Type~Ia supernovae. All three parameters agree with our baseline posteriors to within $1\sigma$.

Applying the~\citetalias{Riess_2021} $\chi^2$ method to the \ac{C22} and \ac{C27} samples yields $\MWH = -5.904 \pm 0.035~\magn$ and $\deltapi = -14.0 \pm 6.3~\uas$~(\cref{tab:results}).
Compared to our \ac{MW}-only forward model with selection ($\MWH = -5.889 \pm 0.040~\magn$, $\deltapi = -9.9 \pm 6.1~\uas$), the $\chi^2$ method recovers a ${\sim}\,0.02~\magn$ brighter zero-point and a ${\sim}\,4~\uas$ more negative parallax offset.
On the other hand, our \ac{MW}-only forward model zero-point is consistent with the $\MWH = -5.903 \pm 0.024~\magn$ reported in table~4 of~\citetalias{Riess_2022} from a global period--luminosity analysis of the~\citetalias{Riess_2021} sample.
Including the \ac{LMC} and N4258 tightens the constraint to $\MWH = -5.909 \pm 0.022~\magn$, consistent with the baseline SH0ES value of $\MWH = -5.894~\magn$ and implies a shift in $H_0$ below $0.2~\kmsecMpc$.
The main difference in~\cref{fig:corner_baseline} is the opposite sign of the $\MWH$--$\bW$ posterior correlation relative to the baseline SH0ES contour.
A possible reason for this difference is that the SH0ES contour is not the same posterior object: it comes from the full three-rung analysis, whereas our contour uses only \ac{MW}, \ac{LMC}, and N4258 Cepheids with geometric distances.

\begin{figure}
    \centering
    \includegraphics[width=\columnwidth]{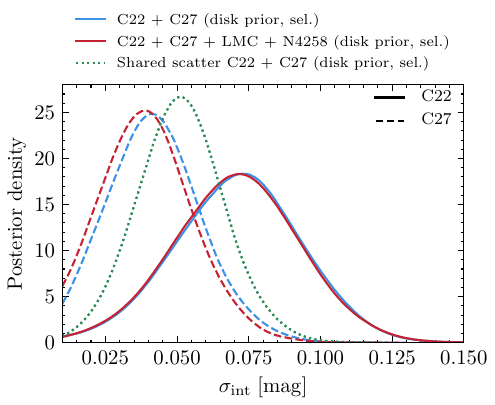}
    \caption{Marginalised posterior of the intrinsic scatter $\sigma_{\rm int}$ for \ac{C22} (solid) and \ac{C27} (dashed), comparing the \ac{MW}-only model with separate per-campaign scatters (blue), the baseline model including the \ac{LMC} and N4258 with separate scatters (red), and the \ac{MW}-only model with a single shared $\sigma_{\rm int}$ for both campaigns (green dotted; $0.051 \pm 0.015~\magn$). Including the \ac{LMC} and N4258 and inferring their joint scatter yields $\sigma_{\rm int} = 0.063 \pm 0.009~\magn$.}
    \label{fig:scatter_comparison}
\end{figure}

\Cref{fig:scatter_comparison} shows the marginalised posterior distributions of the per-campaign intrinsic scatter $\sigma_{\rm int}$.
The \ac{C22} scatter peaks at ${\sim}\,0.07~\magn$, roughly $1.8$ times larger than the \ac{C27} value of ${\sim}\,0.04~\magn$, possibly reflecting residual extinction scatter in the more distant \ac{C22} sample, where dust corrections are larger and the Wesenheit magnitude does not fully remove sightline-to-sightline variations in the reddening law.
Including the \ac{LMC} and N4258 tightens the period--luminosity relation parameters but leaves the \ac{MW} scatter posteriors unchanged, as expected since $\sigma_{\rm int}$ is determined by the spread of \ac{MW} magnitudes about the relation and is largely independent of the global zero-point.

To test whether the \ac{MW}, \ac{LMC}, and N4258 scatters are consistent, we run two additional configurations.
First, we impose a single shared $\sigma_{\rm int}$ across all four populations (with the $0.06~\magn$ already in $\bm{\Sigma}_j$ subtracted in quadrature for the \ac{LMC} and N4258).
The ${\sim}\,500$ \ac{LMC} and N4258 Cepheids dominate the constraint, yielding $\sigma_{\rm int} = 0.063 \pm 0.009~\magn$; without these populations the same shared model gives the less precise $\sigma_{\rm int} = 0.051 \pm 0.015~\magn$.
Second, we allow separate scatters for \ac{C22}, \ac{C27}, and the \ac{LMC}\,+\,N4258 jointly.
The \ac{LMC}\,+\,N4258 scatter is $\sigma_{\rm int}^{\rm LMC+N4258} = 0.070 \pm 0.010~\magn$, roughly 15 per cent larger than the $0.06~\magn$ assumed by~\citetalias{Riess_2022}, while the \ac{MW} campaigns separate cleanly: $\sigma_{\rm int}^{\rm C22} = 0.071 \pm 0.022~\magn$ and $\sigma_{\rm int}^{\rm C27} = 0.041 \pm 0.015~\magn$.
Despite these differences, the period--luminosity parameters are robust to the scatter model: $\MWH$, $\bW$, and $\ZW$ shift by less than $0.3\sigma$ between the shared and per-campaign configurations when the \ac{LMC} and N4258 are included~(\cref{fig:robustness}).
We also run models with a free parallax uncertainty scaling $f_\varpi$ that multiplies the reported \textit{Gaia} parallax uncertainties~(\citetalias{Riess_2021} adopt a fixed factor of $1.1$).
With a single shared $\sigma_{\rm int}$ across \ac{C22} and \ac{C27} in the \ac{MW}-only configuration, we obtain $\sigma_{\rm int} = 0.039 \pm 0.016~\magn$, $f_\varpi = 1.23 \pm 0.16$, and $\deltapi = -8.5 \pm 6.4~\uas$~(\cref{tab:results}).
The inferred $f_\varpi > 1$ indicates that the reported parallax uncertainties may be mildly underestimated.
Analysing each campaign separately with the \ac{LMC} and N4258, we find $f_\varpi = 1.24 \pm 0.18$ for \ac{C22} and $f_\varpi = 0.71 \pm 0.33$ for \ac{C27}. In all cases, $f_\varpi$ has negligible impact on the inferred period--luminosity parameters and parallax offset.

The baseline result is robust to the choice of $\deltapi$ prior: switching to a uniform prior shifts $\deltapi$ by $4.8~\uas$ (from $-12.4$ to $-17.2~\uas$) while $\MWH$ changes by $0.013~\magn$, both well within the posterior uncertainties.
Similarly, replacing the axisymmetric disc prior with the spiral-arm-modulated prior of~\cref{app:spiral_arms} leaves the posteriors unchanged: $\deltapi$ shifts by $0.4~\uas$ and $\MWH$ by $0.001~\magn$.
Posterior means and standard deviations for all model variants are reported in~\cref{tab:results}.

\begin{table*}
    \centering
    \resizebox{\textwidth}{!}{%
    \begin{tabular}{ccccccc}
    \toprule
    Model & $\MWH$ & $\bW$ & $\ZW$ & $\deltapi\;[\uas]$ & $\sigma_{\rm int}^{\rm C22}$ & $\sigma_{\rm int}^{\rm C27}$ \\
    \midrule
    \multicolumn{7}{l}{Bayesian forward model: \textbf{disc prior\,+\,selection modelling}} \\[2pt]
     \midrule
    \makecell{\textbf{Baseline: C22\,+\,C27\,+\,LMC\,+\,N4258}} & $\bm{-5.909 \pm 0.022}$ & $\bm{-3.307 \pm 0.030}$ & $\bm{-0.22 \pm 0.08}$ & $\bm{-12.4 \pm 5.3}$ & $\bm{0.071 \pm 0.022}$ & $\bm{0.040 \pm 0.016}$ \\
    \hdashline[0.5pt/2pt]
    \makecell{C22\,+\,C27} & $-5.889 \pm 0.040$ & $-3.269 \pm 0.064$ & $-0.27 \pm 0.18$ & $-9.9 \pm 6.1$ & $0.071 \pm 0.022$ & $0.043 \pm 0.016$ \\
    \hdashline[0.5pt/2pt]
    \makecell{C22\,+\,C27\,+\,LMC\,+\,N4258 \\ (spiral arms)} & $-5.910 \pm 0.022$ & $-3.307 \pm 0.030$ & $-0.23 \pm 0.08$ & $-12.8 \pm 5.3$ & $0.070 \pm 0.022$ & $0.040 \pm 0.015$ \\
    \hdashline[0.5pt/2pt]
    \makecell{C22\,+\,C27 \\ (shared $\sigma_{\rm int}$, free $f_\varpi$)} & $-5.888 \pm 0.039$ & $-3.262 \pm 0.064$ & $-0.24 \pm 0.18$ & $-8.5 \pm 6.4$ & \multicolumn{2}{c}{$0.039 \pm 0.016$} \\
    \hdashline[0.5pt/2pt]
    \makecell{C22\,+\,C27\,+\,LMC\,+\,N4258 \\ (wide $\deltapi$ prior)} & $-5.922 \pm 0.023$ & $-3.315 \pm 0.031$ & $-0.24 \pm 0.08$ & $-17.2 \pm 6.0$ & $0.068 \pm 0.022$ & $0.040 \pm 0.015$ \\
    \arrayrulecolor{black}\specialrule{1.0pt}{2pt}{2pt}\arrayrulecolor{black}
    \multicolumn{7}{l}{Bayesian forward model: \textbf{uniform-in-volume vs.\ disc prior\,+\,no selection modelling} (\textbf{\textit{misspecified}}; for comparison with~\citetalias{Hogas_2026})} \\[2pt]
     \midrule
    \makecell{\citetalias{Hogas_2026} as reported} & $\bm{-5.959 \pm 0.018}$ & --- & --- & $\bm{-26.2 \pm 5.0}$ & --- & --- \\
    \hdashline[0.5pt/2pt]
    \makecell{ C22\,+\,C27\,+\,LMC\,+\,N4258 \\ (emulating~\citetalias{Hogas_2026}: uniform-in-volume, wide $\deltapi$)} & $-5.982 \pm 0.023$ & $-3.360 \pm 0.031$ & $-0.33 \pm 0.09$ & $-36.3 \pm 6.1$ & $0.075 \pm 0.022$ & $0.041 \pm 0.015$ \\
    \hdashline[0.5pt/2pt]
    \makecell{C22\,+\,C27\,+\,LMC\,+\,N4258 \\ (uniform-in-volume)} & $-5.954 \pm 0.021$ & $-3.341 \pm 0.030$ & $-0.28 \pm 0.08$ & $-26.4 \pm 5.1$ & $0.070 \pm 0.021$ & $0.039 \pm 0.015$ \\
    \hdashline[0.5pt/2pt]
    \makecell{C22\,+\,C27 \\ (uniform-in-volume)} & $-5.984 \pm 0.039$ & $-3.408 \pm 0.065$ & $-0.22 \pm 0.17$ & $-30.1 \pm 6.2$ & $0.075 \pm 0.021$ & $0.037 \pm 0.015$ \\
    \hdashline[0.5pt/2pt]
    \makecell{C22\,+\,C27\,+\,LMC\,+\,N4258 \\ (disc prior)} & $-5.938 \pm 0.021$ & $-3.328 \pm 0.030$ & $-0.26 \pm 0.08$ & $-22.3 \pm 5.0$ & $0.068 \pm 0.021$ & $0.039 \pm 0.015$ \\
    \hdashline[0.5pt/2pt]
    \makecell{C22\,+\,C27 \\ (disc prior)} & $-5.950 \pm 0.039$ & $-3.353 \pm 0.063$ & $-0.23 \pm 0.17$ & $-23.6 \pm 6.0$ & $0.070 \pm 0.021$ & $0.039 \pm 0.015$ \\
    \arrayrulecolor{black}\specialrule{1.0pt}{2pt}{2pt}\arrayrulecolor{black}
    \multicolumn{7}{l}{Frequentist $\chi^2$ \,\,\,\,\,\, (method used in~\citetalias{Riess_2021})} \\[2pt]
     \midrule
    \makecell{\citetalias{Riess_2021} as reported \\ (C22\,+\,C27)} &  $\bm{-5.915 \pm 0.030}$ & $-3.28 \pm 0.06$ & $-0.20 \pm 0.13$ & $-14.0 \pm 6$  & $0.06\,(\mathrm{fixed})$ & $0.06\,(\mathrm{fixed})$ \\
    \hdashline[0.5pt/2pt]
    \makecell{\citetalias{Riess_2022} reported baseline \\ (C22\,+\,C27\,+\,LMC\,+\,N4258)} & $-5.894 \pm 0.017$ & $-3.299 \pm 0.015$ & $-0.217 \pm 0.046$ & --- & $0.06\,(\mathrm{fixed})$ & $0.06\,(\mathrm{fixed})$ \\
    \hdashline[0.5pt/2pt]
    \makecell{\citetalias{Riess_2021} $\chi^2$ emulation \\ (C22\,+\,C27)} & $-5.904^a \pm 0.035$ & $-3.277 \pm 0.066$ & $-0.19 \pm 0.13$ & $-14.0 \pm 6.3$ & --- & --- \\
    \hdashline[0.5pt/2pt]
    \makecell{\citetalias{Riess_2021} $\chi^2$ emulation \\ (C22\,+\,C27\,+\,LMC\,+\,N4258)} & $-5.911 \pm 0.022$ & $-3.309 \pm 0.030$ & $-0.19 \pm 0.08$ & $-15.3 \pm 5.3$ & --- & --- \\
    \bottomrule
    \end{tabular}}%
    \caption{Posterior means and standard deviations for several model variants.
    The baseline configuration is shown in bold.
    A ``$+$'' denotes a joint inference in which the campaigns share period--luminosity parameters but retain separate selection functions and population hyperpriors.
    Parenthetical labels indicate departures from this baseline (distance prior or $\deltapi$ prior width). ``Wide $\deltapi$'' replaces the baseline $\mathcal{N}(0,\, 10^2)~\uas$ prior with a uniform prior over $[-100,\, 100]~\uas$.
    Variants without selection modelling are ordered from the most biased (closest to~\protect\citetalias{Hogas_2026}) to the least.
    The frequentist $\chi^2$ section is shown for comparison with~\protect\citetalias{Riess_2021} and~\protect\citetalias{Riess_2022}. The spiral arm modulation of the distance prior is described in~\cref{app:spiral_arms}. $^a$ The difference between~\protect\citetalias{Riess_2021} as reported and our emulation for \ac{C22}\,+\,\ac{C27} is the result of~\protect\citetalias{Riess_2021} fixing the period--luminosity parameters where they appear in the error terms.}
    \label{tab:results}
\end{table*}

\subsection{Impact of selection modelling and distance prior}\label{sec:selection_impact}

To quantify the impact of selection modelling and the distance prior, we replace the disc prior with a uniform-in-volume prior for the \ac{MW} Cepheids (retaining only the aforementioned per-campaign $\dmin$ and $\dmax$ for normalisation) and disable the selection modelling. The resulting posteriors are incompatible with the baseline model: $\MWH$ shifts to brighter values by ${\sim}\,0.05~\magn$ and $\deltapi$ to more negative values by ${\sim}\,14~\uas$.

\Cref{tab:results} reports further model variants that isolate the individual contributions of the selection modelling and distance prior.
Disabling the selection modelling alone (disc prior, no selection) in the \ac{C22}\,+\,\ac{C27}\,+\,\ac{LMC}\,+\,N4258 configuration shifts $\deltapi$ from $-12.4$ to $-22.3~\uas$ and $\MWH$ from $-5.909$ to $-5.938~\magn$; additionally replacing the disc prior with a uniform-in-volume prior moves $\deltapi$ to $-26.4~\uas$ and $\MWH$ to $-5.954~\magn$.
The selection modelling and the disc prior therefore act in the same direction, each pulling $\deltapi$ towards zero and $\MWH$ towards fainter values.
Their contributions differ substantially, however.
Replacing the disc prior with a uniform-in-volume prior while retaining the selection modelling shifts $\MWH$ by only $0.002~\magn$ and $\deltapi$ by $0.2~\uas$, well within the posterior uncertainties.
Selection modelling alone accounts for ${\sim}\,0.03~\magn$ in $\MWH$ and ${\sim}\,10~\uas$ in $\deltapi$, substantially larger than the effect of changing the distance prior.
Once the selection function is correctly modelled, the shape of the distance prior---disc or uniform-in-volume---is effectively irrelevant: the per-star likelihood constrains each distance to a narrow interval over which the disc geometry varies minimally.
Without the \ac{LMC} and N4258, the shifts are larger still: the \ac{MW}-only inference without selection yields $\deltapi = -23.6 \pm 6.0~\uas$ and $\MWH = -5.950 \pm 0.039~\magn$ with the disc prior, and $\deltapi = -30.1 \pm 6.2~\uas$ and $\MWH = -5.984 \pm 0.039~\magn$ with the uniform-in-volume prior.
These models neglect the known selection criteria and therefore yield biased posteriors; they are reported to quantify the magnitude of the bias, not as reliable inferences.

\subsection{Model validation}\label{sec:validation}

\begin{figure*}
    \centering
    \includegraphics[width=\textwidth]{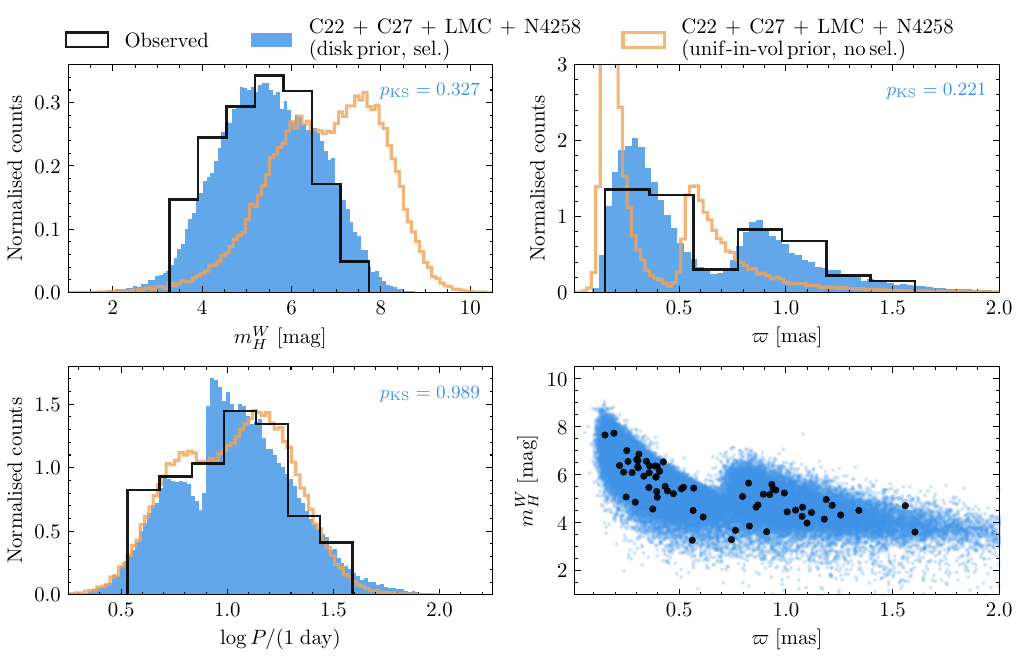}
    \caption{\acp{PPC} for two model variants: the baseline model (C22\,+\,C27\,+\,LMC\,+\,N4258, disc prior, with selection) and the model with a uniform-in-volume prior and no selection, emulating~\protect\citetalias{Hogas_2026}.
    Marginal panels compare normalised histograms of the observed (black) and simulated $\mWH$, $\varpi$, and $\log P$, each annotated with the two-sample KS $p$-value for the baseline model (which is effectively zero for the model without selection); the bottom-right panel shows the joint $\varpi$--$\mWH$ distribution for the baseline model. The effective selection thresholds are inferred jointly with the period--luminosity parameters~(\cref{tab:parameters}), except for the \ac{C27} parallax cut which is fixed.}
    \label{fig:PPC}
\end{figure*}

To verify that the forward model reproduces the observed data, we perform a \ac{PPC}, shown in~\cref{fig:PPC}.
For each of \num{1000} thinned posterior draws, we generate mock \ac{MW} Cepheids by sampling distances from the disc prior and periods and metallicities from the inferred population distributions, then computing apparent magnitudes and parallaxes from the period--luminosity relation and parallax offset.
Observational noise is bootstrapped from the data, and each mock candidate is accepted only if it passes the probabilistic selection function.
The resulting distributions of $\mWH$, $\varpi$, and $\log P$ closely match the observed data, with two-sample Kolmogorov--Smirnov $p$-values of $0.33$, $0.22$, and $0.99$, respectively.
The joint $\piobs$--$\mWH$ distribution~(\cref{fig:PPC}, bottom right) confirms that the forward model also reproduces the observed correlation between parallax and magnitude.
Even when analysing the \ac{C22} or \ac{C27} samples individually, the forward model with selection reproduces the observed distributions for each campaign.
To illustrate the bias introduced by neglecting the selection function and adopting a uniform-in-volume prior for the \ac{MW} Cepheids,~\cref{fig:PPC} also shows the corresponding \ac{PPC}.
This model predicts systematically fainter magnitudes and smaller parallaxes than observed and is clearly incompatible with the data.
The predicted parallax distribution is bimodal, reflecting the two observational campaigns; a single shared distance range would exacerbate the discrepancy further.

In~\cref{app:distance_comparison}, we compare the per-star distance moduli inferred from the forward model with photometric predictions from the baseline SH0ES period--luminosity parameters; the residuals scatter about zero with no significant systematic offset.


\section{Discussion}\label{sec:discussion}

\subsection{Comparison with previous approaches}\label{sec:discussion_interpretation}

We present a forward-modelling framework for the \ac{MW}, \ac{LMC}, and N4258 Cepheid samples that incorporates physically motivated priors---including the thin-disc geometry of the \ac{MW}---and includes a rigorous treatment of the sample selection function, the neglect of which otherwise biases the inference.
\citet{Stiskalek_2026} introduced a related framework for extragalactic Cepheids, inferring $H_0$ without supernovae but compressing the \ac{MW} calibration into a single Gaussian constraint on $\MWH$.
Here we model the \ac{MW} sample star by star, lifting that compression.

The standard treatment, such as that of~\citet{Riess_2016} or~\citetalias{Riess_2021}, minimises a $\chi^2$ between observed parallaxes and the photometric predictions of the period--luminosity relation, with no explicit priors or selection modelling.
The $\chi^2$ specifies a Gaussian likelihood for each observed parallax around its photometric prediction but invokes no distribution for the latent Cepheid properties, so no selection-dependent term enters the inference.
Selection nonetheless biases the result when the cuts correlate with the residuals being minimised, an effect that is negligible for the \ac{MW} Cepheid sample~(\cref{app:similarity}).
In Type~Ia supernova cosmology, where the selection-induced residual bias is severe, the $\chi^2$ inference is supplemented by dedicated bias-correction frameworks such as \acl{BBC}~(\acs{BBC};~\citealt{Kessler_2017, Popovic_2021}).
The forward modelling approach, by contrast, generates the catalogue from an explicit population, distance prior, and detection probability.
The selection therefore enters the inference natively, removing the need for a separate bias correction.

On the Bayesian side, several efforts have tackled the problem.
\citet{Cardona_2017} marginalise over hyperparameters that scale the per-star uncertainties, effectively down-weighting outliers.
\citet{Delgado_2019} construct a hierarchical Bayesian model for period--luminosity relations from \textit{Gaia} parallaxes, placing a joint Gaussian-mixture hyperprior on period, metallicity, and parallax to capture their correlations.
\citet{Feeney_2018} build a hierarchical model of the full distance ladder---from \ac{MW} Cepheids through extragalactic Cepheids to Hubble-flow supernovae---inferring $H_0$ end to end.

What all three Bayesian approaches share is the absence of a principled treatment of the selection function.
\citet{Delgado_2019} implicitly learn the correlation structure of period, metallicity, and parallax through their Gaussian-mixture hyperprior, which acts as a proxy for selection modelling: it allows the model to discover that longer-period Cepheids are brighter and therefore observed at larger distances (smaller parallaxes in their parametrisation).
In reality, this correlation is induced by selection without being present in the intrinsic population; it ought therefore to be captured by an explicit selection treatment, as we construct here.

\citet{Feeney_2018}, by contrast, adopt a uniform prior on the distance modulus, which does not reflect the (approximately) uniform-in-volume distribution of extragalactic objects.
At small redshifts, a uniform-in-distance-modulus prior goes as $1/d$, favouring closer objects, whereas a uniform-in-volume prior goes as $d^2$, favouring more distant ones.
The similarity between results using a uniform-in-distance-modulus prior and those from a uniform-in-volume prior with selection modelling in this case is not coincidental. In~\cref{app:similarity} we show that, for the \ac{MW} Cepheid calibration specifically, the per-star forward-model likelihood is well approximated by the~\citetalias{Riess_2021} $\chi^2$ after marginalising over the latent distance and linearising the distance modulus. This relies on the very small intrinsic scatter of the Cepheid period--luminosity relation, ${\sim}\,0.05~\magn$ (here) or 2--3 per cent in distance; it has not been demonstrated to hold even approximately in other settings, where larger intrinsic scatter or peculiar velocity uncertainties may break the linearisation. For example, in~\cref{app:similarity} we show that an inflated intrinsic scatter of $0.15~\magn$ would have produced a ${\sim}\,0.5\sigma$ bias in $\MWH$ for the combined sample.
\citet{Desmond_2025} lay out this interplay between selection and priors explicitly in a toy model for a distance-ladder inference of $H_0$, showing that the cancellation between volume prior and selection holds exactly in the case of negligible redshift uncertainties and selection on redshift.

In this work we construct the first forward model of the \ac{MW} Cepheid population that explicitly accounts for selection effects, models distance priors accurately, and infers the intrinsic population properties.
The inferred period--luminosity relation parameters and \textit{Gaia} parallax offset are consistent within $1\sigma$ with, and of comparable precision to, both~\citetalias{Riess_2021} and the baseline SH0ES result of~\citetalias{Riess_2022}.
Our results highlight the importance of forward modelling and principled selection treatment, an approach set to become the standard for future distance-ladder (and other) analyses.
Realising this programme requires well-defined observational campaigns with well-characterised selection functions, though the advent of \acl{SBI}~(\ac{SBI};~\citealt{Cranmer_2020, Alsing_2019}) will in principle make even complex selection modelling tractable.
\ac{SBI} does not, however, remove the underlying modelling burden: the simulator must still specify a parent population, a selection function, an extinction model, and a parallax-error model, and misspecification of any of these would propagate into the inferred posterior.
For the current ${\sim}\,60$ \ac{MW} Cepheids split between \ac{C22} and \ac{C27}, the data are unlikely to resolve a Gaussian-\ac{CDF} selection from a more complex shape, and in any case no more realistic alternative is available from the published catalogue information.

That our forward model and the $\chi^2$ should agree so closely was not a priori obvious; the agreement reinforces the SH0ES \ac{MW} Cepheid calibration of the period--luminosity relation~\citepalias{Riess_2021, Riess_2022} and bounds the impact of any alternative selection treatment, including \ac{SBI}, since the $\chi^2$ emulation makes no selection-modelling assumption at all~(\cref{tab:results,app:similarity}) yet $\MWH$ and $\deltapi$ shift by less than $1\sigma$ from the analytic forward model.
For example, the agreement is unlikely to extend to bluer-band period--luminosity calibrations, where the intrinsic scatter is a factor of two or more larger than the Wesenheit $H$-band benchmark of ${\sim}\,0.06~\magn$~\citep{Persson_2004, Soszynski_2015}.
The Bayesian framework additionally supports posterior predictive checks~(\cref{fig:PPC}) as a direct diagnostic of generative-model adequacy, showing that the uniform-in-volume-without-selection variant of~\citetalias{Hogas_2026} is incompatible with the observed Cepheid distributions.

The intrinsic scatter of the period--luminosity relation is another point of departure from~\citetalias{Riess_2021}, who adopt a fixed value of $0.06~\magn$ motivated by the scatter of \ac{LMC} Cepheids.
We instead treat the scatter as a free parameter, inferring $\sigma_{\rm int}^{\rm C22} = 0.071 \pm 0.022~\magn$ and $\sigma_{\rm int}^{\rm C27} = 0.040 \pm 0.016~\magn$ in the baseline model.
The larger \ac{C22} scatter could reflect residual sightline-to-sightline extinction uncertainties in the more distant \ac{C22} sample, although a mean extinction mismatch would instead enter mainly as a systematic zero-point shift.
Consistent with this possibility, the separate \ac{C22}\,+\,\ac{LMC}\,+\,N4258 inference gives $\MWH = -5.865 \pm 0.026~\magn$, offset by $0.051~\magn$ ($1.4\sigma$) from the \ac{C27}\,+\,\ac{LMC}\,+\,N4258 value of $-5.916 \pm 0.024~\magn$~(\cref{tab:individual_results,fig:corner_individual}).
A period-dependent intrinsic scatter could also contribute given the longer periods of the \ac{C22} sample.
As discussed in~\cref{sec:baseline}, allowing a free parallax uncertainty scaling $f_\varpi$ yields $f_\varpi = 1.23 \pm 0.16$ for the joint \ac{MW} sample, consistent with the fixed factor of $1.1$ adopted by~\citetalias{Riess_2021}, with negligible impact on the period--luminosity parameters.

The low scatter of the Cepheid period--luminosity relation and the high precision of the \ac{HST} Wesenheit magnitudes provide a stringent test of the \textit{Gaia} EDR3 parallaxes.
Our tests of rescaling the parallax uncertainties yield $f_\varpi = 1.24 \pm 0.18$ for \ac{C22} and $f_\varpi = 0.71 \pm 0.33$ for \ac{C27}, or $f_\varpi = 1.23 \pm 0.16$ when modelling the two samples jointly (in which case the more distant \ac{C22} sample dominates the $f_\varpi$ constraint), indicating that the \textit{Gaia} uncertainties are reasonably estimated.
This conclusion is consistent with that of~\citetalias{Riess_2021}, who found good agreement between model residuals and (10 per cent inflated) parallax errors, with a best $\chi^2$ of 68 for 66 Cepheids.
A similar result was reported by~\citet{Ripepi_2025}, who found $\chi^2_{\rm dof} \approx 1.04$ for this same \ac{HST} sample.
In contrast,~\citet{Madore_2026} report on noise in the \textit{Gaia} EDR3 parallaxes of \ac{MW} Cepheids based on photometry drawn from heterogeneous ground-based sources, stating ``something subtle is amiss'' regarding \textit{Gaia} parallaxes.
However, they do not quantify the significance of this statement or whether, or by how much, their residuals exceed the combined parallax and photometric uncertainties of the data, precluding a direct comparison with the results from \textit{Gaia} presented here.
We attempted to replicate their ``qualine'' colour analysis, $Q=(V-I)-X(I-{\rm NIR})$, where ${\rm NIR}$ denotes a near-infrared photometric band and $X$ is the relevant reddening ratio.
The measure of $Q$ is noisy, as it combines the noise of four bands with a value of $X>1$ so that it is empirically equivalent to the noise of ${\sim}\,6$ bands added in quadrature. In contrast, the conventional Wesenheit of ${\rm NIR}-R(V-I)$, where $R \leq 0.4$, has noise equivalent to ${\sim}\,1.5$ bands added in quadrature and thus is less demanding to measure. Specifically, we extended the period--luminosity model to include the qualine colour $Q$ as an additional predictor, inferring its coefficient $c_W$ jointly with all other parameters.
For \ac{MW} Cepheids and the \ac{LMC} where $Q$ can be reliably measured, we find $c_W = -0.06 \pm 0.38$, consistent with zero, with no evidence that residual colour information improves the calibration.
At greater distances, the qualine colour becomes too noisy to measure usefully. The period--luminosity zero-point remains unchanged regardless of whether variations around the mean $Q$ are included.

\subsection{Comparison with H\"og\aa s \& M\"ortsell (2026)}\label{sec:discussion_hogas}

\citetalias{Hogas_2026} claim that adopting physically motivated priors in the distance ladder reduces the Hubble tension from $5\sigma$ to $2\sigma$.
They extend the $\chi^2$ model of~\citetalias{Riess_2022} by treating \ac{MW} Cepheids individually and sampling their distances as free parameters, rather than collapsing them into a single Gaussian constraint on the period--luminosity zero-point.
Motivated by~\citet{Desmond_2025}, they impose a uniform-in-volume distance prior, $\pi(d) \propto d^2$, on both the \ac{MW} Cepheids and the nearby Cepheid host galaxies in the SH0ES sample.
An equally fundamental point of~\citet{Desmond_2025}, however, is that modelling selection is crucial: neglecting it biases the inference.
\citetalias{Hogas_2026} entirely neglect this point, arguing that the \ac{MW} Cepheid sample can be considered effectively complete because both \textit{Gaia} and \ac{HST} are capable of detecting Cepheids at much larger distances.
This capability is beside the point.
What matters is how the \ac{MW} Cepheids in the SH0ES sample were actually selected: they are clearly confined to a limited distance range, and hence are not representative of the underlying population, which extends throughout the \ac{MW} disc (and the Universe).
The same argument applies to the nearby Cepheid host galaxies.
Even if Type~Ia supernovae are routinely detected at much larger distances, the question is why these particular nearby galaxies were selected for Cepheid observations---a selection effect that must be modelled, as done by~\citet{Stiskalek_2026}.
Indeed,~\citet{Desmond_2025} showed that even for a volume-limited sample---one that is complete out to a fixed true distance (rather than observed quantities like redshift or magnitude)---the selection must be accounted for by restricting the distance prior to the same range.
An improper $d^2$ prior that extends beyond the true-distance cutoff in this toy setting biases the inference even in that case.
The mechanism is straightforward: a uniform-in-volume prior assigns increasing weight to large distances through the $d^2$ volume factor, but the selection excludes stars beyond a certain range.
Without modelling the selection, the inference overweights distant Cepheids, shifting the period--luminosity zero-point to brighter values and $\deltapi$ to more negative values.

\citetalias{Hogas_2026} report a marginalised posterior of $\MWH = -5.959 \pm 0.018~\magn$ and $\deltapi = -26.2 \pm 5.0~\uas$ (their figure~2).
This brighter zero-point relative to the baseline SH0ES value of $-5.894~\magn$ implies a larger distance scale, yielding $H_0 = 70.6 \pm 1.0~\kmsecMpc$.
This result is driven by their neglect of selection effects.
Of our analysis variants, the one most closely resembling their approach---a uniform-in-volume prior with no selection modelling and a wide prior on $\deltapi$, using \ac{MW}, \ac{LMC}, and N4258 Cepheids---yields $\MWH = -5.982 \pm 0.023~\magn$ and $\deltapi = -36.3 \pm 6.1~\uas$, in reasonable agreement with them.
The residual difference is likely attributable to two factors: we model the Cepheid population hyperprior, and their global fit includes the remainder of the distance ladder, which, while lacking a geometric distance calibration, may still mildly pull the period--luminosity parameters and $\deltapi$.
Modelling selection, as we do in this work, shifts the Cepheid zero-point back towards the baseline SH0ES value and consequently pushes $H_0$ back up~(\cref{sec:discussion_H0}).

In~\cref{fig:PPC} we present \ac{PPC}s for our baseline model.
When selection is included, the predicted distributions of $\mWH$, $\piobs$, and $\log P_{\rm obs}$ closely match the observed data.
Without selection modelling, the predicted distributions of both $\mWH$ and $\piobs$ are incompatible with the observations.
The degree of this incompatibility depends on the range of the distance prior: we adopt $\dmin = 0.1~\kpc$ for both samples and $\dmax = 8.5~\kpc$ and $\dmax = 2.0~\kpc$ for the \ac{C22} and \ac{C27} samples, respectively, somewhat larger than the most distant Cepheid in each.
Extending $\dmax$ further would worsen the disagreement, as the model would predict Cepheids at distances where none are observed.
This connects directly to the selection argument above.
Even for the \ac{C27} sample, where the dominant selection is in parallax and therefore approximately in distance, completeness up to some maximum distance does not obviate the need for selection modelling.
That maximum distance must be explicitly encoded, e.g., in the prior; otherwise the inference assigns non-negligible probability to distances beyond the threshold, allowed by the likelihood but completely excluded by the selection.

The close agreement between our Bayesian forward model and the SH0ES parallax-space $\chi^2$ analysis is interesting.
The two approaches differ fundamentally: SH0ES calibrates directly in parallax space without sampling distances, imposing a spatial prior, or explicitly modelling selection.
In~\cref{app:similarity} we show analytically that the per-star likelihood of the forward model, after marginalising over the latent distance under a power-law prior and locally linearising the distance modulus, reduces to a Gaussian in parallax space whose exponent matches the $\chi^2$.
Depending on whether the selection acts in parallax or in magnitude, different prefactors multiply the exponential, but these are effectively constant or negligible when the intrinsic scatter of the period--luminosity relation is small, i.e.\ when the parallax measurement uncertainty dominates the error budget, as is the case for \ac{MW} Cepheids.
In this regime, the Bayesian forward model reduces effectively to a weighted regression in parallax space conditional on the observed sample.
Mock bias tests~(\cref{app:bias_tests}) confirm that, at the baseline scatter of $\sigma_{\rm int} \approx 0.06~\magn$, consistent with the value inferred from the data, the linearised Gaussian and $\chi^2$ methods are only mildly biased; only when the scatter $\gtrsim0.15~\magn$ would the $\chi^2$ method be significantly biased (the Bayesian forward model is unbiased in all cases).

The consistency of the inferred period--luminosity parameters across methods therefore follows from the small intrinsic scatter of \ac{MW} Cepheids, and demonstrates that the larger shift introduced by a uniform-in-volume prior without selection modelling, as adopted by~\citetalias{Hogas_2026}, is a consequence of a poor generative model of the data.
Because the intrinsic scatter is small, the per-star likelihood sharply constrains the distance, rendering the prior approximately constant over the relevant interval; its effective value is set by the peak location, which depends on the model parameters.
When the selection function is included in the posterior, these per-star prior factors approximately cancel against the selection normalisation, as shown in~\cref{app:similarity} under the linearisation of the distance modulus, and the functional form of the prior has negligible impact on the inferred parameters. Without selection modelling, however, the prior factors evaluated at different parameter-dependent distances for different stars no longer cancel, and the inference becomes biased.
This is shown by the adoption of a uniform-in-volume prior without modelling of the selection function producing posterior predictive distributions of observables clearly discrepant with the data~(\cref{fig:PPC}).

While not important for our main argument, we also note a spurious asymmetry in~\citetalias{Hogas_2026}'s treatment of the second and third rungs of the distance ladder.
They impose a uniform-in-volume prior on \ac{MW} Cepheids and nearby Cepheid host galaxies, yet for the Hubble flow supernovae they retain the $\chi^2$ treatment of~\citetalias{Riess_2022}.
They argue that the \acl{BBC} method (\ac{BBC};~\citealt{Popovic_2021}), while correcting SN~Ia distance moduli for observational selection effects and standardisation biases, operates at the level of SN~Ia standardisation and is not designed to compensate for assumptions about distance priors in the calibration of the local distance ladder.
On this basis they dismiss the suggestion of~\citet{Desmond_2025} that \ac{BBC} may already compensate for the use of a flat prior on distance modulus.
If this is the case, and \ac{BBC} does not effectively account for the distance prior, then a uniform-in-volume prior should have been applied to the Hubble-flow hosts as well, yet~\citetalias{Hogas_2026} do not do so.
They motivate this by noting that Hubble-flow distances can be viewed as derived quantities ($d = cz / H_0$) rather than free parameters, so that a uniform-in-volume prior on them would translate into a prior on $H_0$ itself.
This is unconvincing: by the same reasoning, any distance in the ladder could be recast as a derived quantity, and no distance prior would ever be required. There is no reason why a prior on distances should not translate to an effective prior on $H_0$, which is precisely what happens in the rest of the model anyway when marginalising over the latent distances.

\subsection{Implications for the Hubble constant}\label{sec:discussion_H0}

Our full Bayesian forward model is restricted to the calibration of the first rung of the distance ladder (Cepheids), although the consequences for the full distance ladder inference of $H_0$ are clear.
Given that the Cepheid period--luminosity relation parameters we infer are consistent with the baseline SH0ES values, we expect modifications to the SH0ES $H_0$ value to be marginal and the Hubble tension to remain at ${\sim}\,5\sigma$.
To verify this, we replicate the $\chi^2$ analysis of the full distance ladder of~\citetalias{Riess_2022}, recovering $H_0 = 73.04 \pm 1.01~\kmsecMpc$, matching their reported value.
\citetalias{Riess_2022} constrain the period--luminosity zero-point using both \ac{HST} and \textit{Gaia} parallaxes, obtaining $\MWH = -5.804 \pm 0.082~\magn$ and $\MWH = -5.903 \pm 0.025~\magn$, respectively.
These constraints are derived from partially overlapping sets of \ac{MW} Cepheids observed with two independent astrometric instruments; the per-star error budget is dominated by the parallax uncertainties (${\sim}\,0.14~\magn$ for \textit{Gaia}, ${\sim}\,0.35~\magn$ for \ac{HST}), which are independent, minimising the correlation between the two constraints.
We consider dropping the \ac{HST} constraint and retaining only the \textit{Gaia} one, which yields $H_0 = 72.91 \pm 1.01~\kmsecMpc$, a negligible shift.
Replacing the \textit{Gaia} constraint with the zero-point inferred in this work yields $H_0 = 73.12 \pm 1.08~\kmsecMpc$.
These shifts are at most ${\sim}\,0.1\sigma$, confirming that our inferred Cepheid zero-point leaves the Hubble tension intact.
To isolate the impact of the \textit{Gaia} parallax-offset modelling on $H_0$, we repeat the~\citetalias{Riess_2021} $\chi^2$ inference of the \ac{MW} calibration with $\deltapi$ fixed to zero, obtaining $\MWH = -5.867 \pm 0.015~\magn$, fainter by $0.044~\magn$ than the baseline.
Substituting this constraint into the SH0ES $\chi^2$ shifts $H_0$ from the aforementioned $72.91 \pm 1.01~\kmsecMpc$ to $73.70 \pm 0.95~\kmsecMpc$, an upward shift of about $1\sigma$.
This $\chi^2$-based estimate is only illustrative, as it grafts our Bayesian zero-point onto the frequentist framework of~\citetalias{Riess_2022}; we leave a fully consistent Bayesian reanalysis of the complete distance ladder to future work.
We similarly expect a negligible impact on the two-rung Bayesian distance ladder of~\citet{Stiskalek_2026}, who used the same \ac{MW} constraints as~\citetalias{Riess_2022}.

\subsection{Limitations of this work}\label{sec:discussion_limitations}

For the \ac{C27} sample, the reported selection cuts alone suffice to reproduce the observed distributions in a \ac{PPC}~(\cref{fig:PPC}).
We additionally introduce a mild selection term on the Wesenheit magnitude, in place of a $V$-band cut to avoid \ac{HST} saturation, to remove the brightest Cepheids, though this has a negligible effect on the inferred parameters.

The \ac{C22} sample presents greater challenges.
It was drawn from a parent catalogue of \ac{MW} Cepheids with an implicit completeness limit $V \lesssim 15~\magn$, to which three explicit cuts were applied: a period cut $P > 8~{\rm days}$, an extinction cut $\AH < 0.4~\magn$, and a saturation cut $V > 6~\magn$.
We implement the period and extinction cuts, but these alone do not reproduce the observed distributions of magnitude and parallax.
We therefore introduce additional smooth selection terms on the Wesenheit magnitude and observed parallax.
The parallax selection dominates; results would be near-indistinguishable had we neglected the Wesenheit magnitude term.
This therefore constitutes an effective selection function, not identical to the reported cuts, and can potentially be understood as partially modelling the selection of the parent sample population itself.
The inferred thresholds yield a \ac{PPC} that closely reproduces the observed distributions~(\cref{fig:PPC}).

The parent sample selection is complicated by the fact that it operates on the $V$-band magnitude, whereas our forward model predicts the Wesenheit magnitude.
Modelling the $V$-band selection self-consistently would require predicting $V$-band magnitudes and hence modelling the period--luminosity relation in that band, with its own intrinsic scatter correlated with the Wesenheit scatter (see~\citealt{Yasin_2026} for such a treatment of correlated scatter in galaxy cluster scaling relations).
While in principle straightforward, this would require extending the framework to multiple bands simultaneously.
We instead adopt the simpler approach of modelling an effective selection in Wesenheit magnitude with a smooth cut, even though the data were never selected on this quantity.
Since the $V$-band and Wesenheit magnitudes are strongly correlated, a smooth cut in the latter can effectively mimic a cut in the former.

The choice of smooth selection function is itself an approximation: the true \ac{HST} magnitude completeness is not exactly a Gaussian \ac{CDF} in $\mWH$, and the \textit{Gaia} parallax selection was applied to the photometric parallax computed from fiducial period--luminosity parameters rather than directly to $\piobs$, so our smooth cut on $\piobs$ only approximates the true selection criterion.
The \ac{PPC} of~\cref{fig:PPC} is statistically consistent with the data, leaving no obvious signature of selection-shape misspecification in the observables that enter the cuts.
The effective \ac{C22} and \ac{C27} thresholds are inferred jointly with the period--luminosity parameters and $\deltapi$, and~\cref{fig:corner_thresholds} shows that they are well constrained and not particularly correlated with $(\MWH,\,\bW,\,\ZW,\,\deltapi)$.

A further potential limitation concerns the modelling of selection associated with extinction.
The \ac{C22} sample was subject to an explicit cut $\AH < 0.4~\magn$, but modelling this selection requires knowing the extinction not only along the observed sightlines but in principle at all positions in the Galaxy.
The extinction model described in~\cref{sec:data} is itself a source of systematic uncertainty: the standard deviation of the inter-map differences for the \ac{C22} Cepheids is $0.08~\magn$, and no existing map provides reliable extinctions throughout the Galactic plane.
Hierarchical reddening inference, as in BayeSN~\citep{Mandel_2022}, would require the individual F555W, F814W, and F160W magnitudes, whereas our likelihood uses only the reddening-suppressed $\mWH$.
Moreover, the definition of the Wesenheit magnitude assumes a fixed extinction law;~\citet{Skowron_2026} show that spatial $R_V$ variations across the \ac{MW} disc can shift the \textit{Gaia}-based Wesenheit index $W_G$ by up to ${\sim}\,0.7~\magn$, though the near-infrared $\mWH$ used here is far less sensitive.
Given these systematic uncertainties, the baseline \ac{C22} selection does not include the extinction cut.
When the \ac{LMC} and N4258 are included, enabling the extinction selection has a negligible effect on the period--luminosity parameters ($\Delta\MWH = 0.008~\magn$); without anchors, the effect is larger, but the inference is in any case poorly constrained~(\cref{app:individual_results}).
The \ac{C27} Cepheids are too nearby for extinction selection to matter.
For a single \ac{MW} campaign, the $\MWH$--$\deltapi$ degeneracy is not exact: changing $\MWH$ applies a constant shift in distance modulus, whereas a shift in $\deltapi$ alters the inferred parallax distance modulus by $\Delta\mu \simeq -(5/\ln 10)\Delta\deltapi\, d$.
The \ac{C22} campaign analysed alone gives $\deltapi = 1 \pm 7~\uas$, but this posterior remains prior-dominated and should not be interpreted as a parallax-offset constraint.
Including the \ac{LMC} and N4258 extends the distance baseline and anchors $\MWH$ independently of the \ac{MW} parallaxes.


\section{Conclusion}\label{sec:conclusion}

We have presented a Bayesian forward model of the \ac{MW} Cepheid population that jointly infers the period--luminosity relation parameters, the \textit{Gaia} parallax zero-point offset, and the intrinsic population distributions, while marginalising over the distances and latent true periods and metallicities of individual Cepheids and explicitly accounting for the sample selection function and the disc geometry of the \ac{MW}.
Combined with geometric calibration of the \ac{LMC} and N4258, our baseline model yields a period--luminosity zero-point of $\MWH = -5.909 \pm 0.022~\magn$ and a \textit{Gaia} parallax offset of $\deltapi = -12.4 \pm 5.3~\uas$; the zero-point is consistent within $0.7\sigma$ with the baseline SH0ES value.
Neglecting the selection function and adopting a uniform-in-volume prior shifts $\MWH$ brighter by ${\sim}\,0.05~\magn$ and $\deltapi$ more negative by ${\sim}\,14~\uas$, demonstrating that principled selection modelling is essential for unbiased inference.
The reduced Hubble tension reported by~\citetalias{Hogas_2026} is an artefact of this neglect: applying a uniform-in-volume prior without accounting for the selection that defines the sample biases the zero-point bright and $H_0$ low (see also~\citealt{Desmond_2025}).
The apparent mitigation of the Hubble tension in~\citetalias{Hogas_2026} therefore arises from modelling assumptions that fail to match the generating process of the data, not from new information.

This work is part of a larger programme to forward-model the full distance ladder in a statistically rigorous way.
\citet{Stiskalek_2026} developed the framework for the second rung---the Cepheid host galaxies---and showed that, when coupled with the \texttt{Manticore-Local} model of the local Universe~\citep{McAlpine_2025}, which accounts for galaxy bias and peculiar velocities, $H_0$ can be inferred without supernovae at ${\sim}\,1.8$ per cent precision from as few as 35 host galaxies.
The present work extends the programme to the first rung, treating the \ac{MW} calibration self-consistently.
The remaining piece is the third rung: applying a similar approach to the Type~Ia supernovae and performing end-to-end Bayesian inference from geometric anchors to the Hubble flow.
\citet{Feeney_2018} pursued this goal but adopted a uniform-in-distance-modulus prior and neglected the selection function.
Modelling Type~Ia supernova selection is particularly challenging, as it depends on discovery magnitude, light-curve colour and stretch, host-galaxy properties, and survey-specific targeting strategies, among others~\citep{Kessler_2017, Popovic_2021, Boyd_2024}.

Looking beyond current data, such forward modelling will be particularly compelling when applied to dedicated observational campaigns collecting samples with well-defined selection criteria, rather than to archival data with heterogeneous selection criteria, as is currently the case.
The selection terms computed here are detection probabilities, i.e.\ the average fraction of objects selected from a parent population.
They can therefore be estimated by Monte Carlo from simulated catalogues and then emulated when direct evaluation is expensive or analytically intractable.
\citet{Boyd_2026} present a complementary simulation-based route, which can learn the selection-conditioned likelihood directly from realistic survey simulations.
More broadly, the framework we develop extends beyond the Cepheid--supernova ladder.
Alternative distance indicators such as the tip of the red giant branch, surface brightness fluctuations, and masers can be treated analogously, provided their selection is well understood, and can independently challenge or corroborate the Hubble tension.

\section{Data availability}

The \ac{C22} and \ac{C27} data used in this work were extracted from table 1 of~\citetalias{Riess_2022}.
The SH0ES data are available at \href{https://github.com/PantheonPlusSH0ES/DataRelease}{\texttt{github.com/PantheonPlusSH0ES/DataRelease}}.
The code and all other data will be made available on reasonable request to the authors.

\section*{Acknowledgements}

We thank Pedro G. Ferreira for useful inputs and discussions.
RS acknowledges financial support from STFC Grant No. ST/X508664/1, the Snell Exhibition of Balliol College, Oxford, and a Hintze Fellowship at the Oxford Centre for Astrophysical Surveys, funded through generous support from the Hintze Family Charitable Foundation. HD is supported by a Royal Society University Research Fellowship (grant no. 211046).

The authors would like to acknowledge the use of the University of Oxford Advanced Research Computing (ARC) facility in carrying out this work\footnote{\href{https://doi.org/10.5281/zenodo.22558}{\texttt{doi.org/10.5281/zenodo.22558}}}.

\bibliographystyle{mnras}
\bibliography{ref}

\appendix
\crefalias{section}{appendix}

\section{Individual campaign results}\label{app:individual_results}

\begin{figure*}
    \centering
    \includegraphics[width=0.9\textwidth]{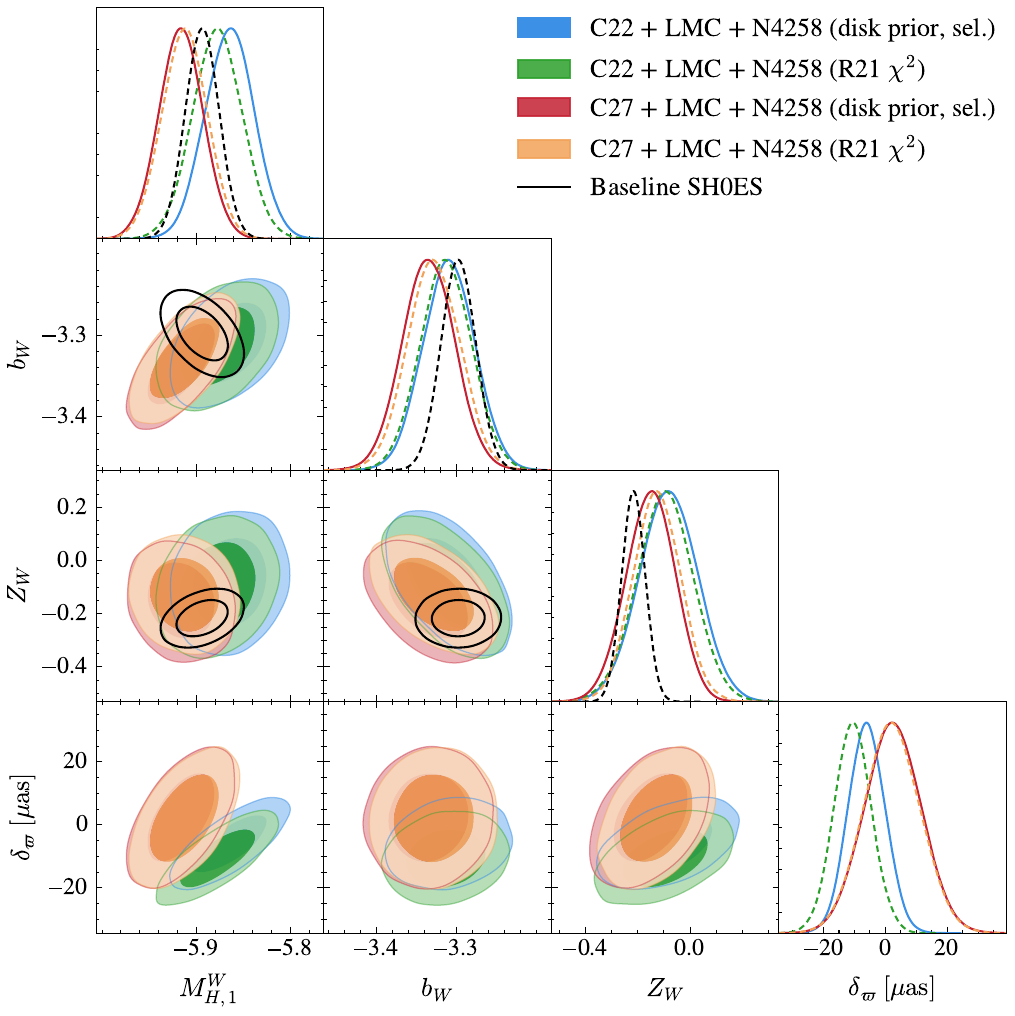}
    \caption{Marginalised posteriors of $(\MWH,\, \bW,\, \ZW,\, \deltapi)$ for the individual \ac{C22}\,+\,\ac{LMC}\,+\,N4258 (blue, filled) and \ac{C27}\,+\,\ac{LMC}\,+\,N4258 (red, filled) configurations with disc prior and selection modelling. The corresponding \protect\citetalias{Riess_2021} $\chi^2$ contours are shown for \ac{C22} (green, filled) and \ac{C27} (orange, filled), and the baseline SH0ES contours from~\protect\citetalias{Riess_2022} in black. All contours show the $1\sigma$ and $2\sigma$ credible regions.}
    \label{fig:corner_individual}
\end{figure*}

\Cref{tab:individual_results} reports posteriors for the \ac{C22} and \ac{C27} campaigns analysed separately, each combined with the \ac{LMC} and N4258; \cref{fig:corner_individual} shows the corresponding corner plots.
The \ac{LMC} and N4258 are required to break the $\MWH$--$\deltapi$ degeneracy inherent in parallax-only data, so we include them throughout.

With the \ac{LMC} and N4258---which dominate the constraint on $\MWH$---the two campaigns yield consistent zero-points: $\MWH = -5.865 \pm 0.026~\magn$ (\ac{C22}) and $-5.916 \pm 0.024~\magn$ (\ac{C27}), differing by $1.4\sigma$.
The slopes and metallicity coefficients, likewise dominated by the \ac{LMC} and N4258 data, agree to within $0.5\sigma$.
Both individual-campaign zero-points are consistent with the joint \ac{C22}\,+\,\ac{C27} result of $\MWH = -5.909 \pm 0.022~\magn$~(\cref{tab:results}), confirming that the two campaigns carry no internal tension.
Selection modelling has a larger effect on \ac{C22} than on \ac{C27}: disabling the selection function shifts $\MWH$ by $0.040~\magn$ and $\deltapi$ by $11~\uas$ for \ac{C22}, compared with $0.008~\magn$ and $6~\uas$ for \ac{C27}.
This is expected, as the more distant \ac{C22} sample is more strongly affected by magnitude truncation.
Similarly, the nearby \ac{C27} Cepheids ($d \lesssim 1.25~\kpc$, $\varpi \gtrsim 800~\uas$) are largely insensitive to the parallax offset: for $\deltapi = -10~\uas$, the induced distance modulus shift is only ${\sim}\,0.02~\magn$, compared with ${\sim}\,0.07~\magn$ for the typical \ac{C22} parallax of ${\sim}\,300~\uas$.
The constraint on $\deltapi$ is therefore driven by the more distant \ac{C22} sample.

The baseline \ac{C22} selection does not include the extinction cut $\AH < 0.4~\magn$, owing to the systematic uncertainties in three-dimensional dust maps discussed in~\cref{sec:discussion_limitations}.
With the \ac{LMC} and N4258 included, enabling the extinction selection has a negligible effect on the period--luminosity parameters ($\Delta\MWH = 0.008~\magn$), but the inferred scatter increases, consistent with residual extinction along the more distant \ac{C22} sightlines.
The~\citetalias{Riess_2021} $\chi^2$ rows apply the frequentist $\chi^2$ method with intrinsic scatter fixed at $0.06~\magn$.

\begin{table*}
    \centering
    \begin{tabular}{ccccc}
    \toprule
    Model & $\MWH$ & $\bW$ & $\ZW$ & $\deltapi\;[\uas]$ \\
    \midrule
    \multicolumn{5}{c}{\textbf{\ac{C22}\,+\,\ac{LMC}\,+\,N4258}} \\[2pt]
    Disc prior, selection modelling & $-5.865 \pm 0.026$ & $-3.309 \pm 0.032$ & $-0.08 \pm 0.11$ & $-6 \pm 6$ \\
    \hdashline[0.5pt/2pt]
    Disc prior, no selection & $-5.905 \pm 0.027$ & $-3.320 \pm 0.032$ & $-0.19 \pm 0.12$ & $-17 \pm 6$ \\
    \hdashline[0.5pt/2pt]
    \citetalias{Riess_2021} $\chi^2$ & $-5.878 \pm 0.027$ & $-3.313 \pm 0.032$ & $-0.10 \pm 0.11$ & $-11 \pm 6$ \\
    \arrayrulecolor{black}\specialrule{1.0pt}{2pt}{2pt}\arrayrulecolor{black}
    \multicolumn{5}{c}{\textbf{\ac{C27}\,+\,\ac{LMC}\,+\,N4258}} \\[2pt]
    Disc prior, selection modelling & $-5.916 \pm 0.024$ & $-3.335 \pm 0.033$ & $-0.15 \pm 0.09$ & $2 \pm 9$ \\
    \hdashline[0.5pt/2pt]
    Disc prior, no selection & $-5.924 \pm 0.024$ & $-3.336 \pm 0.033$ & $-0.18 \pm 0.09$ & $-4 \pm 9$ \\
    \hdashline[0.5pt/2pt]
    \citetalias{Riess_2021} $\chi^2$ & $-5.912 \pm 0.024$ & $-3.328 \pm 0.033$ & $-0.13 \pm 0.09$ & $2 \pm 9$ \\
    \bottomrule
    \end{tabular}
    \caption{Posterior means and standard deviations for the \ac{C22} and \ac{C27} campaigns analysed separately, each combined with the \ac{LMC} and N4258.
    Layout follows~\cref{tab:results}.
    The \ac{C22} selection does not include the extinction cut $\AH < 0.4~\magn$; the effect of including it is discussed in~\cref{sec:discussion_limitations}.}
    \label{tab:individual_results}
\end{table*}

\section{Similarity of the forward model and the $\chi^2$ method}\label{app:similarity}

Here we derive the relation between the forward model of~\cref{sec:methods} and the $\chi^2$ treatment of~\citetalias{Riess_2021}.
For simplicity, we treat the period and metallicity as known quantities with no population prior or selection, so that neither requires marginalisation; since they enter the Gaussian likelihood linearly, they could be marginalised analytically under a Gaussian population prior as in the main text.
The per-star likelihood for a single \ac{MW} Cepheid at distance $d$ then reduces to
\begin{multline}\label{eq:app_perstar}
    \mathcal{L}(m^W_{H,{\rm obs}},\, \piobs \mid d,\, \bm{\theta}) = \mathcal{N}\!\left(m^W_{H,{\rm obs}} \mid M + \mu(d),\; \sigma_{1}^2\right) \\
    \times \mathcal{N}\!\left(\piobs \mid 1/d - \deltapi,\; \sigma_\varpi^2\right),
\end{multline}
where $M \equiv \MWH + \bW(\log P - 1) + \ZW\,[{\rm O/H}]$ is the predicted absolute magnitude, $\sigma_1^2 = \sigma_m^2 + \sigma_{\rm int}^2$ is the total magnitude variance, and $\mu(d) = 5\log(d / 10~{\rm pc})$ is the distance modulus.
Adopting a uniform-in-volume distance prior, $\pi(d) \propto d^2$, the marginal likelihood is
\begin{multline}\label{eq:app_marginal}
    \mathcal{L}^{\rm marg}(m^W_{H,{\rm obs}},\, \piobs \mid \bm{\theta}) \propto \int {\rm d}d\; d^2\, \\
    \times \mathcal{N}\!\left(m^W_{H,{\rm obs}} \mid M + \mu(d),\; \sigma_{1}^2\right)\, \mathcal{N}\!\left(\piobs \mid 1/d - \deltapi,\; \sigma_\varpi^2\right).
\end{multline}
Changing variables to $\varpi = 1/d - \deltapi$, so that $d = 1/(\varpi + \deltapi)$ and $|{\rm d}d| = {\rm d}\varpi / (\varpi + \deltapi)^2$, the parallax likelihood simplifies and the integral becomes
\begin{multline}\label{eq:app_marginal_varpi}
    \mathcal{L}^{\rm marg} \propto \int {\rm d}\varpi\; (\varpi + \deltapi)^{-4}\, \mathcal{N}\!\left(m^W_{H,{\rm obs}} \mid M + \mu(\varpi),\; \sigma_{1}^2\right) \\
    \times \mathcal{N}\!\left(\piobs \mid \varpi,\; \sigma_\varpi^2\right),
\end{multline}
where $\mu(\varpi) = -5\log(\varpi + \deltapi) + 10$.
Following the main text, we define the photometric parallax as the geometric parallax corresponding to the photometric distance,
\begin{equation}\label{eq:app_varpi_phot}
    \varpi_{\rm phot} = 10^{-0.2\,(m^W_{H,{\rm obs}} - M - 10)}.
\end{equation}
In the $\varpi$ coordinate, the magnitude likelihood peaks at $\varpi = \varpi_{\rm phot} - \deltapi$, since $\mu(\varpi) = m^W_{H,{\rm obs}} - M$ when $\varpi + \deltapi = \varpi_{\rm phot}$.

To marginalise~\cref{eq:app_marginal_varpi} analytically, we linearise $\mu(\varpi)$ around $\varpi_{\rm phot} - \deltapi$.
Since
\begin{equation}
    \frac{{\rm d}\mu}{{\rm d}\varpi} = -\frac{5}{\ln 10\,(\varpi + \deltapi)},
\end{equation}
a first-order Taylor expansion gives
\begin{equation}\label{eq:app_linearise}
    \mu(\varpi) \approx \mu(\varpi_{\rm phot} - \deltapi) - \frac{5}{\ln 10}\, \frac{\varpi - \varpi_{\rm phot} + \deltapi}{\varpi_{\rm phot}}.
\end{equation}
Substituting into the magnitude likelihood and using $m^W_{H,{\rm obs}} = M + \mu(\varpi_{\rm phot} - \deltapi)$, the residual is
\begin{equation}
    m^W_{H,{\rm obs}} - M - \mu(\varpi) \approx \frac{5}{\ln 10}\, \frac{\varpi - \varpi_{\rm phot} + \deltapi}{\varpi_{\rm phot}}.
\end{equation}
The magnitude Gaussian becomes
\begin{equation}\label{eq:app_mag_rewrite}
    \mathcal{N}\!\left(m^W_{H,{\rm obs}} \mid M + \mu(\varpi),\; \sigma_1^2\right) \approx \frac{\sigma_{\varpi,m}}{\sigma_1}\, \mathcal{N}\!\left(\varpi \mid \varpi_{\rm phot} - \deltapi,\; \sigma_{\varpi,m}^2\right),
\end{equation}
where the uncertainty in ``parallax space'' after the linearisation is
\begin{equation}\label{eq:app_sigma_m}
    \sigma_{\varpi,m} = \frac{\ln 10}{5}\;\varpi_{\rm phot}\, \sigma_1.
\end{equation}
Expanding the volume-prior factor around $\varpi_{\rm phot} - \deltapi$,
\begin{equation}
    (\varpi + \deltapi)^{-4} \approx \varpi_{\rm phot}^{-4}\left[1 - \frac{4\,(\varpi - \varpi_{\rm phot} + \deltapi)}{\varpi_{\rm phot}}\right].
\end{equation}
The first-order correction is suppressed by the Gaussian likelihood centred on $\varpi_{\rm phot} - \deltapi$ and we drop it.
With this, the marginal likelihood reduces to a product of two Gaussians in $\varpi$,
\begin{multline}\label{eq:app_product}
    \mathcal{L}^{\rm marg} \propto \varpi_{\rm phot}^{-3} \\
    \times \int {\rm d}\varpi\; \mathcal{N}\!\left(\piobs \mid \varpi,\; \sigma_\varpi^2\right)\, \mathcal{N}\!\left(\varpi \mid \varpi_{\rm phot} - \deltapi,\; \sigma_{\varpi,m}^2\right),
\end{multline}
which evaluates to
\begin{equation}\label{eq:app_result}
    \mathcal{L}^{\rm marg} \propto \varpi_{\rm phot}^{-3}\, \mathcal{N}\!\left(\piobs \mid \varpi_{\rm phot} - \deltapi,\; \tilde{\sigma}_\varpi^2\right),
\end{equation}
with
\begin{equation}\label{eq:app_combined_sigma}
    \tilde{\sigma}_\varpi^2 \equiv \left(\frac{\ln 10}{5}\right)^2 \varpi_{\rm phot}^2\, \sigma_1^2 + \sigma_\varpi^2.
\end{equation}
The Gaussian exponent $(\piobs - \varpi_{\rm phot} + \deltapi)^2 / \tilde{\sigma}_\varpi^2$ recovers the~\citetalias{Riess_2021} $\chi^2$ of~\cref{eq:R21_chi2}, and $\tilde{\sigma}_\varpi$ matches~\cref{eq:R21_sigma} upon identifying $\sigma_\varpi$ with $\alpha\,\sigma_{\varpi,{\rm EDR3}}$ and $\sigma_1$ with $\sigma_{m,{\rm tot}}$. We now consider the selection function term.

\subsection{Parallax selection}\label{app:parallax_selection}

For a parallax selection $\mathcal{S}(\piobs) = \Theta(\piobs - \varpi_{\min})$, modelled as a step function,
integrating the parallax likelihood against the selection gives
\begin{equation}
    \int_{\varpi_{\min}}^{\infty} {\rm d}\piobs\; \mathcal{N}\!\left(\piobs \mid v - \deltapi,\; \sigma_\varpi^2\right) = \Phi\!\left(\frac{v - \deltapi - \varpi_{\min}}{\sigma_\varpi}\right),
\end{equation}
where $v = 1/d$ is the geometric parallax.
Since $\mathcal{S}$ depends only on $\piobs$, the magnitude likelihood integrates to unity over $m^W_{H,{\rm obs}}$ for any $\sigma_1$.
Marginalising over the distance with the uniform-in-volume prior ($v^{-4} = d^2\,|{\rm d}d/{\rm d}v|$), the detection probability is
\begin{equation}\label{eq:app_sel}
    p(S = 1 \mid \deltapi) \propto \int {\rm d}v\; v^{-4}\, \Phi\!\left(\frac{v - \deltapi - \varpi_{\min}}{\sigma_\varpi}\right).
\end{equation}
The detection probability is independent of the period--luminosity relation parameters ($\MWH$, $\bW$, $\ZW$), since $M$ has dropped out entirely.
To evaluate the integral analytically, we take the sharp-cut limit $\sigma_\varpi \to 0$, in which the parallax measurement errors are assumed to have a negligible effect on the selection.
Then $\Phi \to \Theta(v - \deltapi - \varpi_{\min})$ and the integral evaluates to
\begin{equation}\label{eq:app_sel_sharp}
    p(S = 1 \mid \deltapi) \propto \int_{\varpi_{\min} + \deltapi}^{\infty} {\rm d}v\; v^{-4} = \frac{1}{3}(\varpi_{\min} + \deltapi)^{-3}.
\end{equation}
Applying this approximation to the detection probability while retaining the full parallax uncertainty in the marginal likelihood, the selection-adjusted likelihood is
\begin{equation}\label{eq:app_corrected_lin}
    \frac{\mathcal{L}^{\rm marg}}{p(S = 1 \mid \deltapi)} \propto \left(\frac{\varpi_{\min} + \deltapi}{\varpi_{\rm phot}}\right)^3 \mathcal{N}\!\left(\piobs \mid \varpi_{\rm phot} - \deltapi,\; \tilde{\sigma}_\varpi^2\right).
\end{equation}

\subsection{Magnitude selection}\label{app:mag_selection}

We now consider a selection on the Wesenheit magnitude, $\mathcal{S}(m) = \Theta(m^W_{H,\max} - m^W_{H,{\rm obs}})$, imposing a faint-end limit.
As before, because $\mathcal{S}$ depends on $m^W_{H,{\rm obs}}$ rather than $\piobs$, the parallax likelihood integrates to unity over $\piobs$.
Integrating the magnitude likelihood against the selection,
\begin{equation}
    \int_{-\infty}^{m^W_{H,\max}} {\rm d}m\; \mathcal{N}\!\left(m \mid M + \mu(d),\; \sigma_1^2\right) = \Phi\!\left(\frac{m^W_{H,\max} - M - \mu(d)}{\sigma_1}\right),
\end{equation}
so the marginalised detection probability becomes
\begin{equation}\label{eq:app_sel_mag}
    p(S = 1 \mid \bm{\theta}) \propto \int_0^{\infty} {\rm d}d\; d^2\, \Phi\!\left(\frac{m^W_{H,\max} - M - \mu(d)}{\sigma_1}\right),
\end{equation}
which, unlike the parallax selection of~\cref{eq:app_sel}, depends on the period--luminosity relation parameters through $M$ (but not $\deltapi$).
In a similar sharp-cut limit ($\sigma_1 \to 0$), the step function imposes an upper distance cut at $d_{\max}$ defined by $\mu(d_{\max}) = m^W_{H,\max} - M$, corresponding to
\begin{equation}
    \varpi_{\min}^{(m)} \equiv 1/d_{\max} = 10^{-0.2\,(m^W_{H,\max} - M - 10)}.
\end{equation}
The integral then evaluates to
\begin{equation}\label{eq:app_sel_mag_sharp}
    p(S = 1 \mid \bm{\theta}) \propto d_{\max}^3 = \left(\varpi_{\min}^{(m)}\right)^{-3}.
\end{equation}
The parallax ratio entering the selection-adjusted likelihood is
\begin{equation}\label{eq:app_mag_ratio}
\begin{split}
    \frac{\varpi_{\min}^{(m)}}{\varpi_{\rm phot}} &= \frac{10^{-0.2\,(m^W_{H,\max} - M - 10)}}{10^{-0.2\,(m^W_{H,{\rm obs}} - M - 10)}} \\
    &= 10^{-0.2\,(m^W_{H,\max} - m^W_{H,{\rm obs}})},
\end{split}
\end{equation}
which is a per-star constant independent of $\bm{\theta}$. The selection-adjusted marginal likelihood under a magnitude cut therefore reduces to
\begin{equation}\label{eq:app_corrected_mag}
    \frac{\mathcal{L}^{\rm marg}}{p(S = 1 \mid \bm{\theta})} \propto \mathcal{N}\!\left(\piobs \mid \varpi_{\rm phot} - \deltapi,\; \tilde{\sigma}_\varpi^2\right).
\end{equation}

\subsection{Summary}\label{app:similarity_summary}

The results above rest on two approximations: the linearisation of $\mu(\varpi)$, which is accurate when the intrinsic magnitude scatter $\sigma_1$ is small, and the sharp-cut evaluation of the detection probabilities, which assumes that observational errors have a negligible effect on the selection.
Had we instead adopted $\pi(d) \propto d^k$ for arbitrary $k$, the prefactor and detection probability would both scale as the $(k+1)$-th power of the geometric parallax: the magnitude-selection cancellation would remain exact, and the parallax-selection ratio would retain the same structure with exponent $k + 1$ in place of $3$.
The Galactic thin-disc prior used in the main text is not a simple power law, so the magnitude-selection cancellation is no longer exact; the ratio $\varpi_{\min}^{(m)}/\varpi_{\rm phot}$ remains a per-star constant, but the prior-dependent prefactor no longer reduces to a simple power of this ratio.
We verify in the main text that the choice of distance prior has a negligible effect on the inferred parameters once the selection is self-consistently accounted for in the forward model, because the per-star likelihood localises each distance to a narrow range over which the disc geometry varies minimally.
In~\cref{app:bias_tests} we show on mock data that for both magnitude- and parallax-selected samples, the linearised Gaussian with selection and the $\chi^2$ approach yield consistent results, with neither exhibiting significant bias in the inferred period--luminosity parameters.

For a magnitude-selected sample, the linearised Gaussian per-star likelihood differs from the $\chi^2$ only by the prefactor
\begin{equation}\label{eq:app_mag_prefactor}
    \frac{1}{\tilde{\sigma}_\varpi} = \frac{1}{\sqrt{\left(\frac{\ln 10}{5}\right)^2 \varpi_{\rm phot}^2\, \sigma_1^2 + \sigma_\varpi^2}}\,.
\end{equation}
Because the parallax measurement uncertainty dominates the error budget, with the photometric contribution $(\ln 10 / 5)\,\varpi_{\rm phot}\,\sigma_1$ being small when the intrinsic scatter $\sigma_1$ is small, we have $\tilde{\sigma}_\varpi \approx \sigma_\varpi$, a per-star constant independent of the model parameters $\bm{\theta}$. The prefactor therefore drops out of the posterior and the two methods yield identical inferences.

For a parallax-selected sample, the linearised Gaussian carries an additional factor
\begin{equation}\label{eq:app_plx_sel_factor}
   \left(\frac{\varpi_{\min} + \deltapi}{\varpi_{\rm phot}}\right)^3
   \frac{1}{\tilde{\sigma}_\varpi}\,.
\end{equation}
The $1/\tilde{\sigma}_\varpi$ term is again negligible for the same reason. The cubic ratio in principle depends on $\bm{\theta}$ through both $\varpi_{\rm phot}$ and $\deltapi$; however, since $|\deltapi| \sim 10~\uas$ is much smaller than the selection threshold $\varpi_{\min} = 800~\uas$, the numerator $\varpi_{\min} + \deltapi \approx \varpi_{\min}$ is effectively constant with respect to $\deltapi$. The ratio then reduces to $(\varpi_{\min}/\varpi_{\rm phot})^3$, which still depends on $\bm{\theta}$ through $\varpi_{\rm phot}$, but for most \ac{C27} stars the photometric parallax is close to the selection threshold ($\varpi_{\rm phot} \approx \varpi_{\min}$), driving this ratio towards unity. The entire correction factor is therefore approximately negligible.
In~\cref{app:bias_tests} we verify these conclusions on mock data, additionally testing the full forward model derived in the main text and probing the effect of higher intrinsic scatter.

\section{Validation on mock data}\label{app:bias_tests}

We validate the inference framework on mock catalogues that mimic the \ac{C22} and \ac{C27} observational campaigns separately, isolating the effect of the magnitude-limited and parallax-limited selection functions.

For each mock realisation, we generate a parent population of Cepheids as follows.
Per-star distances are drawn from a uniform-in-volume prior (adopted for simplicity in this mock example),
\begin{equation}
    \pi(d) \propto d^2, \quad d \in [0.3,\, \dmax]~\kpc,
\end{equation}
where $\dmax = 10~\kpc$ for the \ac{C22} mock and $\dmax = 2~\kpc$ for the \ac{C27} mock; in both cases $\dmax$ is much larger than the distances of any stars retained after selection, so the upper boundary does not affect the selected sample.
The parent population sizes $N_{\rm parent} = \num{2000}$ for \ac{C22} and $N_{\rm parent} = 100$ for \ac{C27} are chosen so that after selection the mock catalogues contain approximately $47$ and $26$ stars, respectively, comparable to the real samples.
True pulsation periods and metallicities are drawn from Gaussian distributions,
\begin{align}
    \log P &\hookleftarrow \mathcal{N}(\mu_{\log P},\, \sigma_{\log P}^2), \nonumber \\
    [{\rm O/H}] &\hookleftarrow \mathcal{N}(0,\, 0.15^2),
\end{align}
with $(\mu_{\log P},\, \sigma_{\log P}) = (0.8,\, 0.3)$ for \ac{C22} and $(0.75,\, 0.2)$ for \ac{C27}.
Given the sampled distances, periods, and metallicities, we compute absolute Wesenheit magnitudes from the period--luminosity relation in~\cref{eq:PL_relation} with baseline parameters $(\MWH,\, \bW,\, \ZW) = (-5.90,\, -3.30,\, -0.22)$ and intrinsic scatter $\sigma_{\rm int} = 0.06~\magn$.
Observed magnitudes and parallaxes are then drawn from the per-star likelihoods,
\begin{align}
    m^W_{H,{\rm obs}} &\hookleftarrow \mathcal{N}\!\left(m^W_H,\, \sigma_m^2 + \sigma_{\rm int}^2\right), \nonumber \\
    \piobs &\hookleftarrow \mathcal{N}\!\left(1/d_i - \deltapi,\, \sigma_\varpi^2\right),
\end{align}
with $\sigma_m = 0.028~\magn$, $\sigma_\varpi = 0.019~\mas$ (matching the median uncertainties of the \ac{MW} Cepheid catalogue), and baseline offset $\deltapi = -0.014~\mas$.

The selection function is then applied to the parent sample.
For \ac{C22}, we impose an upper apparent-magnitude limit $\mWH < 6.5~\magn$ and a lower period cut $P > 8~{\rm days}$.
For \ac{C27}, we impose a lower parallax cut with threshold $\varpi_{\min} = 0.8~\mas$ and transition width $w_\varpi = 0.05~\mas$; we verify that selecting instead on the photometric parallax with period--luminosity parameters slightly offset from the true values yields no significant difference in the bias tests.
We adopt wide uniform priors on all inferred parameters.

We compare three inference approaches applied to each mock: (i) the linearised Gaussian likelihood in~\cref{eq:app_result}, incorporating the magnitude-selection factor for \ac{C22} in~\cref{eq:app_corrected_mag} and the parallax-selection factor for \ac{C27} in~\cref{eq:app_corrected_lin}; (ii) the~\citetalias{Riess_2021} $\chi^2$ method in~\cref{eq:R21_chi2}, which does not model selection explicitly; and (iii) the full forward model derived in the main text, which samples per-star distances from the disc prior and self-consistently accounts for selection.
For each mock realisation and inference method, we record the normalised bias for each parameter $\theta$,
\begin{equation}
    b_\theta = \frac{\hat{\theta} - \theta_{\rm true}}{\sigma_\theta},
\end{equation}
where $\hat{\theta}$ and $\sigma_\theta$ are the posterior mean and standard deviation.
For an unbiased result with correct uncertainty calibration and approximately Gaussian posteriors away from prior boundaries, $b_\theta$ follows a standard normal distribution across realisations; both conditions are satisfied here.
We repeat the test at two values of the intrinsic scatter: $\sigma_{\rm int} = 0.06~\magn$, consistent with the value inferred from the real data, and an artificially inflated $\sigma_{\rm int} = 0.15~\magn$ to probe the regime in which the approximations of~\cref{app:similarity} begin to break down.

\begin{figure*}
\centering
\subcaptionbox{$\sigma_{\rm int} = 0.06~\magn$ (baseline)\label{fig:mock_bias_006}}{\includegraphics[width=\textwidth]{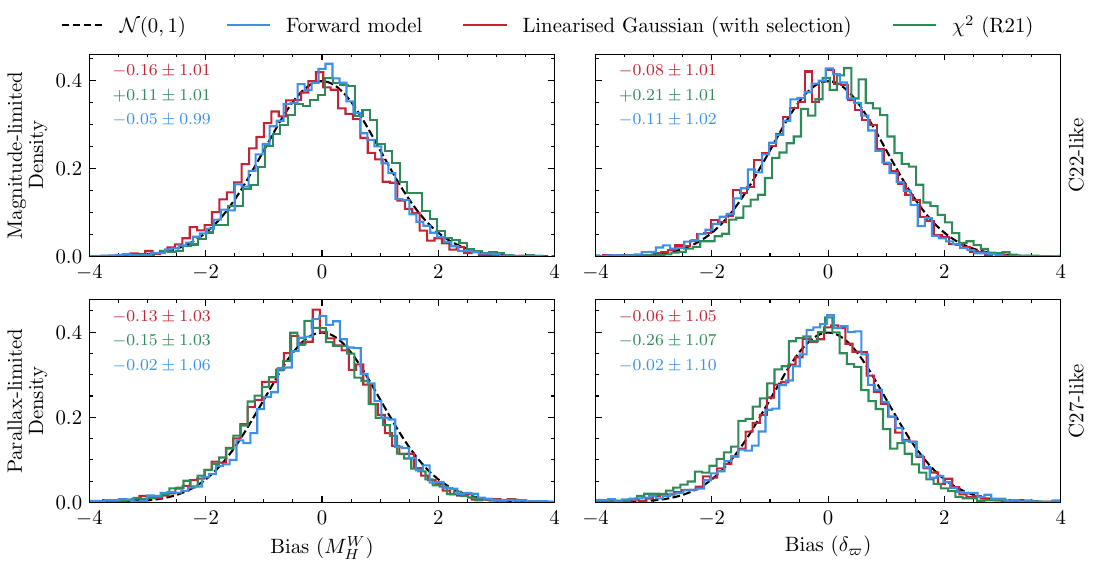}}\\[1ex]
\subcaptionbox{$\sigma_{\rm int} = 0.15~\magn$ (inflated)\label{fig:mock_bias_015}}{\includegraphics[width=\textwidth]{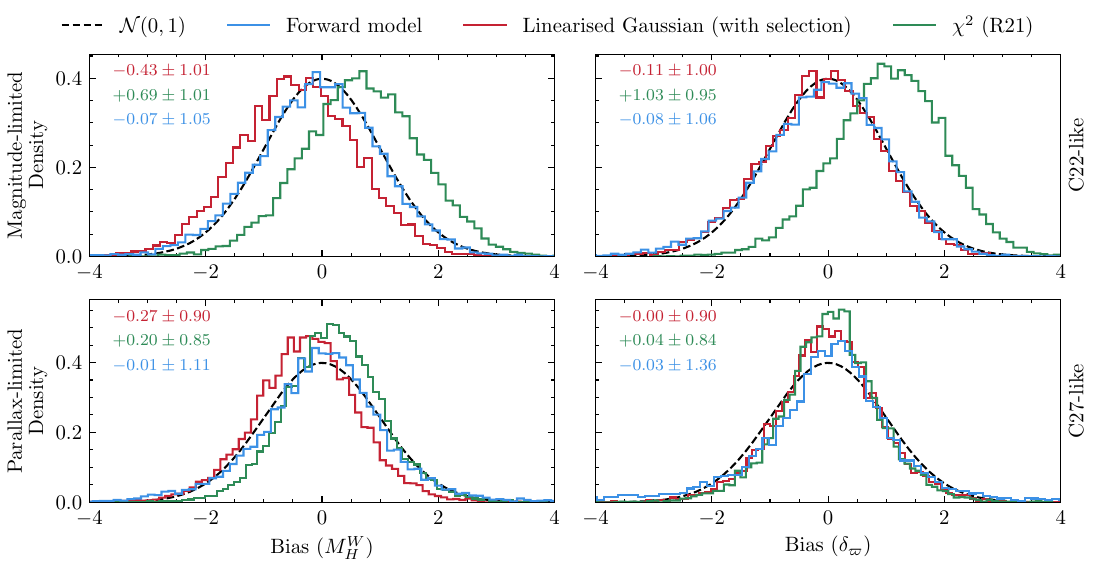}}
    \caption{Normalised bias $b_\theta = (\hat{\theta} - \theta_{\rm true}) / \sigma_\theta$ over \num{10000} mock realisations of \ac{C22} (upper rows) and \ac{C27} (lower rows), for the linearised Gaussian (red), the~\protect\citetalias{Riess_2021} $\chi^2$ (green), and the forward model (blue). The dashed curve is $\mathcal{N}(0,\,1)$. At the baseline scatter (panel~a), all three methods are approximately unbiased~(\cref{app:similarity}). At inflated scatter (panel~b), the $\chi^2$ method develops significant biases in $\MWH$ and $\deltapi$ for \ac{C22}, while the other two remain well calibrated. Annotations give the mean and standard deviation of $b_\theta$.
    }
\label{fig:mock_bias}
\end{figure*}

\Cref{fig:mock_bias} shows the distribution of $b_\theta$ for all inferred parameters over \num{10000} mock realisations.
At the baseline scatter (\cref{fig:mock_bias_006}), all three methods recover the input parameters without significant bias for both \ac{C22} and \ac{C27}, with the $\chi^2$ method showing only a modest $0.26\sigma$ bias in $\deltapi$ for \ac{C27}, consistent with the parallax truncation effect derived in~\cref{app:parallax_selection}.
The slope $\bW$ is similarly unbiased across all configurations.
The metallicity coefficient $\ZW$ shows a ${\sim}\,0.3\sigma$ bias under both the $\chi^2$ and linearised Gaussian methods, because neither accounts for the intrinsic scatter of the metallicity values; the forward model, which does, recovers $\ZW$ without bias.

At the inflated scatter (\cref{fig:mock_bias_015}), the $\chi^2$ method develops mild biases for \ac{C22}: $0.69\sigma$ in $\MWH$ and $1.03\sigma$ in $\deltapi$, while the linearised Gaussian shows a smaller $0.43\sigma$ bias in $\MWH$.
The forward model remains well calibrated throughout.
This demonstrates that, at the baseline scatter of the real data, the approximations underlying the $\chi^2$ and linearised Gaussian methods are reasonable, but would break down were the intrinsic scatter substantially larger.

\section{Posterior robustness checks}\label{app:robustness_checks}

\begin{figure}
\centering
\includegraphics[width=\columnwidth]{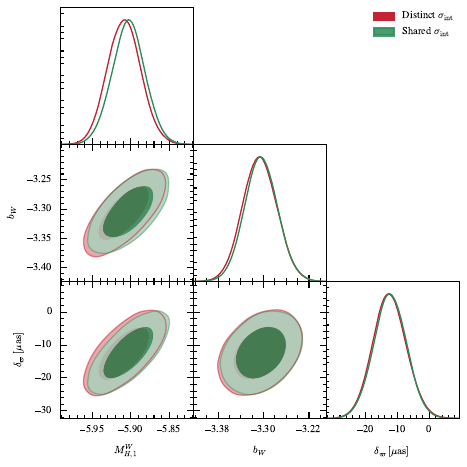}
\caption{Robustness of the period--luminosity calibration to the intrinsic-scatter parametrisation. Red shows the baseline forward model with independent \ac{C22} and \ac{C27} intrinsic scatters, green the variant with a single shared \ac{MW} scatter; both inferences use \ac{C22}\,+\,\ac{C27}\,+\,\ac{LMC}\,+\,N4258 with the disc distance prior and the baseline selection model. Posterior-mean shifts are $0.006~\magn$ in $\MWH$, $0.002$ in $\bW$, and $0.2~\uas$ in $\deltapi$, all small compared to the marginal uncertainties.}
\label{fig:robustness}
\end{figure}

\Cref{fig:robustness} compares the baseline posterior---with separate per-campaign \ac{MW} intrinsic scatters $\sigma_{\rm int}^{\rm C22} = 0.071 \pm 0.022~\magn$ and $\sigma_{\rm int}^{\rm C27} = 0.041 \pm 0.015~\magn$~(\cref{sec:baseline})---with the variant that collapses these into a single shared \ac{MW} value.
The posteriors of $\MWH$, $\bW$, and $\deltapi$ are nearly unchanged: the posterior-mean shifts are $0.006~\magn$ in $\MWH$, $0.002$ in $\bW$, and $0.2~\uas$ in $\deltapi$, corresponding to $0.19$, $0.06$, and $0.02$ times the combined posterior standard deviation, respectively.
The period--luminosity calibration is therefore insensitive to this choice of intrinsic-scatter parametrisation.

\begin{figure*}
\centering
\includegraphics[width=\textwidth]{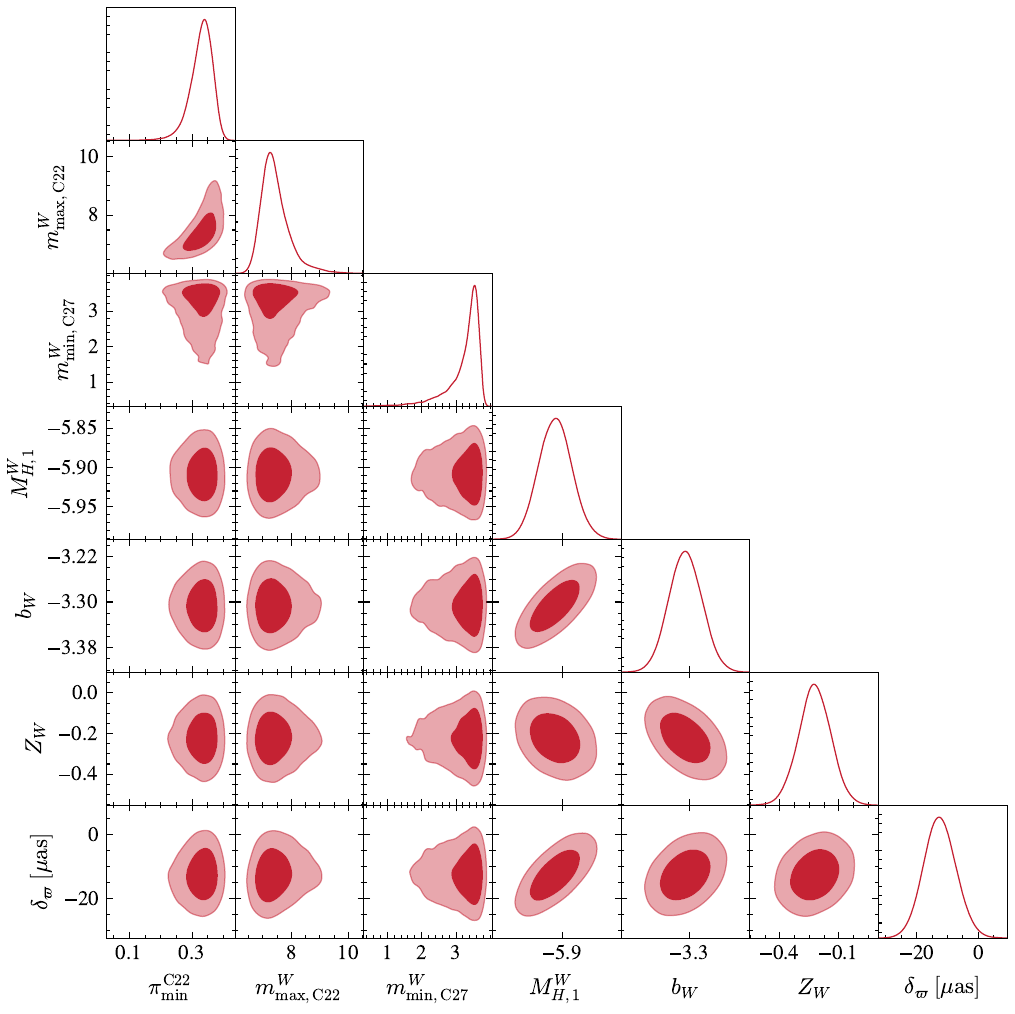}
\caption{Joint posterior of the inferred selection thresholds and the period--luminosity parameters in the baseline forward model (\ac{C22}\,+\,\ac{C27}\,+\,\ac{LMC}\,+\,N4258), extending the red contour of~\cref{fig:corner_baseline}. The parallax offset $\deltapi$ is in $\rm \mu as$; contours show the $1\sigma$ and $2\sigma$ credible regions.}
\label{fig:corner_thresholds}
\end{figure*}

\Cref{fig:corner_thresholds} extends the baseline contour of~\cref{fig:corner_baseline} (\ac{C22}\,+\,\ac{C27}\,+\,\ac{LMC}\,+\,N4258 with the disc prior and selection modelling) by also showing the inferred selection thresholds jointly with the period--luminosity parameters.
The thresholds are only weakly correlated with the physical calibration: the largest linear (Pearson) correlation is $\rho = 0.11$ between the \ac{C22} magnitude cut and $\deltapi$, followed by $\rho = 0.08$ between the \ac{C22} parallax cut and $\deltapi$, while all remaining threshold--calibration coefficients have $|\rho| < 0.03$, though some of the posteriors are non-Gaussian.
The inferred period--luminosity parameters are therefore at most weakly degenerate with the selection thresholds.

\section{Spiral arm modulation of the distance prior}\label{app:spiral_arms}

\ac{MW} Cepheids are young stars that trace the Galactic spiral structure.
The axisymmetric disc prior of~\cref{eq:disc_prior} does not capture this azimuthal dependence and may therefore assign non-negligible probability to inter-arm regions where few Cepheids reside.
We optionally modulate the disc prior with a spiral arm density profile to test whether accounting for this structure affects the inferred parameters.

We adopt the four-arm log-periodic spiral model of~\citet{Drimmel_2025}, derived from \num{2857} classical Cepheids with WISE mid-infrared distances~\citep{Skowron_2025}.
The four arms---Scutum, Sagittarius--Carina, Local (Orion), and Perseus---are each described by a trace in the Galactic plane, parametrised as
\begin{equation}\label{eq:log_spiral}
    \ln R = \ln R_{0,k} - \tan \psi_k \, \phi,
\end{equation}
where $R$ is the Galactocentric radius, $\phi$ the azimuthal angle, $R_{0,k}$ the reference radius, and $\psi_k$ the pitch angle of the $k$\textsuperscript{th} arm.

The spiral-modulated prior is the joint position density of~\cref{eq:disc_prior} scaled by a factor that enhances the density near the arm traces:
\begin{equation}\label{eq:spiral_prior}
    \pi(d_i,\, \ell_i,\, b_i) \propto \pi_{\rm disc}(d_i,\, \ell_i,\, b_i)\; \mathcal{S}_{\rm arm}(d_i,\, \ell_i,\, b_i),
\end{equation}
where the spiral modulation factor is
\begin{equation}\label{eq:spiral_factor}
    \mathcal{S}_{\rm arm} = (1 - f_{\rm arm}) + f_{\rm arm} \sum_{k=1}^{4} \exp\!\left(-\frac{\Delta_k^2}{2\sigma_{\rm arm}^2}\right),
\end{equation}
$\pi_{\rm disc}$ is the joint position density of~\cref{eq:disc_prior}, $\Delta_k(d_i,\, \ell_i,\, b_i)$ is the projected distance in the Galactic plane from the Cepheid to the nearest point on the $k$\textsuperscript{th} arm trace, $f_{\rm arm} \in [0,\, 1]$ is the fraction of the Cepheid surface density attributed to spiral arms, and $\sigma_{\rm arm}$ is the Gaussian arm width.
When $f_{\rm arm} = 0$, the spiral factor reduces to unity and the axisymmetric disc prior is recovered.
At the opposite limit $f_{\rm arm} = 1$, the prior is concentrated entirely on the arm traces.
Both $f_{\rm arm}$ and $\sigma_{\rm arm}$ are sampled as free parameters: $f_{\rm arm}$ with a uniform prior on $[0,\, 1]$ and $\sigma_{\rm arm}$ with a half-normal prior $\mathcal{N}^+(0.3,\, 0.2^2)~\kpc$, truncated at zero.

The nearest-point distances $\Delta_k$ are computed by converting the heliocentric coordinates $(d_i,\, \ell_i,\, b_i)$ to Galactocentric Cartesian coordinates $(x_{{\rm GC}},\, y_{{\rm GC}})$ in the midplane and querying a KD-tree built from the densified arm traces of~\citet{Drimmel_2025}.
Because the arm traces are fixed, the squared distances $\Delta_k^2$ for each sightline on a fine distance grid are precomputed and cached; at inference time, the spiral factor is evaluated by interpolating $\Delta_k^2$ to the sampled distance $d_i$.
The prior of~\cref{eq:spiral_prior} is normalised numerically per sightline via Simpson's rule on the same grid.
An analogous precomputation is performed for the Monte Carlo sightlines entering the detection probability of~\cref{eq:prob_detection_mc}, ensuring that the spiral modulation propagates consistently into both the per-star prior and the selection modelling.
As reported in~\cref{tab:results}, the spiral-arm-modulated prior leaves the posteriors virtually unchanged relative to the axisymmetric disc prior.

\section{Per-star distance comparison}\label{app:distance_comparison}

\begin{figure*}
    \centering
    \includegraphics[width=\textwidth]{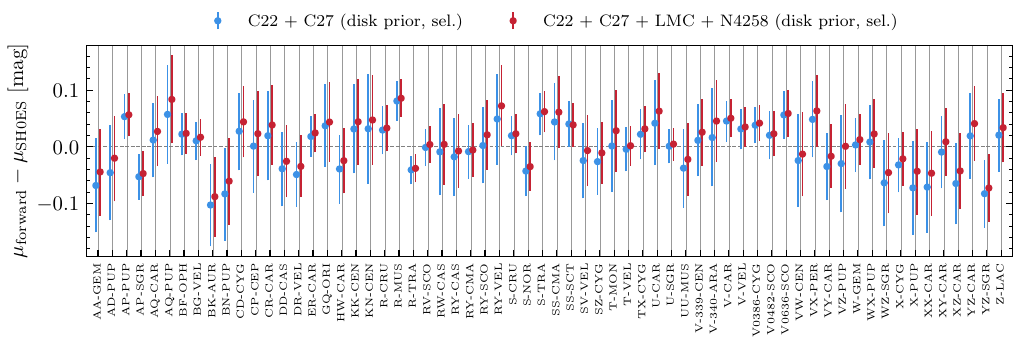}
    \caption{Per-star distance modulus residuals $\langle \mu_{\rm forward} \rangle - \mu_{\rm SH0ES}$ for each \ac{MW} Cepheid, comparing the \ac{MW}-only model (blue) and the baseline model including the \ac{LMC} and N4258 (red).
    The reference $\mu_{\rm SH0ES}$ is computed from the observed Wesenheit magnitude and the baseline SH0ES period--luminosity parameters.
    Error bars span the 16th--84th percentiles of the posterior.
    Stars are sorted alphabetically by name; vertical grid lines indicate individual Cepheids.}
    \label{fig:distance_comparison}
\end{figure*}

The forward model samples a distance $d_i$ for each \ac{MW} Cepheid. Converting these to distance moduli $\mu_{{\rm forward},i} = 5\log(d_i / \kpc) + 10$ provides a direct comparison with the photometric distance moduli $\mu_{\rm SH0ES}$ predicted from the baseline SH0ES period--luminosity parameters ($\MWH = -5.894~\magn$, $\bW = -3.299$, $\ZW = -0.217$).
\Cref{fig:distance_comparison} shows the per-star residuals $\langle \mu_{{\rm forward}} \rangle - \mu_{\rm SH0ES}$ for both the \ac{MW}-only and baseline models.
The residuals scatter about zero with only a minor systematic offset.
Including the \ac{LMC} and N4258 tightens the period--luminosity relation but does not shift the per-star distance posteriors.

\begin{figure}
    \centering
    \includegraphics[width=\columnwidth]{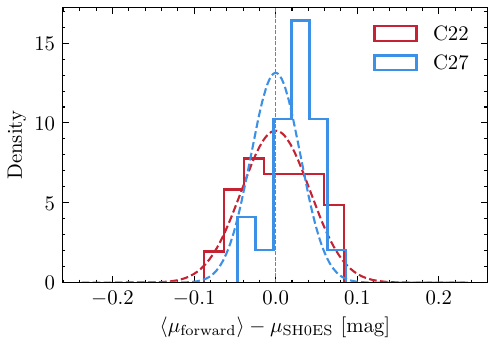}
    \caption{Distribution of per-star posterior-mean distance modulus residuals $\langle \mu_{\rm forward} \rangle - \mu_{\rm SH0ES}$ for \ac{C22} (red) and \ac{C27} (blue) from the baseline model.
    Dashed curves show zero-mean Gaussians with standard deviations matched to each campaign.}
    \label{fig:distance_residual_hist}
\end{figure}

\Cref{fig:distance_residual_hist} shows the distribution of the posterior-mean residuals from the baseline \ac{C22}\,+\,\ac{C27}\,+\,\ac{LMC}\,+\,N4258 inference, split by campaign.
The \ac{C22} residuals have a mean offset of $+0.004~\magn$ with a standard deviation of $0.042~\magn$, while the \ac{C27} residuals have a mean of $+0.025~\magn$ and a standard deviation of $0.030~\magn$.
The positive \ac{C27} offset is expected: the reference $\mu_{\rm SH0ES}$ is computed from the baseline SH0ES zero-point ($\MWH = -5.894~\magn$), which is $0.015~\magn$ fainter than our baseline value ($\MWH = -5.909~\magn$), predicting systematically shorter SH0ES distances and hence positive residuals.

\bsp
\label{lastpage}
\end{document}